\pdfoutput=1
\documentclass[%
 reprint,
 amsmath,amssymb,
 aps,
]{revtex4-1}

\usepackage{multirow}
\usepackage{graphicx}
\usepackage{dcolumn}
\usepackage{bm}
\usepackage{hhline}
\usepackage{subfigure}
\usepackage{xcolor}
\usepackage{hyperref}

\allowdisplaybreaks

\begin{document}


\title{Competing orders in pyrochlore magnets from a $\mathbb{Z}_2$ spin liquid perspective}

\author{Chunxiao Liu}
\affiliation{Department of Physics, University of California, Santa
Barbara, CA 93106-9530, USA}
\author{G\'abor B. Hal\'asz}
\affiliation{Materials Science and Technology Division, Oak Ridge
National Laboratory, Oak Ridge, TN 37831, USA} \affiliation{Kavli
Institute for Theoretical Physics, University of California, Santa
Barbara, CA 93106-4030, USA}
\author{Leon Balents}
\affiliation{Kavli Institute for Theoretical Physics, University of
  California, Santa Barbara, CA 93106-4030, USA}
\affiliation{Canadian Institute for Advanced Research, 661 University Ave., Toronto, ON M5G 1M1 Canada}





\date{\today}

\begin{abstract}

The pyrochlore materials have long been predicted to harbor a
quantum spin liquid, an intrinsic long-range-entangled state
supporting fractionalized excitations. Existing pyrochlore
experiments, on the other hand, have discovered several weakly
ordered states and a tendency of close competition amongst them.
Motivated by these facts, we give a complete classification of
spin-orbit-coupled $\mathbb{Z}_2$ spin-liquid states on the
pyrochlore lattice by using the projective symmetry group (PSG)
approach for bosonic spinons. For each spin liquid, we construct a
mean-field Hamiltonian that can be used to describe phase
transitions out of the spin liquid via spinon condensation. Studying
these phase transitions, we establish phase diagrams for our
mean-field Hamiltonians that link magnetic orders to specific spin
liquids. In general, we find that seemingly unrelated magnetic
orders are intertwined with each other and that the conventional
spin orders seen in the experiments are accompanied by more exotic
hidden orders. Our critical theories are categorized into $z=1$ and
$z=2$ types, based on their spinon dispersion and Hamiltonian
diagonalizability, and are shown to give distinct signatures in the
heat capacity and the spin structure factor. This study provides a
clear map of pyrochlore phases for future experiments and
variational Monte Carlo studies in pyrochlore materials.

\end{abstract}

\pacs{Valid PACS appear here}
\maketitle


\section{Introduction}

Quantum spin liquids (QSLs) \cite{savary2016quantum} are
zero-temperature phases of interacting spin systems which possess
intrinsic long-range entanglement and support nonlocal excitations
carrying fractionalized quantum numbers.  Typically, they respect
all symmetries of the underlying lattice, i.e., they exhibit a lack
of conventional symmetry-breaking order.  The theoretical
understanding of QSLs is largely in terms of {\em emergent} gauge
theory, which provides a convenient mathematical framework to
describe long-range entanglement, along with the nonlocal nature of
the fractionalized excitations.

In frustrated magnetic systems \cite{balents2010spin}, QSL ground
states may control the physics even at (small) finite temperatures,
as long as energy dominates over entropy.  For two-dimensional spin
liquids, this statement is purely asymptotic; at any nonzero
temperature $T>0$, the putative QSL is adiabatically connected to a
high-temperature paramagnet.  However, some three-dimensional spin
liquids, particularly the so-called $\mathbb{Z}_2$ states with
Ising-like emergent gauge fields, are more robust, and can persist
in the form of a distinct low-temperature phase up to a nonzero
critical temperature.

While QSLs are extremely interesting from a conceptual perspective,
it is far from obvious to realize them in experimental materials, or
even realistic spin Hamiltonians.  Traditionally, most studies
considered spin-rotation-invariant Heisenberg systems on
geometrically frustrated two-dimensional lattices.  However, it has
recently been recognized that magnetic systems with strong
spin-orbit coupling provide a promising alternative avenue to QSLs
\cite{witczak2014correlated, gingras2014quantum, rau2016spin,
SciPostPhys.4.1.003}.  In general, these systems have a large number
of magnetically anisotropic terms, leading to exchange frustration
as well as an extended parameter space, and are thus expected to
harbor QSL ground states on a wide range of two- and
three-dimensional lattices.

The most widely studied such three-dimensional structure is the
pyrochlore lattice, consisting of periodically arranged
corner-sharing tetrahedra.  Experimentally, two large families of
materials, the pyrochlore spinels and the rare-earth pyrochlores,
provide vast real-world possibilities \cite{gardner2010magnetic} to
test theoretical predictions on the pyrochlore lattice.  In the
2000s, it was predicted that certain antiferromagnetic pyrochlore
models could support a U(1) QSL phase \cite{hermele2004pyrochlore}
[the ``U(1)'' means that the gauge field belongs to the Lie algebra
of the U(1) group and that the emergent charges are characterized by
integers related to the generating charge of U(1)], which is a
simulacrum of electromagnetic gauge theory in high-energy physics.
In 2011/2012, theoretical applications of this idea to realistic
models emerged, suggesting the presence of a U(1) spin liquid in the
so-called ``quantum spin ice'' pyrochlore materials
\cite{ross2011quantum, savary2012coulombic}.  So far, these
predictions remain to be confirmed in experiments, even though there
are some promising recent developments \cite{Hirschberger106,hirschberger2019enhanced,gaudet2019quantum,
PhysRevB.94.024436, PhysRevLett.115.097202, 
PhysRevLett.119.127201,PhysRevLett.119.057203,
sibille18:_exper_pr2hf,
doi:10.7566/JPSJ.87.064702}.

Another thread recurring in the experimental study of rare-earth
pyrochlores is the close competition amongst several weakly ordered
states \cite{hallas2017experimental}.  Several hints at this
competition are present in the family of Yb pyrochlores,
Yb$_2$B$_2$O$_7$, which have a systematic structural evolution
across the series B = Ge, Ti, Pt, Sn.  While the germanate orders
antiferromagnetically, the remaining members of the family have
ferromagnetic ground states, suggesting the close proximity of at
least these two phases.  In each material, the specific heat is
peaked at a temperature of 2-4K, while the maximum ordering
temperature is 0.6K in the germanate and half or less than that in
the rest of the family.  These findings indicate the onset of strong
spin correlations well above the ordering temperature, but an
inability of the system to decide upon its ground state.  The weak
ferromagnetic ground state in Yb$_2$Ti$_2$O$_7$ is also famously
mercurial, changing its character substantially with sample
variations \cite{bowman2019role}.  Theoretically, a classical analysis indeed finds close
competition amongst several distinct phases
\cite{PhysRevB.95.094422}, but a quantum picture of this phase
competition is not yet available.

In this work, we combine the two threads of phase competition and
QSL physics by utilizing the connection of symmetry to emergent
gauge structure.  This connection is mathematically described by the
projective symmetry group (PSG), proposed by Wen in 2002
\cite{wen2002quantum}, which encapsulates the fact that, in a QSL,
the group operations of the physical symmetry group are interleaved
with those of the emergent gauge group.  The embedding of the
physical symmetries into the PSG can then lead to a unification of
distinct symmetry-breaking orders that are unrelated in classical
physics. Such a unified description of seemingly unrelated magnetic
orders is the main motivation behind the present study.

The PSG also offers a straightforward method to classify QSLs in the
presence of symmetry. Concretely, the PSG specifies a distinct set
of transformation rules for the emergent matter and gauge fields in
each QSL phase, corresponding to a given PSG class. Employing the
PSG method, an entire zoo of QSLs has been found on the square
\cite{reuther2014classification}, triangular \cite{lu2016symmetric},
kagome \cite{lu2011z}, honeycomb \cite{you2012doping}, star
\cite{choy2009classification}, and hyperkagome
\cite{PhysRevB.95.054404} lattices, to give a few notable examples.
Generally, these QSLs can be connected to magnetically ordered
states by considering the condensation patterns that emerge when the
energy of a bosonic QSL excitation is brought to zero
\cite{bergman2006ordering, PhysRevB.94.205107, PhysRevB.95.041106}.

In this paper, we employ the PSG method to obtain a full
classification of QSLs with $\mathbb{Z}_2$ gauge structure on the
pyrochlore lattice using Schwinger bosons \cite{sachdev1992kagome,
PhysRevB.74.174423, PhysRevB.82.024419, PhysRevB.94.035160}.  While
standard parton constructions also allow U(1) and SU(2) gauge
structures, we consider the $\mathbb{Z}_2$ gauge structure for two
reasons.  First, it is the simplest one: quasiparticles in a
$\mathbb{Z}_2$ QSL are weakly interacting because the gauge field
itself is gapped. Second, it is also the richest one: a single U(1)
PSG class can be further split into several $\mathbb{Z}_2$ PSG
classes if the gauge symmetry is lowered from U(1) to
$\mathbb{Z}_2$.  We use Schwinger bosons rather than Abrikosov
fermions \cite{PhysRevB.99.134425} to immediately obtain a bosonic excitation, the elementary
Schwinger boson itself, that can condense at the phase transition
out of the QSL.

As a result of our PSG analysis, we find $16$ different
$\mathbb{Z}_2$ QSLs on the pyrochlore lattice.  We use a standard
mean-field description to study the $0$-flux QSLs, in which
translation symmetry acts linearly (i.e., as in classical physics)
on the Schwinger bosons. The PSG method also allows us to describe
phase transitions from these QSLs to magnetically ordered phases.
Condensing the Schwinger bosons, we identify $15$ different ordering
patterns, and call them ``paraphases'', since each of them actually
unifies several distinct symmetry-breaking orders.  We find that,
generically, these orders are intertwined, necessarily appearing
together at the phase transition out of the QSL, and that
conventional spin orders are in many cases accompanied by
inversion-breaking ``hidden'' orders.

The phase transitions corresponding to these $15$ paraphases fall
into two dynamical classes of $z=1$ and $z=2$ quantum criticality,
exhibiting critical modes with linear and quadratic dispersions,
respectively. We uncover the mathematical structure discriminating
between these two classes, related to Hamiltonian diagonalizability,
and derive their effective field theories, along with their most
important experimental signatures. In particular, we use mean-field
theory to compute static and dynamic spin structure factors for each
of the $15$ paraphases. Finally, by comparing the magnetic orders
associated with each paraphase to those observed in experiments, we
identify a set of likely QSL phases that might be relevant to
real-world pyrochlore materials.

The rest of the paper is organized as follows. First, in
Sec.~\ref{sec:MR}, we summarize our main results on the different
QSL phases and the corresponding phase transitions out of them
(``paraphases''). In Sec.~\ref{sec:level1}, we employ the PSG
method, deriving the PSG classes, and constructing a mean-field
theory for each PSG class. In Sec.~\ref{secMF}, we analyze the
mean-field theories of our QSL phases, describing phase transitions
out of them, and establishing the two dynamical classes with
critical exponents $z=1,2$. In Sec.~\ref{sec:exp}, we move on to the
experimental signatures of our phase transitions, describing the
heat capacity and the spin structure factors, and also introducing
the concept of intertwined and hidden orders. Finally, in
Sec.~\ref{sec:discussion}, we discuss our results and connect them
to existing experimental data. Detailed derivations and lengthy
formulas are given in the Appendices for reference.

\section{Main Results\label{sec:MR}}

\begin{table*}[t]
\begin{tabular*}{0.99\textwidth}{@{\extracolsep{\fill}} c | c | c | c | c | c | c | c | c | c }
\hline \hline
Spin-                         & \multicolumn{4}{c |}{Critical ``paraphases''}                                                                                                                                                                 & \multicolumn{5}{c}{Magnetically ordered phases}                                                                       \\
\cline{2-10}
liquid                        & Condensation                & Dynamical                 & Heat capacity:                      & \multirow{2}{*}{Dynamic spin structure factor}                                                                & \multicolumn{4}{c |}{Spin orders}                                                             & Hidden                \\
\cline{6-9}
phases                        & momenta                     & exponent                  & $C_V \propto T^x$                   &                                                                                                               & AIAO                  & AFM                   & FM                    & PC                    & orders                \\
\hline
\multirow{6}{*}{$0$-$(001)$}  & \multirow{2}{*}{$\Gamma$}   & \multirow{2}{*}{$z = 2$}  & \multirow{2}{*}{$x = \frac{3}{2}$}  & Gapless at $\Gamma$                                                                                           & \multirow{2}{*}{$0$}  & \multirow{2}{*}{$0$}  & \multirow{2}{*}{$+$}  & \multirow{2}{*}{$0$}  & \multirow{2}{*}{$+$}  \\
                              &                             &                           &                                     & Weak in the low-energy limit                                                                                  &                       &                       &                       &                       &                       \\
\cline{2-10}
                              & \multirow{2}{*}{$\mathrm{L}$}    & \multirow{2}{*}{$z = 2$}  & \multirow{2}{*}{$x = \frac{3}{2}$}  & Gapless at $\Gamma$ and $\mathrm{X}$                                                                             & \multirow{2}{*}{$+$}  & \multirow{2}{*}{$+$}  & \multirow{2}{*}{$+$}  & \multirow{2}{*}{$+$}  & \multirow{2}{*}{$+$}  \\
                              &                             &                           &                                     & Weak in the low-energy limit                                                                                  &                       &                       &                       &                       &                       \\
\cline{2-10}
                              & \multirow{2}{*}{$\Lambda$}  & \multirow{2}{*}{$z = 2$}  & \multirow{2}{*}{$x = 1$}            & Gapless along $\Gamma \rightarrow \mathrm X$ and $\mathrm K \rightarrow \Gamma \rightarrow \mathrm L \rightarrow \mathrm U$                   & \multicolumn{4}{c |}{\multirow{2}{*}{Unclear at NN level}}                                    & \multirow{2}{*}{$+$}  \\
                              &                             &                           &                                     & Low-energy weight at all momenta                                                                              & \multicolumn{4}{c |}{}                                                                        &                       \\
\hline
\multirow{4}{*}{$0$-$(010)$}  & \multirow{2}{*}{$\Gamma$}   & \multirow{2}{*}{$z = 2$}  & \multirow{2}{*}{$x = \frac{3}{2}$}  & Gapless at $\Gamma$                                                                                           & \multirow{2}{*}{$0$}  & \multirow{2}{*}{$0$}  & \multirow{2}{*}{$+$}  & \multirow{2}{*}{$+$}  & \multirow{2}{*}{$0$}  \\
                              &                             &                           &                                     & Weak in the low-energy limit                                                                                  &                       &                       &                       &                       &                       \\
\cline{2-10}
                              & \multirow{2}{*}{$\Lambda$}  & \multirow{2}{*}{$z = 1$}  & \multirow{2}{*}{$x = 2$}            & \multirow{2}{*}{Gapless along $\Gamma \rightarrow \mathrm X$ and $\mathrm K \rightarrow \Gamma \rightarrow \mathrm L \rightarrow \mathrm U$}  & \multicolumn{4}{c |}{\multirow{2}{*}{Unclear at NN level}}                                    & \multirow{2}{*}{$0$}  \\
                              &                             &                           &                                     &                                                                                                               & \multicolumn{4}{c |}{}                                                                        &                       \\
\hline
\multirow{4}{*}{$0$-$(100)$}  & \multirow{2}{*}{$\Gamma$}   & \multirow{2}{*}{$z = 2$}  & \multirow{2}{*}{$x = \frac{3}{2}$}  & Gapless at $\Gamma$                                                                                           & \multirow{2}{*}{$+$}  & \multirow{2}{*}{$+$}  & \multirow{2}{*}{$0$}  & \multirow{2}{*}{$0$}  & \multirow{2}{*}{$0$}  \\
                              &                             &                           &                                     & Weak in the low-energy limit                                                                                  &                       &                       &                       &                       &                       \\
\cline{2-10}
                              & \multirow{2}{*}{$\Lambda$}  & \multirow{2}{*}{$z = 1$}  & \multirow{2}{*}{$x = 2$}            & \multirow{2}{*}{Gapless along $\Gamma \rightarrow \mathrm X$ and $\mathrm K \rightarrow \Gamma \rightarrow \mathrm L \rightarrow \mathrm U$}  & \multicolumn{4}{c |}{\multirow{2}{*}{Unclear at NN level}}                                    & \multirow{2}{*}{$0$}  \\
                              &                             &                           &                                     &                                                                                                               & \multicolumn{4}{c |}{}                                                                        &                       \\
\hline
\multirow{6}{*}{$0$-$(101)$}  & \multirow{2}{*}{$\Gamma$}   & \multirow{2}{*}{$z = 2$}  & \multirow{2}{*}{$x = \frac{3}{2}$}  & Gapless at $\Gamma$                                                                                           & \multirow{2}{*}{$+$}  & \multirow{2}{*}{$+$}  & \multirow{2}{*}{$0$}  & \multirow{2}{*}{$0$}  & \multirow{2}{*}{$+$}  \\
                              &                             &                           &                                     & Weak in the low-energy limit                                                                                  &                       &                       &                       &                       &                       \\
\cline{2-10}
                              & \multirow{2}{*}{$\mathrm{W}$}    & \multirow{2}{*}{$z = 1$}  & \multirow{2}{*}{$x = 3$}            & Gapless at $\Gamma$, $\mathrm{X}$, and $\frac{2}{3} \mathrm K$                                                                 & \multirow{2}{*}{$+$}  & \multirow{2}{*}{$+$}  & \multirow{2}{*}{$+$}  & \multirow{2}{*}{$+$}  & \multirow{2}{*}{$+$}  \\
                              &                             &                           &                                     & Singular in the low-energy limit                                                                              &                       &                       &                       &                       &                       \\
\cline{2-10}
                              & \multirow{2}{*}{$\mathrm{X}$}    & \multirow{2}{*}{$z = 1$}  & \multirow{2}{*}{$x = 3$}            & Gapless at $\Gamma$ and $\mathrm{X}$                                                                                   & \multirow{2}{*}{$+$}  & \multirow{2}{*}{$+$}  & \multirow{2}{*}{$0$}  & \multirow{2}{*}{$+$}  & \multirow{2}{*}{$+$}  \\
                              &                             &                           &                                     & Singular in the low-energy limit                                                                              &                       &                       &                       &                       &                       \\
\hline
\multirow{4}{*}{$0$-$(110)$}  & \multirow{2}{*}{$\Gamma$}   & \multirow{2}{*}{$z = 1$}  & \multirow{2}{*}{$x = 3$}            & Gapless at $\Gamma$                                                                                           & \multirow{2}{*}{$0$}  & \multirow{2}{*}{$+$}  & \multirow{2}{*}{$0$}  & \multirow{2}{*}{$0$}  & \multirow{2}{*}{$0$}  \\
                              &                             &                           &                                     & Characteristic lower edge of the spectrum                                                                     &                       &                       &                       &                       &                       \\
\cline{2-10}
                              & \multirow{2}{*}{$\Lambda$}  & \multirow{2}{*}{$z = 1$}  & \multirow{2}{*}{$x = 2$}            & \multirow{2}{*}{Gapless along $\Gamma \rightarrow \mathrm X$ and $\mathrm K \rightarrow \Gamma \rightarrow \mathrm L \rightarrow \mathrm U$}  & \multicolumn{4}{c |}{\multirow{2}{*}{Unclear at NN level}}                                    & \multirow{2}{*}{$0$}  \\
                              &                             &                           &                                     &                                                                                                               & \multicolumn{4}{c |}{}                                                                        &                       \\
\hline
\multirow{6}{*}{$0$-$(111)$}  & \multirow{2}{*}{$\Gamma$}   & \multirow{2}{*}{$z = 2$}  & \multirow{2}{*}{$x = \frac{3}{2}$}  & Gapless at $\Gamma$                                                                                           & \multirow{2}{*}{$+$}  & \multirow{2}{*}{$+$}  & \multirow{2}{*}{$0$}  & \multirow{2}{*}{$0$}  & \multirow{2}{*}{$+$}  \\
                              &                             &                           &                                     & Weak in the low-energy limit                                                                                  &                       &                       &                       &                       &                       \\
\cline{2-10}
                              & \multirow{2}{*}{$\mathrm{W}$}    & \multirow{2}{*}{$z = 1$}  & \multirow{2}{*}{$x = 3$}            & Gapless at $\Gamma$, $\mathrm{X}$, and $\frac{2}{3} \mathrm K$                                                                 & \multirow{2}{*}{$+$}  & \multirow{2}{*}{$+$}  & \multirow{2}{*}{$+$}  & \multirow{2}{*}{$+$}  & \multirow{2}{*}{$+$}  \\
                              &                             &                           &                                     & Singular in the low-energy limit                                                                              &                       &                       &                       &                       &                       \\
\cline{2-10}
                              & \multirow{2}{*}{$\mathrm{X}$}    & \multirow{2}{*}{$z = 1$}  & \multirow{2}{*}{$x = 3$}            & Gapless at $\Gamma$ and $\mathrm{X}$                                                                                   & \multirow{2}{*}{$+$}  & \multirow{2}{*}{$+$}  & \multirow{2}{*}{$+$}  & \multirow{2}{*}{$0$}  & \multirow{2}{*}{$+$}  \\
                              &                             &                           &                                     & Singular in the low-energy limit                                                                              &                       &                       &                       &                       &                       \\
\hline \hline
\end{tabular*}
\caption{Most important characteristics of the $15$ critical
``paraphases'' corresponding to phase transitions between the six
$0$-flux $\mathbb{Z}_2$ spin-liquid phases [labeled as
$0$-$(n_{\overline{C}_6S} \, n_{ST_1} \, n_{\overline{C}_6})$] and a
rich variety of magnetically ordered phases. Each paraphase is
labeled by the condensation momenta (see Table IV for notation)
where the spinons become gapless and condense at the phase
transition. For each critical theory, experimental signatures are
given in terms of the dynamical critical exponent, the
low-temperature behavior of the heat capacity, and the low-energy
features of the dynamic spin structure factor (see
Fig.~\ref{DSF_NN_1En9_shifted}). For each magnetically ordered phase
obtained by spinon condensation, it is specified whether various
orders are generically present ($+$) or absent ($0$), including
conventional spin orders, such as all-in-all-out (AIAO) order, XY
antiferromagnetic (AFM) order [$\Psi_{2,3}$], ferromagnetic (FM)
order, and Palmer-Chalker (PC) order [$\Psi_4$], as well as
inversion-breaking ``hidden'' orders. \label{six0fluxclasses}}
\end{table*}

From our PSG classification scheme, we find that there are $16$
different $\mathbb{Z}_2$ PSG classes of Schwinger bosons,
corresponding to $16$ inequivalent $\mathbb{Z}_2$ QSL phases, on the
pyrochlore lattice. Out of these $16$ different QSLs, there are
eight $0$-flux QSLs and eight $\pi$-flux QSLs. For each QSL, we
construct a general quadratic mean-field Hamiltonian for the
Schwinger bosons containing all onsite, nearest-neighbor (NN), and
next-nearest-neighbor (NNN) terms allowed by symmetry. However, for
simplicity, we focus on the $0$-flux QSLs and restrict the
mean-field Hamiltonian to onsite and NN terms. At such a NN level,
two out of eight $0$-flux Hamiltonians have an enlarged U(1) gauge
symmetry, and we thus concentrate on the remaining six $0$-flux
Hamiltonians with $\mathbb{Z}_2$ gauge symmetry.

In each of the six corresponding $\mathbb{Z}_2$ QSL phases, the
Schwinger bosons can be identified as elementary spinon excitations
carrying fractionalized quantum numbers. If the chemical potential
is tuned to its critical value, there is a phase transition driven
by the condensation of these bosonic spinons. Depending on the
particular patterns of spinon condensation, we describe $15$
different critical ``paraphases'' out of the six QSL phases. The
most important characteristics of these paraphases, labeled by their
parent QSL phases and the condensation momenta of the spinons, are
tabulated in Table~\ref{six0fluxclasses}.

For each paraphase, the spinon spectrum is gapless at the critical
point by construction. The effective field theory of the critical
point is characterized by the low-energy spinon dispersion, $\omega
\sim k^z$, in terms of the dynamical critical exponent, which is
either $z=1$ or $z=2$. These two dynamical classes give rise to
distinct sets of experimental signatures. For a start, the power-law
exponent $x$ of the low-temperature heat capacity, $C_V \sim T^x$,
is determined by the dynamical exponent $z$ and the dimensionality
of the condensation manifold, i.e., if the spinons condense at
points or along lines in the Brillouin zone (BZ). Also, the
dynamical exponent gives rise to universal features in the static
and dynamic spin structure factors, which appear on top of more
detailed characteristics specific to given paraphases. In
particular, when approaching zero energy, the spectral weight in the
dynamic structure factor vanishes for $z=2$ but diverges for $z=1$;
the divergence in the $z=1$ case is also observable as a nonanalytic
behavior in the static structure factor.

To establish a connection between spinon condensation and the
resulting magnetic orders, restricted to zero momentum for
simplicity, we investigate the transformation rules of the possible
order parameters under the point group $O_{\text{h}}$ of the
pyrochlore lattice. For each paraphase, we determine which magnetic
orders generically appear, concentrating in particular on the
conventional spin orders seen in the experiments: the
all-in-all-out, antiferromagnetic, ferromagnetic, and Palmer-Chalker
orders. In doing so, we learn two important general lessons on
magnetic orders obtained by spinon condensation. First, several
distinct orders may be intertwined, i.e., they necessarily accompany
each other, even though they are completely unrelated on the
classical level. Second, the conventional spin orders may emerge
together with more exotic inversion-breaking ``hidden'' orders.

\section{\label{sec:level1}Projective symmetry group}

\subsection{Lattice symmetries}

We first introduce the symmetries of the pyrochlore lattice; the
convention and notation we establish here is used throughout the
rest of the paper. The pyrochlore lattice consists of four FCC-type
sublattices, which we label by $\mu =0,1,2,3$. To index the sites of
the lattice, we use two coordinate systems: the global cartesian
coordinates (GCCs) and the sublattice-indexed pyrochlore coordinates
(SIPCs). The GCCs are the standard frame coordinates for the FCC
cube of edge length $a = 1$. The SIPCs are spanned by the lattice
vectors $\hat{e}_1$, $\hat{e}_2$, and $\hat{e}_3$, which are
expressed in GCCs as
\begin{subequations}
\begin{eqnarray}
\hat{e}_1 & = & \frac{1}{2}(0,1,1),\\
\hat{e}_2 & = & \frac{1}{2}(1,0,1),\\
\hat{e}_3 & = & \frac{1}{2}(1,1,0).
\end{eqnarray}
\end{subequations}
We define $\hat{\epsilon}_i = \frac{1}{2} \hat{e}_i$ ($i=1,2,3$) to
be the displacement vectors of the $\mu=1,2,3$ sublattices from the
$\mu=0$ sublattice, where we understand $\hat{\epsilon}_0 =
\hat{e}_0 = 0$. The relation between the SIPCs and the GCCs is then
\begin{align}
(r_1,r_2,r_3)_\mu &= \vec{r}_\mu = \vec{r} +\hat{\epsilon}_\mu \tag*{SIPC}\\
&=\frac{1}{2}(r_2+r_3,r_3+r_1,r_1+r_2)+\frac{1}{2}\hat{e}_\mu. \tag*{GCC}
\end{align}

The space group of the pyrochlore lattice is the cubic space group
$Fd\overline{3}m$ (No.~227), minimally generated by the translations
$T_1$, $T_2$, and $T_3$ along the lattice vectors $\hat{e}_1$,
$\hat{e}_2$, and $\hat{e}_3$, a sixfold rotoreflection
$\overline{C}_6$ around the $[111]$ axis (i.e., around $\hat{e}_1 +
\hat{e}_2 + \hat{e}_3$), and a non-symmorphic screw operation $S$,
which is the composition of a twofold rotation around $\hat{e}_3$
and a translation by $\hat{\epsilon}_3$. These space-group
generators transform the SIPCs according to
\begin{eqnarray}
&T_1 \colon &(r_1,r_2,r_3)_\mu\rightarrow (r_1+1,r_2,r_3)_\mu,\nonumber\\
&T_2 \colon &(r_1,r_2,r_3)_\mu\rightarrow (r_1,r_2+1,r_3)_\mu, \nonumber\\
&T_3 \colon &(r_1,r_2,r_3)_\mu\rightarrow (r_1,r_2,r_3+1)_\mu, \nonumber\\
&\overline{C}_6\colon&(r_1,r_2,r_3)_0\rightarrow (-r_3,-r_1,-r_2)_0, \nonumber\\
&               &(r_1,r_2,r_3)_1\rightarrow (-r_3,-r_1-1,-r_2)_2, \nonumber\\
&               &(r_1,r_2,r_3)_2\rightarrow (-r_3,-r_1,-r_2-1)_3, \nonumber\\
&               &(r_1,r_2,r_3)_3\rightarrow (-r_3-1,-r_1,-r_2)_1, \nonumber\\
&S\colon   &(r_1,r_2,r_3)_0 \rightarrow (-r_1,-r_2,r_1+r_2+r_3)_3, \nonumber\\
&     &(r_1,r_2,r_3)_1 \rightarrow (-r_1-1,-r_2,r_1+r_2+r_3+1)_1, \nonumber\\
&     &(r_1,r_2,r_3)_2 \rightarrow (-r_1,-r_2-1,r_1+r_2+r_3+1)_2, \nonumber\\
&     &(r_1,r_2,r_3)_3 \rightarrow (-r_1,-r_2,r_1+r_2+r_3+1)_0,
\end{eqnarray}
Note that we can write the rotoreflection as $\overline{C}_6=I
{C}_3$, where $I$ is an inversion with respect to the origin and
${C}_3$ is a threefold rotation around the $[111]$ axis. The
generators $\{I, C_3\}$ are therefore equivalent to the generator
$\overline{C}_6$; we choose a single generator $\overline{C}_6$ to
reduce the number of generators and group relations.

The point group of the pyrochlore lattice, formally defined as the
quotient group of the space group and the group of pure
translations, is the cubic group $O_\text{h}$. This group is
minimally generated by $\overline{C}_6$ and $S'$, where $S'$ is a
twofold rotation around $\hat{e}_3$, distinguished from the
space-group generator $S$ by the lack of a subsequent translation
along $\hat{\epsilon}_3$. A detailed description of the point-group
structure is given in Appendix~\ref{app:A}.

In addition to the pyrochlore space-group symmetries, time-reversal
symmetry is also present in the pyrochlore materials. The
corresponding time-reversal operation $\mathcal{T}$ commutes with
all space-group operations and satisfies $\mathcal{T}^2 = -1$ when
acting on a half-integer spin state. The complete list of
independent group relations defining the symmetry group is then
\begin{eqnarray}
T_iT_{i+1} T_i^{-1} T_{i+1}^{-1} &=& 1,\quad i=1,2,3, \nonumber\\
\overline{C}^6_6 &=& 1,  \nonumber\\
S^2 T^{-1}_3 &=& 1, \nonumber\\
\overline{C}_6 T_i \overline{C}^{-1}_6 T_{i+1} &=&1,\quad i=1,2,3, \nonumber\\
ST_iS^{-1} T^{-1}_3T_i&=& 1, \quad i=1,2, \nonumber\\
ST_3 S^{-1} T_3^{-1} &=&1, \nonumber\\
(\overline{C}_6S)^4&=&1, \nonumber\\
(\overline{C}_6^3S)^2 &=&1, \nonumber\\
\mathcal{T}^2 &=& -1, \nonumber\\
\mathcal{T}\mathcal{O}\mathcal{T}^{-1}\mathcal{O}^{-1}&=&1,\quad \mathcal{O}\in\{T_1,T_2,T_3,\overline{C}_6, S\}.\label{SGtr}
\end{eqnarray}
The notation in Eq.~\eqref{SGtr} is understood as $i+3 \equiv i$.

\subsection{Projective symmetry group}

In this subsection, we classify all possible $\mathbb{Z}_2$ quantum
spin liquids that are compatible with the symmetries of the
pyrochlore lattice. We first write the spins in terms of Schwinger
boson bilinears as
\begin{equation}\label{parton}
\hat{S}^\alpha_{\vec{r}_\mu} = \frac{1}{2} b^\dag_{\vec{r}_\mu}
\sigma^{\alpha} b^{\vphantom\dagger}_{\vec{r}_\mu}, \qquad \alpha =
x,y,z,
\end{equation}
where $b_{\vec{r}_\mu} = \left(\begin{array}{c}
b_{\vec{r}_\mu,\uparrow}\\b_{\vec{r}_\mu,\downarrow}\end{array}\right)$,
and $\sigma^{x,y,z}$ are the Pauli matrices (also denoted by
$\sigma^{1,2,3}$, respectively). Physically, the Schwinger bosons
$b_{\vec{r}_\mu}$ describe the deconfined spinon excitations of the
quantum spin liquid and, on the mean-field level, they are governed
by a quadratic Hamiltonian, commonly known as the mean-field ansatz.

It is important to emphasize that the transformation in
Eq.~\eqref{parton} is not faithful as it enlarges the local Hilbert
space at each site $\vec{r}_\mu$. Consequently, there is a local
gauge redundancy for the Schwinger bosons. Indeed, any
site-dependent U(1) phase transformation
\begin{equation}
G \colon b_{\vec{r}_\mu} \rightarrow e^{i \phi (\vec{r}_\mu)}
b_{\vec{r}_\mu} \label{op-G}
\end{equation}
leaves the spins $\hat{S}^\alpha_{\vec{r}_\mu}$ invariant. The
physical Hilbert space can in principle be retained by enforcing the
constraint
\begin{equation}
\sum\limits_{\sigma=\uparrow,\downarrow}
b^\dag_{\vec{r}_\mu,\sigma}b^{\vphantom\dagger}_{\vec{r}_\mu,\sigma}
= 1
\end{equation}
at each site $\vec{r}_\mu$ of the lattice.

Under a space-group operation $\mathcal{O}$, the spins transform as
$\mathcal{O}\colon \hat{S}^\alpha_{\vec{r}_\mu}\rightarrow
U_{\mathcal{O}} \hat{S}^\alpha_{\mathcal{O}(\vec{r}_\mu)}
U^\dag_{\mathcal{O}} = \frac{1}{2} b^\dag_{\mathcal{O}(\vec{r}_\mu)}
U_{\mathcal{O}} \sigma^{\alpha} U_{\mathcal{O}}^\dag
b_{\mathcal{O}(\vec{r}_\mu)}$, where $U_{\mathcal{O}}$ is the SU(2)
rotation matrix associated with the operation $\mathcal{O}$. We
therefore na\"ively expect that the spinons transform as
\begin{equation}
\mathcal{O}\colon b_{\vec{r}_\mu} \rightarrow U^\dag_{\mathcal{O}}
b_{\mathcal{O} (\vec{r}_\mu)}. \label{op-O}
\end{equation}
However, due to the U(1) gauge redundancy, any operation
$\mathcal{O}$ is generally accompanied by a site-dependent U(1)
phase transformation
\begin{equation}
G_{\mathcal{O}} \colon b_{\vec{r}_\mu} \rightarrow e^{i
\phi_{\mathcal{O}} (\vec{r}_\mu)} b_{\vec{r}_\mu}, \label{op-GO}
\end{equation}
and the spinons thus actually transform as
\begin{equation}\label{b_under_o}
\widetilde{\mathcal{O}} = G_{\mathcal{O}} \circ \mathcal{O} \colon
b_{\vec{r}_\mu} \rightarrow e^{i
\phi_{\mathcal{O}}[\mathcal{O}(\vec{r}_\mu)]} U^\dag_{\mathcal{O}}
b_{O(\vec{r}_\mu)},
\end{equation}
where the symbol ``$\circ$'' indicates that the gauge-enriched
operation $\widetilde{\mathcal{O}}$ is a composition of the pure
symmetry operation $\mathcal{O}$ and the gauge transformation
$G_{\mathcal{O}}$.

Under a time reversal $\mathcal{T}$ of the system, the spins
transform as $\mathcal{T}\colon
\hat{S}^\alpha_{\vec{r}_\mu}\rightarrow \mathcal{K}^\dag
U_{\mathcal{T}} \hat{S}^\alpha_{\vec{r}_\mu} U^\dag_{\mathcal{T}}
\mathcal{K}$, where $U_{\mathcal{T}} = i \sigma^2$, while
$\mathcal{K}=\mathcal{K}^\dag=\mathcal{K}^{-1}$ applies complex
conjugation to everything on its right. Once again, combining the
na\"ive transformation rule for the spinons,
\begin{equation}
\mathcal{T} \colon b_{\vec{r}_\mu} \rightarrow \mathcal{K}
U^\dag_{\mathcal{T}} b_{\vec{r}_\mu}, \label{op-T}
\end{equation}
and the accompanying U(1) phase transformation,
\begin{equation}
G_{\mathcal{T}} \colon b_{\vec{r}_\mu} \rightarrow e^{i
\phi_{\mathcal{T}} (\vec{r}_\mu)} b_{\vec{r}_\mu}, \label{op-GT}
\end{equation}
the spinons are found to transform as
\begin{equation}\label{b_under_tr}
\widetilde{\mathcal{T}} = G_{\mathcal{T}} \circ \mathcal{T} \colon
b_{\vec{r}_\mu}\rightarrow
e^{i\phi_{\mathcal{T}}(\vec{r}_\mu)}\mathcal{K}U^\dag_{\mathcal{T}}
b_{\vec{r}_\mu}.
\end{equation}
Note that $[\mathcal{K}, U_{\mathcal{T}}] = 0$ because
$U_{\mathcal{T}}$ is real.

For a quantum spin liquid, the gauge-enriched operations
$\widetilde{\mathcal{O}}$ and $\widetilde{\mathcal{T}}$ generate the
symmetry group of the mean-field ansatz, commonly known as the
projective symmetry group (PSG). To enumerate all quantum spin
liquids, we need to find all distinct PSG solutions, i.e., all
gauge-inequivalent solutions for the gauge transformations
$G_{\mathcal{O}}$ and $G_{\mathcal{T}}$ that are consistent with the
symmetry group of the lattice, including space-group symmetries and
time-reversal symmetry. In particular, for each group relation [see
Eq.~\eqref{SGtr}] taking the general form of
\begin{equation}
\mathcal{O}_1 \circ \mathcal{O}_2 \circ \dots = 1, \label{grprlt0}
\end{equation}
we consider the gauge-enriched group relation
\begin{equation}\label{grprlt1}
\widetilde{\mathcal{O}}_1 \circ \widetilde{\mathcal{O}}_2 \circ
\dots = (G_{\mathcal{O}_1}\circ \mathcal{O}_1)\circ
(G_{\mathcal{O}_2}\circ \mathcal{O}_2)\circ\dots = \mathcal{G},
\end{equation}
where $\mathcal{G}$ is a pure gauge transformation, thus
corresponding to the identity operation for the spins. Being an
element of the PSG by definition, $\mathcal{G}$ is also an element
of the invariant gauge group (IGG), the group of all gauge
transformations that leave the mean-field ansatz invariant. In most
cases, such gauge transformations are exclusively ``global'' (i.e.,
site independent), and the IGG is thus a subgroup of U(1), typically
$\mathbb{Z}_2$ or U(1), corresponding to $\mathbb{Z}_2$ and U(1)
spin liquids, respectively. Since we are interested in classifying
$\mathbb{Z}_2$ spin liquids, we consider IGG $= \mathbb{Z}_2$ in the
following. The only two elements of the IGG are then $\mathcal{G} =
e^{i n \pi}$ with $n = \{ 0,1 \}$.

For any group relation in terms of exclusively space-group
operations, taking the form of Eq.~\eqref{grprlt0}, the
gauge-enriched group relation in Eq.~\eqref{grprlt1} can be
rewritten as
\begin{eqnarray}
G_{\mathcal{O}_1} &\circ& (\mathcal{O}_1 \circ G_{\mathcal{O}_2}
\circ \mathcal{O}^{-1}_1) \label{grprlt2} \\
&\circ& (\mathcal{O}_1 \circ \mathcal{O}_2 \circ G_{\mathcal{O}_3}
\circ \mathcal{O}^{-1}_2 \circ \mathcal{O}^{-1}_1) \circ \dots =
\mathcal{G}. \nonumber
\end{eqnarray}
Using the general conjugation rule
\begin{equation}\label{OGO}
\mathcal{O}_i \circ G_{\mathcal{O}_j} \circ \mathcal{O}^{-1}_i
\colon b_{\vec{r}_\mu} \rightarrow e^{i \phi_{\mathcal{O}_j}
[\mathcal{O}^{-1}_i (\vec{r}_\mu)]} b_{\vec{r}_\mu},
\end{equation}
following directly from Eqs.~\eqref{op-O} and \eqref{op-GO}, this
group relation then becomes a pure phase equation:
\begin{eqnarray}
\phi_{\mathcal{O}_1}(\vec{r}_\mu) &+&
\phi_{\mathcal{O}_2}[\mathcal{O}_1^{-1}(\vec{r}_\mu)] \label{phaseequationO} \\
&+&
\phi_{\mathcal{O}_3}\{\mathcal{O}_2^{-1}[\mathcal{O}_1^{-1}(\vec{r}_\mu)]\}
+ \dots = n \pi \mod 2\pi. \nonumber
\end{eqnarray}
For group relations involving time reversal, special care must be
taken due to the presence of the complex conjugation $\mathcal{K}$.
Using the modified conjugation rule
\begin{eqnarray}\label{TGT}
&& \mathcal{T} \circ G_{\mathcal{O}} \circ \mathcal{T}^{-1} \colon
\\
&& \qquad b_{\vec{r}_\mu} \rightarrow \mathcal{K}
U^\dag_{\mathcal{T}} e^{i \phi_{\mathcal{O}} (\vec{r}_\mu)}
U_{\mathcal{T}} \mathcal{K}^{\dag} b_{\vec{r}_\mu} = e^{-i
\phi_{\mathcal{O}} (\vec{r}_\mu)} b_{\vec{r}_\mu}, \nonumber
\end{eqnarray}
the last group relation in Eq.~\eqref{SGtr} translates into
\begin{equation}
\phi_{\mathcal{T}}(\vec{r}_\mu)-\phi_{\mathcal{T}}[\mathcal{O}^{-1}(\vec{r}_\mu)]
-2\phi_{\mathcal{O}}(\vec{r}_\mu) = n\pi \mod 2\pi,
\label{phaseequationT}
\end{equation}
while the penultimate group relation $\mathcal{T}^2 = -1$ gives rise
to a trivial equation due to the cancellation between the phase
factors $e^{i \phi_{\mathcal{T}} (\vec{r}_\mu)}$ and $e^{-i
\phi_{\mathcal{T}} (\vec{r}_\mu)}$.

The PSG classification is obtained by listing all group relations
and finding all solutions of the corresponding phase equations [see
Eqs.~\eqref{phaseequationO} and \eqref{phaseequationT}] for the
$\mathbb{Z}_2$ parameters $n$ as well as the phases
$\phi_{\mathcal{O}} (\vec{r}_\mu)$ and $\phi_{\mathcal{T}}
(\vec{r}_\mu)$. We emphasize that distinct solutions, describing
distinct spin liquids, must be gauge inequivalent. Indeed, by means
of a general gauge transformation $G$ [see Eq.~\eqref{op-G}], the
gauge-enriched group relations in Eq.~\eqref{grprlt1} can be
rewritten as
\begin{equation}
(G\circ G_{\mathcal{O}_1} \circ \mathcal{O}_1 \circ G^{-1}) \circ
(G\circ G_{\mathcal{O}_2}\circ \mathcal{O}_2\circ G^{-1})\circ\dots
= \mathcal{G},
\end{equation}
transforming the phases $\phi_{\mathcal{O}_i} (\vec{r}_\mu)$
according to
\begin{equation}
\phi_{\mathcal{O}_i} (\vec{r}_\mu) \rightarrow \phi_{\mathcal{O}_i}
(\vec{r}_\mu) + \phi(\vec{r}_\mu) - \phi[\mathcal{O}^{-1}_i
(\vec{r}_\mu)],
\end{equation}
and thus indicating that two seemingly distinct solutions for the
phases might in fact be equivalent.

The detailed solution of the PSG equations is presented in Appendix
\ref{app:B}. The PSG results for the phases are
\begin{subequations}\label{PSGsolution}
\begin{eqnarray}
\phi_{T_1}(\vec{r}_\mu) &=& 0,\qquad\label{psg_phi_1}\\
\qquad \phi_{T_2}(\vec{r}_\mu) &=& n_1 \pi r_1,\qquad\label{psg_phi_2}\\
\phi_{T_3}(\vec{r}_\mu) &=& n_1\pi (r_1+r_2),\qquad\label{psg_phi_3}\\
\phi_{\mathcal{T}}(\vec{r}_\mu) &=& 0,\qquad\label{psg_phi_t}\\
\phi_{\overline{C}_6}(\vec{r}_\mu) &=& \left[\frac{n_{\overline{C}_6}}{2}+(n_1+n_{ST_1})\delta_{\mu=1,2,3}\right]\pi  \qquad\notag\\
&&+ n_1\delta_{\mu=2,3} \pi r_1+ n_1\delta_{\mu=2} \pi r_3\qquad\notag\\
&&+n_1 (r_1r_2+r_1r_3),\qquad\label{psg_phi_c}\\
\phi_S(\vec{r}_\mu) &=& \left[(-)^{\delta_{\mu=1,2,3}}\frac{n_1+n_{ST_1}}{2}+\delta_{\mu=2}n_{\overline{C}_6S}\right]\pi\qquad\notag\\
&&+(n_1 \delta_{\mu=1,2}-n_{ST_1}) \pi r_1\qquad\notag\\
&& +(n_1 \delta_{\mu=2}-n_{ST_1}) \pi r_2
+ n_1 \delta_{\mu=1,2} \pi r_3\notag\qquad\\
&&-\frac{1}{2}n_1 \pi (r_1+r_2)(r_1+r_2+1),\qquad\label{psg_phi_s}
\end{eqnarray}
\end{subequations}
where $n_1$, $n_{\overline{C}_6S}$, $ n_{ST_1}$, and
$n_{\overline{C}_6}$ are four $\mathbb{Z}_2$ parameters, each being
either $0$ or $1$. Therefore, we find that there are $16$
gauge-inequivalent $\mathbb{Z}_2$ PSG classes, corresponding to
distinct $\mathbb{Z}_2$ quantum spin liquids, which we label by the
notation $n_1\pi$-$(n_{\overline{C}_6S} \, n_{ST_1} \,
n_{\overline{C}_6})$. The four $\mathbb{Z}_2$ parameters have
concrete interpretations:
\begin{itemize}
\item The parameter $n_1$ comes from the three PSG equations corresponding to $T_iT_{i+1}T_i^{-1}T_{i+1}^{-1}=1$, which are required by the PSG to share the same $\mathbb{Z}_2$ parameter. Physically, it quantifies the Aharonov-Bohm (AB) phase a spinon accumulates while moving on the closed edge of a plaquette, which is traversed by such a sequence of translations. In the case of $n_1=1$ ($n_1=0$), the AB phase is $\pi$ ($0$), corresponding to a $\pi$-flux ($0$-flux) spin liquid.
\item The parameter $n_{\overline{C}_6}$ comes from the PSG equation corresponding to $\overline{C}_6^6=1$. Physically, it describes the AB phase a spinon accumulates after completing six subsequent sixfold rotoreflections. Together with $n_{ST_1}$, it determines whether or not the sixfold rotoreflection $\overline{C}_6$ acts projectively.
\item The parameter $n_{ST_1}$ comes from the PSG equation corresponding to $ST_1S^{-1}T_3^{-1}T_1 = 1$. Physically, it describes the AB phase a spinon accumulates after completing the operation sequence $ST_1S^{-1}T_3^{-1}T_1$. Together with $n_1$ and $n_{\overline{C}_6S}$ it determines whether or not the screw operation $S$ acts projectively.
\item The parameter $n_{\overline{C}_6S}$ comes from the PSG equation corresponding to $(\overline{C}_6S)^4=1$. Physically, it describes the AB phase a spinon accumulates after completing the operation sequence $(\overline{C}_6S)^4$.
\end{itemize}

\subsection{Construction of mean-field ans\"atze}

We are now in the position to construct the mean-field ansatz for
each PSG class. The most general mean-field ansatz for bosonic
spinons can be written as
\begin{equation}
H = \sum_{\vec{r}_\mu,\vec{r}'_\nu} b^\dag_{\vec{r}_\mu}
u^h_{\vec{r}_\mu,\vec{r}'_\nu} b^{\vphantom\dagger}_{\vec{r}'_\nu} +
b^\dag_{\vec{r}_\mu} u^p_{\vec{r}_\mu,\vec{r}'_\nu} \left(
b^\dag_{\vec{r}'_\nu} \right)^T + h.c., \label{H-mf}
\end{equation}
where $u^h_{\vec{r}_\mu,\vec{r}'_\nu}$ and
$u^p_{\vec{r}_\mu,\vec{r}'_\nu}$ are $2\times 2$ matrices acting on
spin space, and the labels ``h'' and ``p'' indicate hopping and
pairing terms, respectively.

The PSG operators $\widetilde{\mathcal{O}}$ and
$\widetilde{\mathcal{T}}$ are the symmetry operators of the
Hamiltonian $H$, meaning $\widetilde{\mathcal{O}} \colon
H\rightarrow H$ and $\widetilde{\mathcal{T}} \colon H\rightarrow H$.
Since the spinons transforms under $\widetilde{\mathcal{O}}$ and
$\widetilde{\mathcal{T}}$ according to Eqs.~\eqref{b_under_o} and
\eqref{b_under_tr}, the matrices $u^h$ and $u^p$ must transform as
\begin{subequations}\label{bonds_under_o}
\begin{eqnarray}
G^\dag_{\mathcal{O}}[{\mathcal{O}}(\vec{r}_\mu)] U_{\mathcal{O}}
u^h_{\vec{r}_\mu,\vec{r}'_\nu}
U^\dag_{\mathcal{O}}G_{\mathcal{O}}[{\mathcal{O}}(\vec{r}'_\nu)]&=&u^h_{{\mathcal{O}}(\vec{r}_\mu),{\mathcal{O}}(\vec{r}'_\nu)},
\qquad \label{GUuh} \\
G^\dag_{\mathcal{O}}[{\mathcal{O}}(\vec{r}_\mu)] U_{\mathcal{O}}
u^p_{\vec{r}_\mu,\vec{r}'_\nu}
U^T_{\mathcal{O}}G^\dag_{\mathcal{O}}[{\mathcal{O}}(\vec{r}'_\nu)]&=&u^p_{{\mathcal{O}}(\vec{r}_\mu),{\mathcal{O}}(\vec{r}'_\nu)}
\qquad \label{GUup}
\end{eqnarray}
\end{subequations}
for space-group elements $\mathcal{O} \in
\{T_1,T_2,T_3,\overline{C}_6,S\}$ and as
\begin{subequations}\label{GT}
\begin{eqnarray}
G^\dag_{\mathcal{T}}(\vec{r}_\mu) U_{\mathcal{T}}
\left(u^h_{\vec{r}_\mu,\vec{r}'_\nu}\right)^*
U^\dag_{\mathcal{T}}G_{\mathcal{T}}(\vec{r}'_\nu)&=&u^h_{\vec{r}_\mu,\vec{r}'_\nu},\\
G^\dag_{\mathcal{T}}(\vec{r}_\mu) U_{\mathcal{T}}
\left(u^p_{\vec{r}_\mu,\vec{r}'_\nu}\right)^*
U^T_{\mathcal{T}}G^\dag_{\mathcal{T}}(\vec{r}'_\nu)&=&u^p_{\vec{r}_\mu,\vec{r}'_\nu}
\end{eqnarray}
\end{subequations}
for time reversal $\mathcal{T}$. The respective SU(2) matrices are
\begin{equation}\label{su2_under_o}
\begin{aligned}
&U_{T_1}=U_{T_2}=U_{T_3} = \sigma^0,\quad U_{\mathcal{T}} = i \sigma^2,\\
&U_{\overline{C}_6} = U_{C_3}=e^{-\frac{i}{2} \frac{2\pi}{3}\frac{(1,1,1)}{\sqrt{3}}\cdot \vec{\sigma}},\quad
U_S = e^{-\frac{i}{2}\pi \frac{(1,1,0)}{\sqrt{2}}\cdot \vec{\sigma}}.
\end{aligned}
\end{equation}
where $\sigma^0= 1_{2\times 2}$ is the identity matrix. Suppressing
the site indices for simplicity, we parameterize the matrices $u^h$
and $u^p$ in the general forms
\begin{subequations}\label{uhup}
\begin{eqnarray}
u^h &=&  a\sigma^0 + i(b\sigma^1+c \sigma^2+d\sigma^3),\\
u^p &=& \left(a'\sigma^0 + i (b'\sigma^1+c'\sigma^2+d'\sigma^3)\right)\cdot i\sigma^2,
\end{eqnarray}
\end{subequations}
where $a, b, c, d, a', b', c', d'$ are all complex. The additional
factor $i\sigma^2$ appearing in $u^p$ ensures that $(a,b,c,d)$ and
$(a',b',c',d')$ transform in the same way under the respective
unitary conjugations $u^h \rightarrow U u^h U^{\dag}$ and $u^p
\rightarrow U u^p U^T$ for any $U \in$ SU(2). In both cases, the
singlet parameters $a$ and $a'$ transform as scalars, while the
triplet parameters $\vec{b} = (b,c,d)$ and $\vec{b}' = (b',c',d')$
transform as SO(3) vectors. Indeed, any SU(2) rotation leaves the
singlet parameters invariant and performs the corresponding SO(3)
rotation on the triplet vectors: $\vec{b} \rightarrow \mathcal{R}
\vec{b}$ and $\vec{b}' \rightarrow \mathcal{R} \vec{b}'$. For the
generators $\overline{C}_6$ and $S$, these SO(3) rotations are
\begin{equation}
\mathcal{R}^{\overline{C}_6}=\left(\begin{array}{ccc}&&1\\1&&\\&1&\end{array}\right),\quad
\mathcal{R}^S = \left(\begin{array}{ccc}&1&\\1&&\\&&-1\end{array}\right),
\end{equation}
while the translations $T_{1,2,3}$ correspond to trivial SO(3)
rotations: $\mathcal{R}^{T_{1,2,3}} = 1_{3 \times 3}$.

To reduce the number of parameters in the mean-field ansatz, we
first consider the effect of time reversal. Substituting
Eq.~\eqref{uhup} into Eq.~\eqref{GT}, and taking
$G_{\mathcal{T}}(\vec{r}_\mu) = 1$ from Eq.~\eqref{psg_phi_t}, we
obtain $(a,b,c,d)=(a^*,b^*,c^*,d^*)$ as well as
$(a',b',c',d')=(a'^*,b'^*,c'^*,d'^*)$ and deduce that all $8$
parameters of $u^h$ and $u^p$ are real.

Turning to space-group symmetries and using
Eq.~\eqref{bonds_under_o}, we can then establish relations between
the respective parameters of $u^h_{\vec{r}_\mu,\vec{r}'_\nu}$ and
$u^p_{\vec{r}_\mu,\vec{r}'_\nu}$ that correspond to different bonds
$\langle \vec{r}_\mu, \vec{r}'_\nu \rangle$ of the lattice. In fact,
the entire mean-field ansatz in Eq.~\eqref{H-mf} can be constructed
up to next-nearest-neighbor level by specifying the $8$ real
parameters for each of the following three representative bonds:
\begin{itemize}
\item onsite ``bond'' $\vec{0}_0 \rightarrow \vec{0}_0$:
\begin{equation} \label{uhupOS}
\begin{aligned}
u^h_{\vec{0}_0,\vec{0}_0}&=\alpha\sigma^0+i(\beta\sigma^1+\gamma\sigma^2+\delta\sigma^3),\\
u^p_{\vec{0}_0,\vec{0}_0}&=\left(\alpha'\sigma^0+i(\beta'\sigma^1+\gamma'\sigma^2+\delta'\sigma^3)\right)\cdot i \sigma^2,
\end{aligned}
\end{equation}
\item nearest-neighbor (NN) bond $\vec{0}_0\rightarrow \vec{0}_1$:
\begin{equation} \label{uhupNN}
\begin{aligned}
u^h_{\vec{0}_0,\vec{0}_1} &= a\sigma^0+i(b\sigma^1+c\sigma^2+d\sigma^3),\\
u^p_{\vec{0}_0,\vec{0}_1} &= \left(a'\sigma^0+i(b'\sigma^1+c'\sigma^2+d'\sigma^3)\right)\cdot i \sigma^2,
\end{aligned}
\end{equation}
\item next-nearest-neighbor (NNN) bond $\vec{0}_1\rightarrow \vec{0}_2-\hat{e}_2$:
\begin{equation} \label{uhupNNN}
\begin{aligned}
u^h_{\vec{0}_1,\vec{0}_2-\hat{e}_2}&= A\sigma^0+i(B\sigma^1+C\sigma^2+D\sigma^3),\\
u^p_{\vec{0}_1,\vec{0}_2-\hat{e}_2}&=\left(A'\sigma^0+i(B'\sigma^1+C'\sigma^2+D'\sigma^3)\right)\cdot
i \sigma^2.
\end{aligned}
\end{equation}
\end{itemize}

\subsection{Nontrivial parameter constraints}

\begin{table*}
\begin{ruledtabular}
\begin{tabular}{c|lll|lll|l}
Class&\multicolumn{3}{c|}{Independent nonzero parameters}&\multicolumn{3}{c|}{Constraints}&\multirow{ 2}{*}{Note}\\
\cline{2-7}
$n_1\pi$-$(n_{\overline{C}_6S}n_{ST_1}n_{\overline{C}_6})$
&Onsite&NN&NNN&Onsite&NN&NNN&\\
\hline
0-$(000)$&$\mu$        &$a$,$c$      &$A,B,D,B'$      &                            & $c=-d$                  & $B=C$, $B'=-C'$ & U(1) at NN\\
0-$(001)$&$\mu$        &$a$,$c$,$b'$ &$A,B,D,B'$      &                            & $c=-d$,                 & $B=C$, $B'=-C'$ &\\
0-$(010)$&$\mu$, $\nu$ &$a$,$c$,$b'$ &$A,B,D,A',B',D'$&$\beta'=\delta'=\gamma'\equiv\nu$& $c=-d$,                & $B=C$, $B'=C'$  &\\
0-$(011)$&$\mu$,       &$a$,$c$      &$A,B,D,A',B',D'$&                            & $c=-d$,                 & $B=C$, $B'=C'$  & U(1) at NN\\
0-$(100)$&$\mu$,       &$b$,$c'$&$A,B,D,B'$      &                            & $c'=d'$,                & $B=C$, $B'=-C'$ &\\
0-$(101)$&$\mu$,       &$b$,$a'$,$c'$&$A,B,D,B'$      &                            & $c'=-d'$,               & $B=C$, $B'=-C'$ &\\
0-$(110)$&$\mu$, $\nu$ &$b$,$a'$,$c'$&$A,B,D,A',B',D'$&$\beta'=\delta'=\gamma'\equiv\nu$& $c'=-d'$,              & $B=C$, $B'=C'$  &\\
0-$(111)$&$\mu$,       &$b$,$c'$     &$A,B,D,A',B',D'$&                            & $c'=d'$,                & $B=C$, $B'=C'$  &\\
\hline
$\pi$-$(000)$&$\mu$, $\nu$        &$a$,$c$,$b'$      &$B,B'$      &                            $\beta'=\delta'=\gamma'=\nu$& $c=-d$                  & $B=-C$, $B'=-C'$ &\\
$\pi$-$(001)$&$\mu$        &$a$,$c$ &$B,B'$      &                            & $c=-d$         & $B=-C$, $B'=-C'$ &  U(1) at NN\\
$\pi$-$(010)$&$\mu$ &$a$,$c$ &$B,A',B',D'$&& $c=-d$         & $B=-C$, $B'=C'$  &\\
$\pi$-$(011)$&$\mu$,       &$a$,$c$,$b'$      &$B,A',B',D'$&                            & $c=-d$                 & $B=-C$, $B'=C'$  & \\
$\pi$-$(100)$&$\mu$, $\nu$       &$b$,$a'$,$c'$&$B,B'$      &                            $\beta'=\delta'=\gamma'=\nu$& $c'=-d'$& $B=-C$, $B'=-C'$ &U(1) at NN\\
$\pi$-$(101)$&$\mu$,       &$b$,$c'$&$B,B'$      &                            & $c'=d'$& $B=-C$, $B'=-C'$ &\\
$\pi$-$(110)$&$\mu$ &$b$,$c'$&$B,A',B',D'$&& $c'=d'$& $B=-C$, $B'=C'$  &\\
$\pi$-$(111)$&$\mu$,       &$b$,$a'$,$c'$     &$B,A',B',D'$&                            & $c'=-d'$& $B=-C$, $B'=C'$  &\\
\end{tabular}
\end{ruledtabular}
\caption{Independent mean-field parameters and constraints for the
sixteen PSG classes. The parameters not mentioned in this table are
enforced to be zero by the constraints. The mean-field Hamiltonians
for some PSG classes appear to be U(1) on the NN level, as indicated
by the comment ``U(1) at NN'', but recover their $\mathbb{Z}_2$
character upon including NNN terms. Note that a nonzero onsite
chemical potential $\mu=\alpha$ is allowed in all PSG
classes.}\label{MFTparameters}
\end{table*}

When constructing the entire mean-field ansatz from the
representative bonds in Eqs.~\eqref{uhupOS}--\eqref{uhupNNN}, the
significance of using Eq.~\eqref{bonds_under_o} is twofold. On the
one hand, most space-group elements map the representative bonds
onto different bonds, thereby determining the matrices
$u^h_{\vec{r}_\mu,\vec{r}'_\nu}$ and
$u^p_{\vec{r}_\mu,\vec{r}'_\nu}$ for all symmetry-related bonds. On
the other hand, some space-group elements map the representative
bonds onto themselves, thereby leading to nontrivial constraints on
the original $24$ parameters.

For simplicity, we first concentrate on the $0$-flux PSG classes.
Since translation is trivial [see
Eqs.~\eqref{psg_phi_1}--\eqref{psg_phi_3}], we can restrict our
attention to a single unit cell, within which bonds are mapped onto
each other by elements of the \emph{point group}. Since the point
group $O_{\text{h}}$ consists of $48$ elements, and there are $4$
onsite, $12$ NN, and $24$ NNN bonds within a single unit cell, which
can be viewed as three orbits in the point group, the
orbit-stabilizer theorem implies that the onsite, NN, and NNN
representative bonds are mapped onto themselves by $12$, $4$, and
$2$ point-group elements, respectively. When a bond is mapped onto
itself by such a point-group element, nontrivial constraints are
obtained on the parameters by comparing the new and the old
expressions for $u^h_{\vec{r}_\mu,\vec{r}'_\nu}$ and
$u^p_{\vec{r}_\mu,\vec{r}'_\nu}$. These constraints are solved in
Appendix \ref{app:D}.

In Table \ref{MFTparameters}, we present the nonzero parameters of
the mean-field ansatz for each of the eight $0$-flux and each of the
eight $\pi$-flux PSG classes up to NNN level, along with any
constraints between the parameters. From these nonzero parameters,
the entire mean-field ansatz can be constructed via
Eq.~\eqref{bonds_under_o}. Note that some of the mean-field
ans\"atze in Table \ref{MFTparameters} have an enlarged U(1) gauge
symmetry at the NN level which only breaks down to $\mathbb{Z}_2$
when nonzero NNN terms are included.

\section{Analysis of the mean-field ans\"atze\label{secMF}}

The previous section explains how the method of PSG can be used to
obtain classes of 0-flux and $\pi$-flux mean-field ans\"atze, which
describe distinct phases of $\mathbb{Z}_2$ quantum spin liquids on
the mean-field level. In this section, we focus on the 0-flux
mean-field ans\"atze and study their physical properties in great
detail. Since our main goal is to explore the relationship between
spin liquids and magnetic orders adjacent to them, we primarily
concentrate on the critical field theories and the condensation
patterns (i.e., the resulting magnetic orders).

In each mean-field ansatz, we neglect the NNN terms for simplicity,
restricting our attention to onsite and NN terms. Since we are
interested in $\mathbb{Z}_2$ spin liquids, and two out of eight
0-flux mean-field ans\"atze have U(1) gauge symmetry at the NN
level, we only consider the remaining \emph{six} mean-field
ans\"atze in the rest of the paper.

\subsection{Symmetry properties}

The PSG method is rooted in symmetry analysis, and it is important
to understand how the PSG governs the symmetry of the mean-field
Hamiltonians. By means of a Fourier transformation, a general
mean-field Hamiltonian [see Eq.~(\ref{H-mf})] can be written in
momentum space as
\begin{equation}\label{H}
H = \sum\limits_{\vec{k}\in \text{BZ}} B^\dag_{\vec{k}} \mathcal{H}(\vec{k}) ^{\vphantom\dagger}B_{\vec{k}},
\end{equation}
where $B_{\vec{k}} = \left(
b^{\phantom{\dag}}_{\vec{k},0},b^{\phantom{\dag}}_{\vec{k},1},b^{\phantom{\dag}}_{\vec{k},2},b^{\phantom{\dag}}_{\vec{k},3},
b^\dag_{-\vec{k},0},b^\dag_{-\vec{k},1},b^\dag_{-\vec{k},2},b^\dag_{-\vec{k},3}\right)^T$
is a $16$-component vector of operators. The matrix
$\mathcal{H}(\vec{k})$ has the standard Bogoliubov form
\begin{equation}\label{HH}
\mathcal{H}(\vec{k}) =
\left(\begin{array}{cc}
U_h(\vec{k})& U_p(\vec{k})\\
U^\dag_p(\vec{k}) & U^T_h(-\vec{k})
\end{array}\right),
\end{equation}
where $U_h(\vec{k})=U^\dag_h(\vec{k})$ and $U_p(\vec{k}) =
U_p^T(-\vec{k})$, corresponding to hopping and pairing terms,
respectively.

The Hamiltonian matrix $\mathcal{H}(\vec{k})$ combines momenta $\pm
\vec{k}$ and thus assigns a full set of physical degrees of freedom
to only half of the BZ. This redundancy in the description leads to
an effective charge-conjugation ``symmetry'', corresponding to the
matrix-level constraint
\begin{equation}
U^{-1}_{\mathcal{C}}\mathcal{H}^*(\vec{k})U_{\mathcal{C}} = \mathcal{H}(-\vec{k}),
\end{equation}
where we define $U_{\mathcal{C}} = \sigma^1\otimes 1_{8\times 8}$.
The anti-unitary charge-conjugation operator is then given by
$U_{\mathcal{C}}\mathcal{K}$, where $\mathcal{K}$ denotes complex
conjugation.

Considering physical symmetries, time reversal $\mathcal{T}$ gives
rise to an analogous matrix-level constraint
\begin{equation}
U^{-1}_{\mathcal{T}}\mathcal{H}^*(\vec{k})U_{\mathcal{T}} = \mathcal{H}(-\vec{k}),
\end{equation}
where we define $U_{\mathcal{T}} = 1_{8\times 8}\otimes
(i\sigma^2)$. Correspondingly, the anti-unitary time-reversal
operator is $U_{\mathcal{T}} \mathcal{K}$. Note that time reversal
acts non-projectively in all PSG classes because we use gauge
freedom to fix $\phi_{\mathcal{T}} (\vec{r}_\mu) = 0$.

In contrast, inversion $I = \overline{C}_6^3$ acts projectively on
the spinons and generates the matrix-level constraint
\begin{equation}
U_I^{-1}(\vec{k}) \mathcal{H}(\vec{k}) U_I(\vec{k}) = \mathcal{H}(-\vec{k}),
\end{equation}
where $U_I(\vec{k})= \left(\sigma^3\right)^{n_{\overline{C}_6}}
\otimes \left(U_J\cdot I^2(\vec{k})\right)\otimes \sigma^0$, in
terms of the $4 \times 4$ diagonal form-factor matrix
\begin{equation}\label{formfactormatrix}
I(\vec{k}) = \text{Diag}\left(1,e^{i\vec{k}\cdot
\hat{\varepsilon}_1},e^{i\vec{k}\cdot
\hat{\varepsilon}_2},e^{i\vec{k}\cdot \hat{\varepsilon}_3}\right),
\end{equation}
and the diagonal matrix $U_J
=\text{Diag}\left((-1)^{n_{ST_1}},1,1,1\right)$.

The symmetries $\mathcal{C}$, $\mathcal{T}$, and $I$ result in
important general spectral features. First, the symmetry $I \circ
\mathcal{T}$ guarantees that each energy level is doubly degenerate,
according to Kramers theorem. Second, the symmetry $I \circ
\mathcal{C}$ leads to an additional double degeneracy for any
non-zero-energy level, which is connected to the redundant
description in Eqs.~(\ref{H}) and (\ref{HH}). The two symmetries
together thus result in a generic four-fold degeneracy at each
energy level $E>0$ shared by momenta $\pm \vec{k}$. Note that the
degeneracy may be smaller or larger at special
time-reversal-invariant momenta ($\vec{k} = -\vec{k}$) because there
are half as many physical degrees of freedom but, on the other hand,
pure point-group symmetries (e.g., inversion) may lead to additional
degeneracy.

The degeneracy of zero-energy levels is more subtle as it may be
affected by the diagonalizability of the Hamiltonian matrix
$\mathcal{H} (\vec{k})$. Since the low-energy physics is the main
focus of our study, this issue will be addressed in a separate
section (see Sec.~\ref{HD}).

\subsection{Condensation domains: a ``phase diagram'' for paraphases\label{subsection: phasediagram}}

\begin{figure*}[t]
\centering
\includegraphics[width=\textwidth]{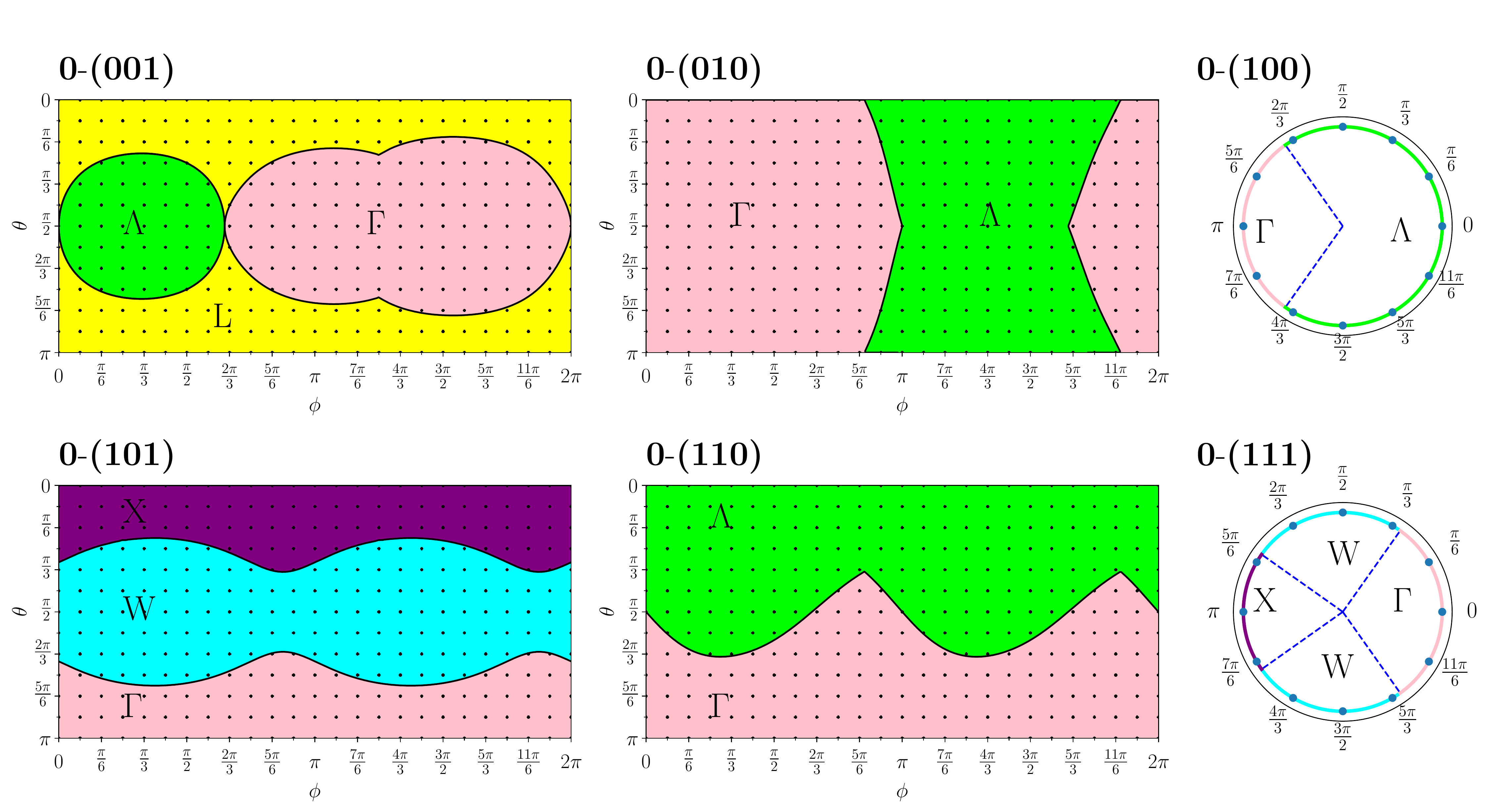}
\caption{Condensation ``phase diagrams'' for the NN mean-field
ans\"atze of the six 0-flux PSG classes 0-(001), 0-(010), 0-(100),
0-(101), 0-(110), and 0-(111). The complete phase diagram at NN
level is 1D for classes 0-(100) and 0-(111), 2D for classes 0-(001)
and 0-(101), and 3D for classes 0-(010) and 0-(110). The parameters
$(\psi,\theta,\phi)$ are related to the mean-field parameters
according to Table~\ref{table3}. For the classes 0-(010) and
0-(110), only a 2D slice with $\psi=0$ is shown. The slices for
other values of $\psi$ share the same qualitative behavior as the
$\psi=0$ slice, e.g., they also consist of two phases $\Gamma$ and
$\Lambda$.}\label{condensation_phase}
\end{figure*}

The use of bosonic mean-field Hamiltonians, obtained from the spinon
decomposition in Eq.~(\ref{parton}), facilitates the study of phase
transitions between spin liquids and magnetically ordered phases.
Indeed, by lowering the chemical potential $\mu$, there is a
critical chemical potential $\mu_c$ at which the bosonic spinons
undergo Bose-Einstein condensation at some critical momenta
$\vec{k}_c$ and the system thus develops magnetic order.

For the mean-field Hamiltonian in each PSG class, the critical
chemical potential $\mu_c$ is a function of the mean-field
parameters (see Table \ref{tablemu} in Appendix \ref{sectablemu} for details). While the value of $\mu_c$ changes continuously with the
mean-field parameters, and this variation of $\mu_c$ is thus locally
analytic, it globally separates into domains across which the
variation of $\mu_c$ is non-analytic. These domains of analyticity
of $\mu_c$ are reminiscent of the domains of analyticity of the free
energy, which define phases in thermodynamics. However, the analogy
is not perfect as each such domain may give rise to several true
phases on crossing the phase transition into magnetic order (i.e.,
when taking $\mu<\mu_c$). We therefore coin the word
\emph{paraphases} to describe the distinct domains of analyticity of
$\mu_c$. Restated, each paraphase is a connected region of phase
space in which the unstable manifold of condensation modes varies
smoothly.

Following this logic, the six mean-field Hamiltonians are further
divided into $15$ paraphases (see Fig.~\ref{condensation_phase}).
The analytical expressions for the paraphase boundaries are given in
Table \ref{table4}, while the distinct critical momenta $\vec{k}_c =
\Gamma, \mathrm{L}, \Lambda, \mathrm{X}, \mathrm{W}$ characterizing
the various paraphases are explained in Table \ref{table1}. Finally,
the distinct expressions for the critical chemical potentials
$\mu_c$ in the $15$ paraphases are specified in Appendix
\ref{sectablemu}.

\begin{table}
\begin{tabular}{c|c|c}
\hhline{===}Class&Adjacency&Paraphase boundary\\
\hline \multirow{4}{*}{0-$(001)$} &$\Lambda$ vs
$\mathrm{L}$&$2ac+c^2-b'^2=0$\\
\cline{2-3}
&\multirow{2}{*}{$\Gamma$ vs $\mathrm{L}$}
&$2a^2+ac-b'^2=0$ for $c>a$ and\\
&& $ac-4c^2+b'^2=0$ for
$c<a$\\
\cline{2-3} &$\Lambda$ vs $\Gamma$& $2a+c=b'=0$\\
\hline
\multirow{2}{*}{0-$(010)$}&\multirow{2}{*}{$\Gamma$ vs $\Lambda$}&
$\sqrt{4 (a-c)^2+3
(\nu- b')^2}$\\
&&$=-2 a-4 c+\sqrt{3}|\nu+b'|$\\\hline 0-$(100)$&$\Gamma$ vs $\Lambda$ &
$\sqrt{2}b=\pm c'$\\\hline
\multirow{2}{*}{0-$(101)$}&$\Gamma$ vs $\mathrm{W}$& $-2 b^2 + 3 c'^2 + 2 c'
a' + a'^2=0$, $b>0$\\\cline{2-3}
&$\mathrm{X}$ vs $\mathrm{W}$ &$-2 b^2 + 3 c'^2 + 2 c' a' + a'^2=0$, $b<0$\\
\hline
\multirow{2}{*}{0-$(110)$}&\multirow{2}{*}{$\Gamma$ vs
$\Lambda$}&
$\sqrt{(\nu+a'-2c')^2+2(\nu-a')^2}$\\
&&
$=4b+\sqrt{3}|\nu +a'+2 c'|$\\
\hline
\multirow{2}{*}{0-$(111)$}&$\Gamma$ vs $\mathrm{W}$&$\sqrt{2}b=\pm
c'$\\\cline{2-3}
&$\mathrm{X}$ vs $\mathrm{W}$&$b=\pm \sqrt{2}c'$\\
\hhline{===}
\end{tabular}
\caption{Paraphase boundaries of the NN ans\"atze.}\label{table4}
\end{table}

Note that, for the PSG classes with $n_{ST_1}=1$, the PSG result for
the screw operation $S$ depends on the spatial coordinates, and it
is convenient to shift the entire BZ by the translation
$\vec{k}\rightarrow \vec{k}-\pi(1,1,1)$. Such a shift of the BZ can
be thought of as a gauge transformation of the spinons, which does
not modify any physical quantities on the spin level. This shift is
assumed throughout the paper and is already taken into account when
specifying the condensation momenta in
Fig.~\ref{condensation_phase}.

Note also that the region $\Lambda$ supports a one-dimensional
manifold of condensation momenta. Since the only physical symmetries
are discrete space-group and time-reversal symmetries, this
ground-state continuum must be accidental, i.e., the result of
restricting the mean-field Hamiltonians to NN level. Indeed, when
including infinitesimal NNN parameters, we see that the condensation
regions are reduced from $\Lambda$ to either $\Gamma$ or
$\mathrm{L}$.

\begin{table}[b]
\begin{ruledtabular}
\begin{tabular}{cc}
Label  & Description \\
\hline $\Gamma$ & $(0,0,0)$\\
$\mathrm{L}$      &  $\pi(\delta_1,\delta_2,\delta_3)$, where
$\delta_1,\delta_2,\delta_3\in
\{1,-1\}$\\
$\Lambda$& $k(\delta_1,\delta_2,\delta_3)$, where $k \in [-\pi,\pi]$ and $\delta_1,\delta_2,\delta_3\in \{1,-1\}$ \\
$\mathrm{X}$      & $\mathrm X^1=2\pi(1,0,0)$, $\mathrm X^2=2\pi(0,1,0)$, $\mathrm X^3=2\pi(0,0,1)$ \\
$\mathrm{W}$      & $\pi(2,\pm1,0)$ and all permutations of the $3$ components \\
\end{tabular}
\end{ruledtabular}
\caption{\label{table1}%
Possible sets of condensation momenta.}
\end{table}

\subsection{Critical spectra}

The critical spectra of the $15$ paraphases, corresponding to $\mu =
\mu_c$ in each case, are shown in Fig.~\ref{DSF_NN_1En9_shifted},
along with the associated dynamical spin structure factors, obtained
on the mean-field level. Generically, each of these spectra consists
of four bands, which is consistent with the fourfold degeneracy of
each band. While certain spectra have distinguishing features, not
all paraphases can be {\em fully} distinguished by their spectra, as
some spectral characteristics are shared by multiple paraphases.
Among other features, several spectra show a quasi-mirror-reflection
symmetry (in terms of energy) between two bands, which accounts for
certain high-energy features in the dynamic spin structure factor
(see Sec.~\ref{subsec:criticalSF}).

Most importantly, however, the critical paraphases can be divided
into two classes, characterized by linear and quadratic dispersions
at low energies. In terms of the dynamical critical exponent $z$,
defined by $\omega \sim |k-k_c|^z$ and specified for each paraphase
in Table~\ref{six0fluxclasses}, these two classes are labeled by
$z=1$ and $z=2$, respectively. As we later show, paraphases with
$z=1$ and $z=2$ correspond to different critical field theories,
which determine the critical exponents of various physical
observables, such as the heat capacity and the magnetic
susceptibility, and thus lead to distinct experimental signatures.

\subsection{Hamiltonian diagonalizability}\label{HD}

From a technical point of view, the distinction between $z=1$ and
$z=2$ theories becomes evident when we try to diagonalize the
Hamiltonian matrix in Eq.~\eqref{HH}. In general, we seek a change
of basis for the bosonic operators,
\begin{equation}\label{diagonalh}
B_{\vec{k}} = V(\vec{k}) \widetilde{B}_{\vec{k}},
\end{equation}
such that the Hamiltonian in Eq.~\eqref{H} is of the form
\begin{equation}
H = \sum\limits_{\vec{k} \in \text{BZ}} \widetilde{B}^\dag_{\vec{k}} \Lambda(\vec{k}) \widetilde{B}_{\vec{k}},
\end{equation}
where $\Lambda(\vec{k}) = V^\dag(\vec{k}) \mathcal{H}(\vec{k})
V(\vec{k})$ is a $16 \times 16$ diagonal matrix, and $V(\vec{k}) \in
\text{SU}(8,8)$ satisfies
\begin{equation}
V(\vec{k}) J V^\dag(\vec{k}) = J, \qquad J = \sigma^3\otimes
1_{8\times 8},
\end{equation}
ensuring that this change of basis is a canonical transformation.
The matrices $\Lambda(\vec{k})$ and $V(\vec{k})$ can be found by
solving the generalized eigenvalue problem
\begin{equation}\label{jhala}
J\mathcal{H}(\vec{k})\bm{a}_{\vec{k},i}= \lambda_{\vec{k},i}
\bm{a}_{\vec{k},i},
\end{equation}
where the eigenvalues $\lambda_{\vec{k},i}$ give the diagonal
elements of the matrix $J\Lambda(\vec{k})$ and the eigenvectors
$\bm{a}_{\vec{k},i}$ form the columns of the matrix $V(\vec{k})$.
However, since $J\mathcal{H}(\vec{k})$ is not necessarily Hermitian
(or even normal), it is not guaranteed that such a matrix
$V(\vec{k})$ actually exists.

In particular, it may happen at the critical chemical potential $\mu
= \mu_c$ that there are not enough independent eigenvectors for the
zero eigenvalues $\lambda_{\vec{k}_c}=0$. Physically, this scenario
means that we cannot diagonalize our critical Hamiltonian by a
canonical transformation of bosonic creation and annihilation
operators, and instead we must decompose our complex operators
$B_{\vec{k}}$ according to
\begin{equation}\label{bxp11}
B_{\vec{k}} = \frac{1}{\sqrt{2}}(\hat X_{\vec{k}}+ i \hat
P_{\vec{k}}),
\end{equation}
where $\hat X_{\vec{k}}$ and $\hat P_{\vec{k}}$ are $16$-dimensional
vectors of real operators, analogous to the position and momentum
operators in first quantization. In terms of these new operators,
the analog for the change of basis in Eq.~(\ref{diagonalh}) is
\begin{equation}
\left(\begin{array}{c} \hat X_{\vec{k}} \\ \hat P_{\vec{k}}
\end{array}\right) = W(\vec{k}) \left(\begin{array}{c} \hat Y_{\vec{k}} \\ \hat
Q_{\vec{k}} \end{array}\right),
\end{equation}
where the $32 \times 32$ matrix $W(\vec{k})$ satisfies
\begin{equation}
W(\vec{k}) \mathcal{E} W^T(\vec{k}) = \mathcal{E}, \qquad
\mathcal{E} = i\sigma^2\otimes 1_{16 \times 16}.
\end{equation}
Using this canonical change of basis, the Hamiltonian can then be
brought to the diagonal form
\begin{equation}
H = \sum_{\vec{k}, i} \alpha_{\vec{k},i}\hat y^2_{\vec{k},i} +
\beta_{\vec{k},i}\hat q^2_{\vec{k},i},
\end{equation}
where $\hat y_{\vec{k},i}$ and $\hat q_{\vec{k},i}$ are the
components of $\hat Y_{\vec{k}}$ and $\hat Q_{\vec{k}}$,
respectively, and the new eigenvalues are related to the original
ones by
\begin{equation}\label{lambdaalphabeta}
\lambda^2_{\vec{k},i} =  \alpha_{\vec{k},i}\beta_{\vec{k},i}.
\end{equation}
Importantly, however, unlike the original method of diagonalization,
$B_{\vec{k}} \rightarrow \widetilde{B}_{\vec{k}}$, which may fail if
$J\mathcal{H}(\vec{k})$ is a defective matrix, the alternative
method of diagonalization, $(\hat X_{\vec{k}},\hat P_{\vec{k}})
\rightarrow (\hat Y_{\vec{k}}, \hat Q_{\vec{k}})$, always works.

For any zero mode $i$ at a critical momentum $\vec{k}_c$, we have
$\alpha_{\vec{k}_c,i}\beta_{\vec{k}_c,i}=0$ from
Eq.~\eqref{lambdaalphabeta}. The diagonalizability of the critical
Hamiltonian $\mathcal{H}(\vec{k}_c)$ is then determined by the
following simple criterion:
\begin{itemize}
\item if $\alpha_{\vec{k}_c,i}=\beta_{\vec{k}_c,i}=0$, the Hamiltonian can be diagonalized in the original basis of creation and annihilation operators;
\item otherwise, either $\alpha_{\vec{k}_c,i}=0$, $\beta_{\vec{k}_c,i}\neq 0$ or $\alpha_{\vec{k}_c,i}\neq 0$, $\beta_{\vec{k}_c,i}= 0$; the Hamiltonian is not diagonalizable in any creation-annihilation-operator basis, meaning that the SU($8$,$8$) transformation is singular.
\end{itemize}

To understand how these two scenarios for the diagonalizability lead
to theories of $z=2$ and $z=1$ types, respectively, we now switch to
the language of path integrals and consider the critical low-energy
actions.

\subsection{Effective low-energy theories}

Our phase transitions from spin liquids to magnetic orders, driven
by a change in the chemical potential $\mu$, are prototypes of
quantum critical points (QCPs). At such a QCP, one can write down an
effective theory in terms of the low-energy degrees of freedom. We
assume a single condensing eigenmode obtained from the Hamiltonian
$\mathcal{H}(\vec{k}_c)$, denoted by $\widetilde{b}_{\vec{k}_c}$.
Including spatial fluctuations, we promote this eigenmode to a field
$\phi(\tau,\vec{x})$ and consider the imaginary-time action $S =
\int\mathcal{L} \,d^3x d\tau$. If the Hamiltonian is diagonalizable,
the critical Lagrangian becomes
\begin{equation}
\label{FTz_2}\mathcal{L} = \overline{\phi} \left(\partial_\tau -
\mu_{ij}\partial_i\partial_j \right)\phi,
\end{equation}
describing a massless field $\phi$ at the QCP. The corresponding
action is invariant under the rescaling
\begin{equation}
\tau\rightarrow \tau e^{-l}, \qquad  x\rightarrow x e^{-l/2},\qquad
\phi\rightarrow \phi e^{3l/4},
\end{equation}
from which we can immediately deduce that the dynamical critical
exponent is $z=2$.

However, the mass of $\phi$ should be generally considered as a
tensor of real fields $\chi$ and $\pi$, which are the real and
imaginary components of $\phi$, such that
\begin{equation}
\phi = \chi + i \pi.
\end{equation}
In the Hamiltonian language, these two components correspond to the
``position'' and ``momentum'' operators in Eq.~\eqref{bxp11}.
Consequently, if the Hamiltonian is not diagonalizable, only one of
these components is massless at the QCP. Assuming without loss of
generality that $\chi$ is massive and $\pi$ is massless, the
critical Lagrangian becomes
\begin{equation}
\mathcal{L}= 2i \chi\partial_\tau \pi + r^2\chi^2 - \pi
\nu_{ij}\partial_i\partial_j  \pi.
\end{equation}
By integrating out the massive field $\chi$ and rescaling the
massless field as $\pi \rightarrow r \pi$, we finally obtain
\begin{equation}\label{FTz_1}
\mathcal{L}_{\text{eff}} = \pi\left(\partial^2_\tau - \widetilde
\nu_{ij}\partial_i\partial_j\right)\pi.
\end{equation}
This effective action is invariant under the rescaling
\begin{equation}
\tau \rightarrow \tau e^{-l},\qquad x\rightarrow x e^{-l},\qquad
\pi\rightarrow \pi e^l,
\end{equation}
from which we can immediately deduce that the dynamical critical
exponent is $z=1$.

These two distinct QCPs, characterized by critical exponents $z=2$
and $z=1$, respectively, are reminiscent of the QCPs governing phase
transitions from quantum paramagnets to XY antiferromagnets
\cite{zapf2014bose}. If such a transition is induced by an external
magnetic field, the QCP is described by the $z=2$ critical theory in
Eq.~\eqref{FTz_2}, while if the transition is induced by pressure
and is thus time-reversal symmetric, the QCP is described by the
$z=1$ critical theory in Eq.~\eqref{FTz_1}.

\begin{table}
\begin{ruledtabular}
\begin{tabular}{c|l}
\multirow{2}{*}{Class}  & Independent nonzero parameters up to NN terms \\
& and parameterized by\\
\hline
0-(001)& $(a,c,b') = (\sin \theta \cos \phi,\sin\theta \sin \phi,\cos \theta)$\\
\multirow{2}{*}{0-(010)}& $(\nu,a,c,b') =$\\
& $\quad(\cos \psi,\sin\psi\sin \theta \cos \phi,\sin\psi\sin\theta \sin \phi,\sin\psi\cos \theta)$\\
0-(100)& $(b,c')= (\cos \phi,\sin\phi)$\\
0-(101)& $(b,a',c') = (\cos \theta, \sin\theta \cos\phi,\sin\theta \sin \phi)$\\
\multirow{2}{*}{0-(110)}& $(\nu,b,a',c') = $\\
& $\quad(\cos \psi,\sin\psi\cos \theta,\sin\psi\sin \theta \cos \phi,\sin\psi\sin\theta \sin \phi)$\\
0-(111)& $(b,c')= (\cos \phi,\sin\phi)$
\end{tabular}
\end{ruledtabular}
\caption{\label{table3}
Parametrization of the NN mean-field ans\"atze using generalized
spherical coordinates $(\psi,\theta,\phi)$ for the phase diagram in
Fig.~\ref{condensation_phase}.}
\end{table}

\subsection{Spin condensation: order patterns}\label{spincond_1}

We are now ready to describe the spin orders obtained by condensing
the spinons in each of the $15$ critical paraphases. When the
chemical potential $\mu$ reaches its critical value $\mu_c$, certain
spinons $\widetilde{b}_{\vec{k}_c}$ at critical momenta $\vec{k}_c$
condense and thereby acquire macroscopic occupation numbers $\langle
\widetilde{b}_{\vec{k}_c} \rangle$. We can then use these $\langle
\widetilde{b}_{\vec{k}_c} \rangle$ as order parameters and detect
spin orders by looking at order parameter bilinears, which,
according to the spinon decomposition in Eq.~\eqref{parton}, recover
spin expectation values.

So far, several types of orders have been successfully identified in
pyrochlore materials, most of which do not break translation
symmetry. These zero-momentum orders correspond to representations
of the point group $O_{\text{h}}$ and can thus be analyzed by the
standard representation theory of groups. We will defer such an
effort to the next section. In this subsection, we select several
paraphases with definite ordering signatures and explicitly
calculate the spin expectation values via condensing spinons. This
way, we capture a limited set of orders, which correspond to
irreducible representations (irreps) of the tetrahedral group
$T_{\text{d}}$ (see Appendix~\ref{app:Yan}), and show that all such orders can be obtained from
at least one of the six $\mathbb{Z}_2$ spin liquids. We mainly
restrict ourselves to NN terms in the spinon Hamiltonian but include
NNN terms whenever necessary.

One must bear in mind that the simplified irrep analysis on these
explicit spin-condensation orders may be incomplete. For example, we
will find from such an analysis that pure all-in-all-out order may
be obtained in the paraphase 0-$(100)\Gamma$, while a full
representation-theory analysis in
Sec.~\ref{subsec:hiddenintertwined} leads to Table~\ref{tableGL},
which indicates that all-in-all-out order is always intertwined with
some hidden orders (i.e., it can never appear alone). Still, the
na\"ive spin-condensation analysis in this subsection is a good
starting point to build some insight into how the six spin liquids
are physically distinct from each other.

\subsubsection{All-in-all-out order}

We consider the paraphase 0-$(100)\Gamma$, but also remark that the
paraphases 0-(101)$\Gamma$ and 0-(111)$\Gamma$ give similar results.
At the critical chemical potential $\mu_c$, the zero-energy subspace
is twofold degenerate. The zero-energy eigenvectors are obtained
from Eq.~\eqref{jhala} and are given by the time-reversal partners
$\bm{a}$ and $U_{\mathcal{T}}\bm{a}^*$. After condensing these two
modes, the corresponding operators $\widetilde{b}_{1,2}$ acquire
macroscopic occupation numbers $\langle \widetilde{b}_i\rangle = r_i
e^{i\phi_i}$, with $i=1,2$, implying $\langle B_{\vec{k}_c}\rangle =
\sum_{i=1,2}\bm{a}_i r_i e^{i \phi_i}$ at the critical momentum
$\vec{k}_c = \Gamma$. In terms of the $12$-component vector
$\mathbf{S} = (\vec{S}_0,\vec{S}_1,\vec{S}_2,\vec{S}_3)$ of the spin
components on the four sublattices, we have, up to a global
coefficient,
\begin{equation}\label{Srcs}
\mathbf{S} = r \mathbf{S}^r+\cos(\phi_1-\phi_2)\mathbf{S}^c+\sin(\phi_1-\phi_2)\mathbf{S}^s,
\end{equation}
where $r=(r_1^2-r_2^2)/(2r_1r_2)$, and $\mathbf{S}^{r,c,s}$ are
three equimodular and mutually orthogonal vectors (see Appendix \ref{sec:D1} for detail). Using the basis for the irreducible representations of $T_{\text{d}}$ (see Appendix~\ref{app:Yan}), it can be shown that this paraphase generically supports two orders:
the all-in-all-out order and the AFM order. One can obtain
pure all-in-all-out order [see Fig.~\ref{fig:2a} for illustration] by setting particular values for the
condensation parameters $r_{1,2}$ and $\phi_{1,2}$.

\subsubsection{XY antiferromagnetic order}

The paraphase 0-(110)$\Gamma$ has a non-diagonalizable critical
Hamiltonian, because $\mathcal{H}$ has four zero-energy eigenvalues,
but the nullspace of $J\mathcal{H}$ is only two dimensional, spanned
by the time-reversal partners $\bm{a}$ and
$U_{\mathcal{T}}\bm{a}^*$. We therefore switch to the
position-momentum representation $(\hat{x},\hat{p})$, according to
Eq.~\eqref{bxp11}. The critical Hamiltonian is then diagonalized by
a basis change $(\hat{x},\hat{p})\rightarrow(\hat{y},\hat{q})$ and
takes the low-energy form
\begin{equation}
H = \hat{q}_1^2+\hat{q}_2^2+0\cdot\hat{y}_1^2+0\cdot \hat{y}_2^2,
\end{equation}
which contains two gapless modes $\hat{y}_1$ and $\hat{y}_2$. To
minimize the energy, we must have $\langle \hat q_i\rangle = 0$ and,
due to the uncertainty principle, $\hat y_i$ must fluctuate
maximally. In terms of $y_i = \langle \hat y_i \rangle$, we then
find $\langle B_{\vec{k}_c}\rangle =\bm{b} y_1+
U_{\mathcal{T}}\bm{b} y_2$ at $\vec{k}_c = \Gamma$ for some vector
$\bm{b}$ determined by $\bm{a}$, and the final result for spin
configuration becomes
\begin{eqnarray}
\mathbf{S} &\propto&
\left(C,-S_2,-S_1,C,S_2,S_1,-C,-S_2,S_1,-C,S_2,-S_1\right),
\nonumber \\
&& S_1 = \sin\left(\frac{\pi}{12}-\theta\right), \quad S_2 =
\sin\left(\frac{\pi}{4}+ \theta\right), \\
&& C = \cos\left(\frac{\pi}{12}+\theta\right), \nonumber
\end{eqnarray}
where $\cos \theta = y_1^2-y_2^2$ and $\sin \theta = y_1 y_2/2$.
This spin configuration, shown in Fig.~\ref{fig:2b},
corresponds to the ``XY'' order of the irrep $E$ obtained in
Eq.~(39) of Ref.~\cite{PhysRevB.95.094422} after a redefinition
$\theta \rightarrow \theta-\frac{\pi}{12}$.

\subsubsection{Ferromagnetic order: collinear and non-collinear\label{subsubsec:ferri}}

\begin{figure*}[t]
\centering
	\centering
	\subfigure[]{\label{fig:2a}\includegraphics[height=0.22\textwidth]{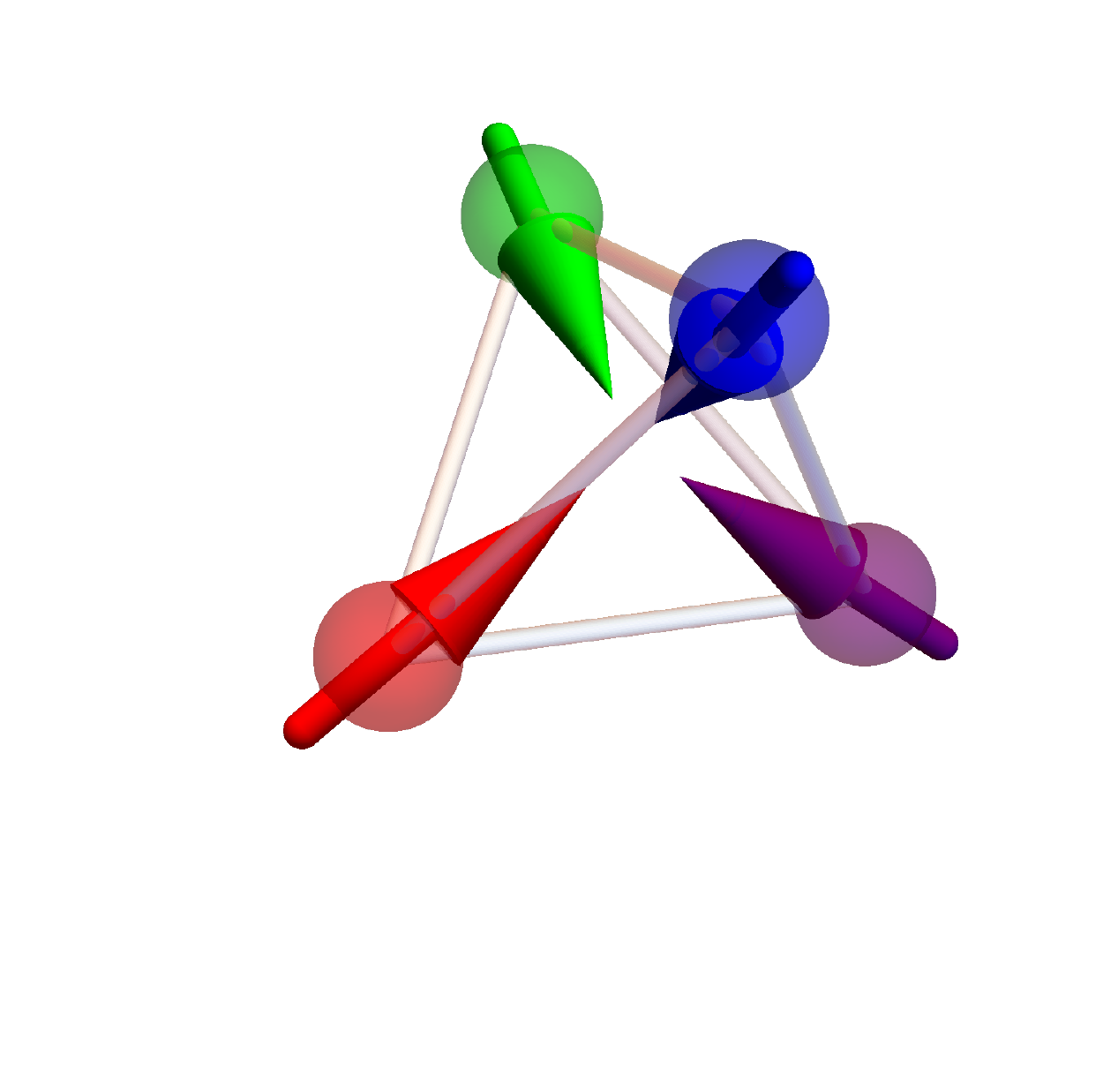}}
	\subfigure[]{\label{fig:2b}\includegraphics[height=0.22\textwidth]{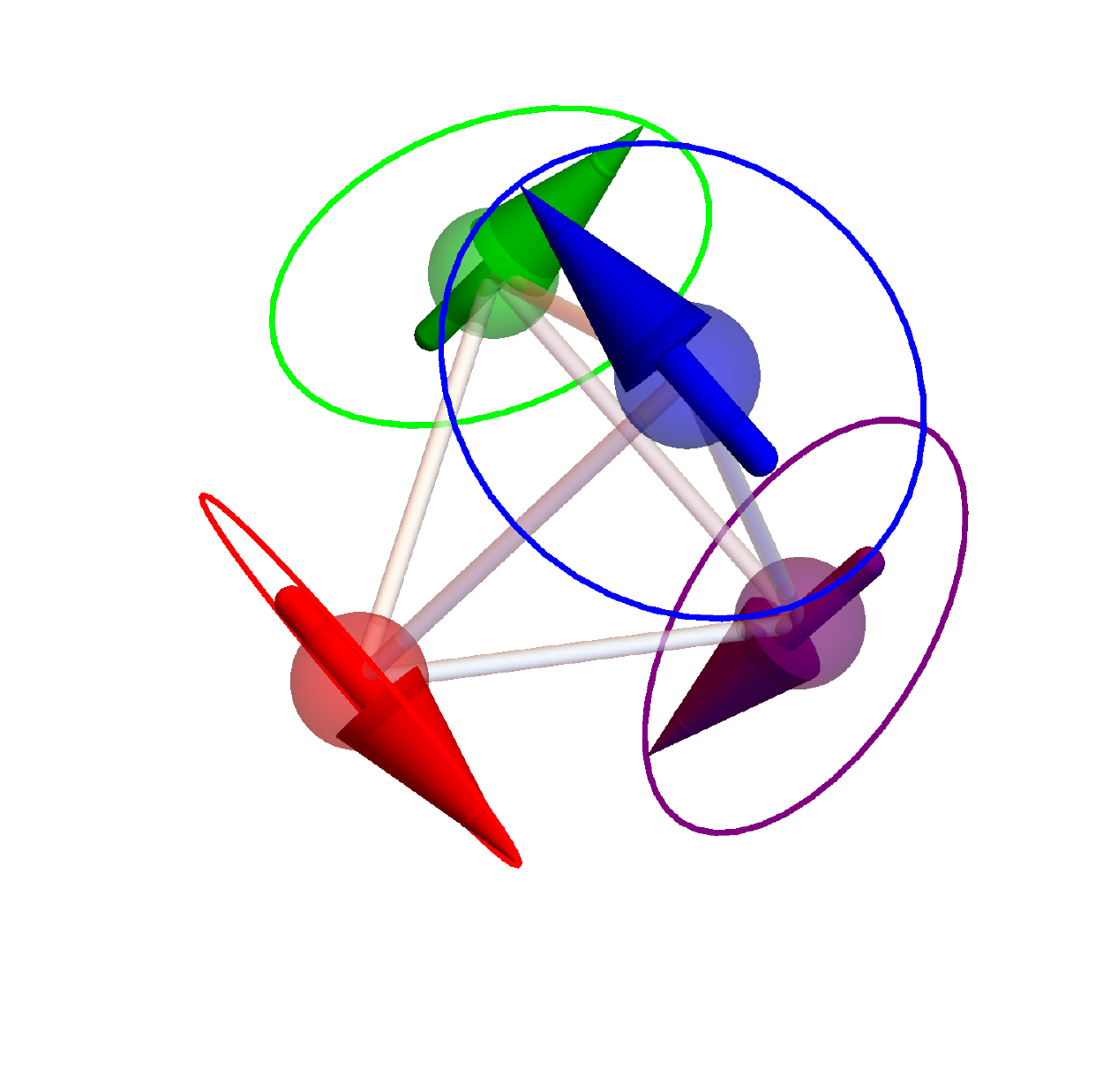}}
	\subfigure[]{\label{fig:2c}\includegraphics[height=0.22\textwidth]{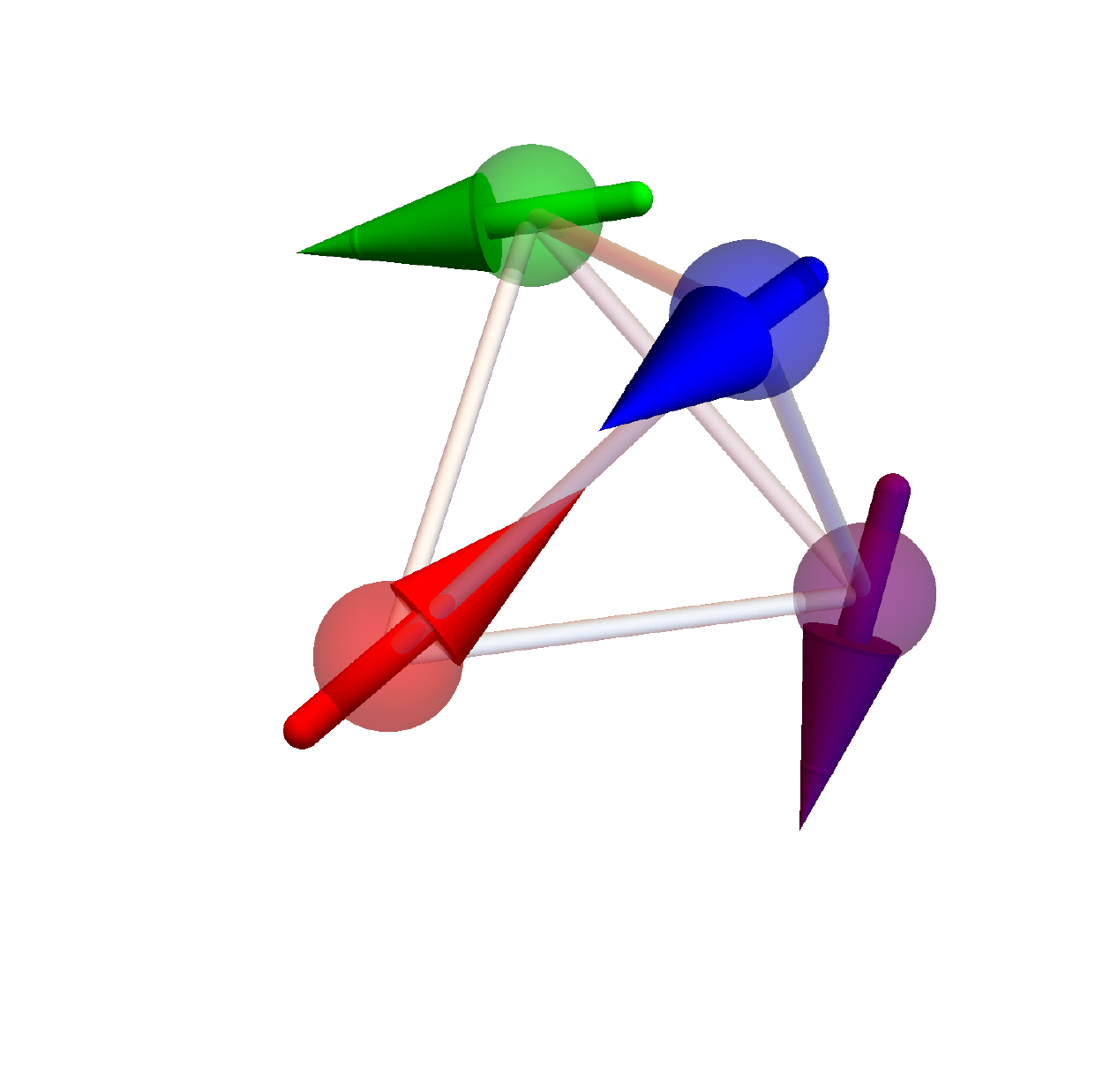}}
	\subfigure[]{\label{fig:2d}\includegraphics[height=0.22\textwidth]{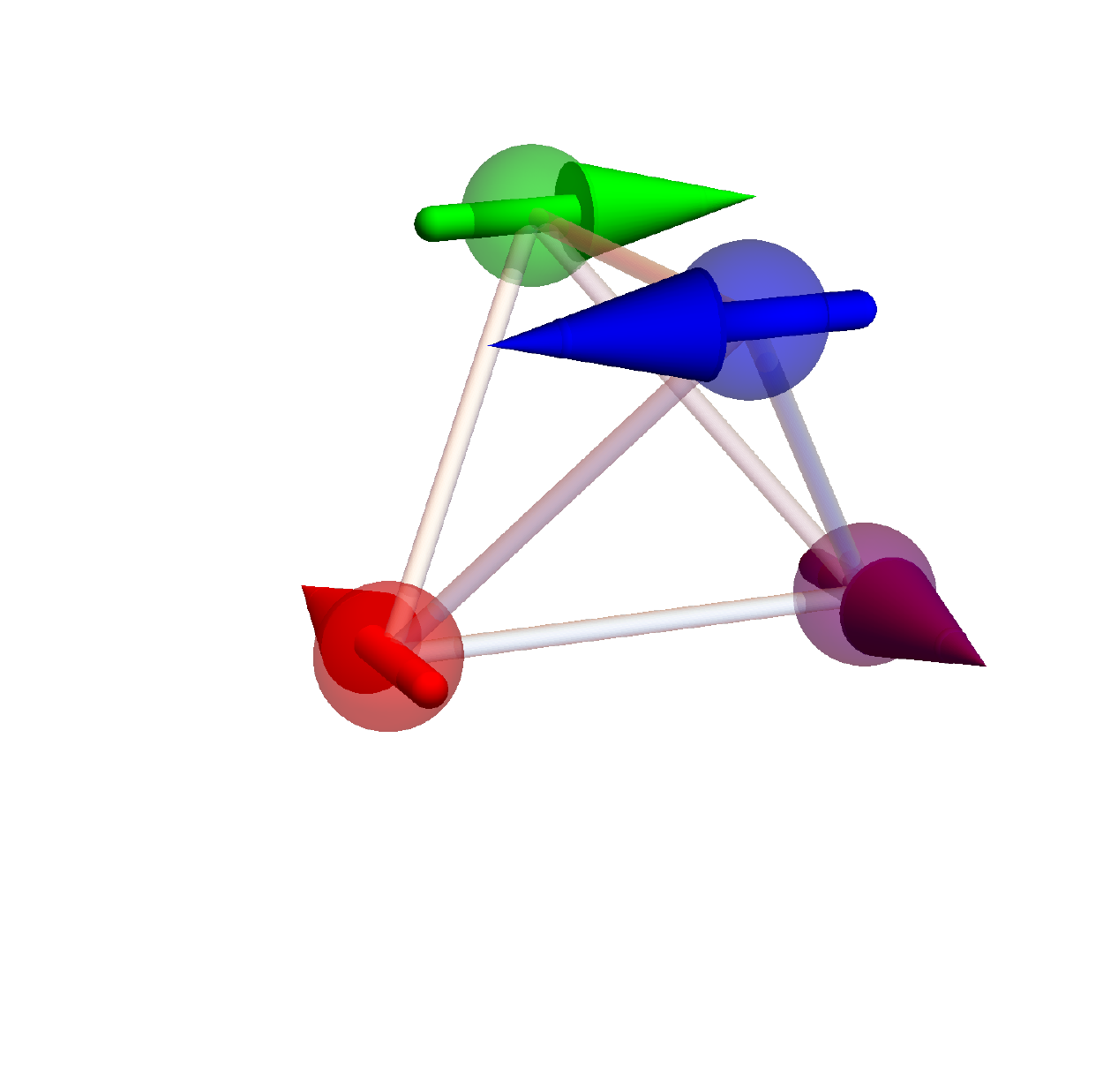}}
\caption{Typical spin order for  (a) the paraphases 0-(100)$\Gamma$, 0-(101)$\Gamma$, and 0-(111)$\Gamma$ (all-in all-out order), (b) the paraphase 0-(110)$\Gamma$ (the $XY$ order), (c) the paraphase 0-$(001)\Gamma$ (ferrimagnetic order), and (d) the paraphase 0-(101)$\mathrm{X}$ (the Palmer-Chalker order).}\label{spinconfs}
\end{figure*}

For the paraphase 0-(001)$\Gamma$, all pairing terms vanish at the
$\Gamma$ point at the NN level and, solving the hopping part at
$\mu_c$, we find that the zero-energy subspace is spanned by the
time-reversal partners $\bm{a}$ and $U_{\mathcal{T}}\bm{a}^*$. There
are two cases depending on the expression for $\mu$ in terms of the
mean field parameters. When $\mu=-6a$, all four spins point in the
same direction, which is the collinear FM order. When $\mu = 2a-8c$,
the spin vector $\mathbf{S}$ follows Eq.~\eqref{Srcs}, where
$\mathbf{S}^{r,c,s}$ are three equimodular and mutually
orthogonal vectors (see Appendix \ref{sec:D1} for detail). A typical spin configuration of such
ferrimagnetic nature is shown in Fig.~\ref{fig:2c}.

\subsubsection{Palmer-Chalker order}

The paraphase 0-(101)$\mathrm{X}$ has a non-diagonalizable critical
Hamiltonian, because $\mathcal{H}$ has eight zero-energy
eigenvalues, but the nullspace of $J\mathcal{H}$ is only four
dimensional. Switching to the $(\hat x,\hat p)$ representation and
diagonalizing the Hamiltonian via the basis change $(\hat x,\hat
p)\rightarrow(\hat y,\hat q)$, we find that there are four gapless
modes $\hat y_{1,2,3,4}$ for each of the three critical momenta
$\vec{k}_c = \mathrm X^{1,2,3}$. The expression for $\langle
B_{\vec{k}_c}\rangle$ thus contains $12$ real parameters: the
expectation values of the maximally fluctuating modes $\hat
y_{1,2,3,4}$ at each critical momentum. Although most choices of
these condensation parameters give an order with an enlarged unit
cell, some special cases respect translation symmetry. For instance,
if condensation is restricted to $X^1$, the spin configuration,
shown in Fig.~\ref{fig:2d}, corresponds to a Palmer-Chalker
order, transforming under the irrep $T_2$.

\subsubsection{Non-uniform spinon condensations and partial orders}\label{nuscpo}

The spin expectation values $\vec{S}_{0,1,2,3}$ for the paraphase
0-(010)$\Gamma$ have different amplitudes on different sublattices, invalidating the irrep analysis that presupposed classically
ordered states of fixed-length spins. There is no a priori reason to
rule out such a non-uniform spin-amplitude state. It does, however,
correspond to a more ``exotic'' ordered phase in which the spin is
more ordered on some sublattices than others. This type of partially
ordered state has been proposed in the material
Gd${}_2$Ti${}_2$O${}_7$~\cite{PhysRevB.64.140407,
PhysRevLett.114.130601}\ and in various theoretical models.

\subsubsection{Spinon line orders}

The line orders $\Lambda$ appearing in classes 0-(001), 0-(010),
0-(100), and 0-(110) have accidental degeneracies, higher than
demanded by the lattice symmetry. This extra degeneracy is an
artifact of the restriction to NN ans\"atze, and should reduce to
discrete condensation momenta in the presence of further-neighbor
terms. If we include infinitesimal NNN terms to the mean-field
ansatz, using Table~\ref{MFTparameters}, we indeed see that line
condensation along $\Lambda$ shrinks to point condensation at either
$\Gamma$ or $\mathrm{L}$. However, if we increase these NNN terms,
the condensation points are shifted away from these high-symmetry
points.

Due to the large NNN parameter space, we were unable to exhaustively
study the effect of NNN terms on the NN mean-field ansatz. However,
for some purposes, the NN level ans\"atze may be adequate. For
example, as we explore in the next section, the line minima
contribute to substantial low-energy continua in the dynamical spin
structure factor. This feature should persist at intermediate
energies when small NNN terms are included.

\subsubsection{Multi-spinon condensation orders}

Spinon condensation at multiple critical momenta, in the paraphases
0-(001)$\mathrm{L}$, 0-(101)$\mathrm X/\mathrm W$, and
0-(111)$\mathrm X/\mathrm W$, allows for richer physics and is often
accompanied by an enlargement of the unit cell. As an example, we
look at the paraphase 0-(001)$\mathrm{L}$: there are two independent
zero-energy modes at each critical momentum $\mathrm{L}$, and the
four inequivalent $\mathrm{L}$ momenta thus give rise to an
eight-dimensional zero-energy subspace. The $16$-component
zero-energy modes at these critical momenta have a complicated
expression and do not form a representation of $T_{\text{d}}$,
thereby leading to non-uniform spinon condensation, as discussed
above. Indeed, if condensation is restricted to one of the four
inequivalent $\mathrm{L}$ momenta, we find that three of the four
sublattices have the same spin amplitude, while the fourth
sublattice has a different one.

\section{Experimental signatures\label{sec:exp}}

\subsection{Critical behavior of the heat capacity}

For each critical paraphase, the low-temperature heat capacity is
expected to follow a power law whose exponent is determined by the
low-energy spinon density of states in the critical theory. Indeed,
depending on the dynamical exponent $z$ and the spinon condensation
manifold (i.e., if spinons condense at points or along lines), this
low-energy density of states follows different power laws $g
(\epsilon) \sim \epsilon^{\alpha}$, where the possible values of
$\alpha$ are listed in Table \ref{tableHC}. The thermal energy due
to spinon excitations at temperature $T$ is then given by
\begin{equation}
E \sim \int d\epsilon \, g(\epsilon) \, \frac{\epsilon} {\exp
(\epsilon / T) - 1} \propto T^{2 + \alpha},
\end{equation}
and the heat capacity takes the form
\begin{equation}
C_V = \frac{dE} {dT} \propto T^{1 + \alpha}.
\end{equation}
We remark that line condensation is not stable against generic
perturbations, corresponding to further-neighbor terms in the
mean-field ansatz. Consequently, at the lowest temperatures, we
expect that the line-condensation paraphases are governed by the
same exponents as their point-condensation counterparts.
Nevertheless, if the NN mean-field ansatz is a good first
approximation, there is an intermediate temperature range in which
the approximate line condensation in such paraphases becomes
manifest and therefore the line-condensation exponents in Table
\ref{tableHC} are experimentally observable.

\begin{table}[t]
\begin{tabular*}{0.47\textwidth}{@{\extracolsep{\fill}} c | c | c | c }
\hline \hline
Condensation               & Dynamical                 & Density of states:                        & Heat capacity:     \\
manifold                   & exponent                  & $g (\epsilon) \propto \epsilon^{\alpha}$  & $C_V \propto T^x$  \\
\hline
\multirow{4}{*}{Point(s)}  & \multirow{2}{*}{$z = 2$}  & \multirow{2}{*}{$\alpha = \frac{1}{2}$}   & \multirow{2}{*}{$x = \frac{3}{2}$}  \\
                           &                           &                                           &                                     \\
\cline{2-4}
                           & \multirow{2}{*}{$z = 1$}  & \multirow{2}{*}{$\alpha = 2$}             & \multirow{2}{*}{$x = 3$}            \\
                           &                           &                                           &                                     \\
\hline
\multirow{4}{*}{Line(s)}   & \multirow{2}{*}{$z = 2$}  & \multirow{2}{*}{$\alpha = 0$}             & \multirow{2}{*}{$x = 1$}            \\
                           &                           &                                           &                                     \\
\cline{2-4}
                           & \multirow{2}{*}{$z = 1$}  & \multirow{2}{*}{$\alpha = 1$}             & \multirow{2}{*}{$x = 2$}            \\
                           &                           &                                           &                                     \\
\hline \hline
\end{tabular*}
\caption{Power-law exponents of the low-energy spinon density of
states and the corresponding low-temperature heat capacity for
critical theories of dynamical exponents $z = 1,2$ where spinons
condense at points or along lines. \label{tableHC}}
\end{table}

\subsection{Critical spin structure factors\label{subsec:criticalSF}}

\begin{figure*}[t]
\centering
\includegraphics[width=\textwidth]{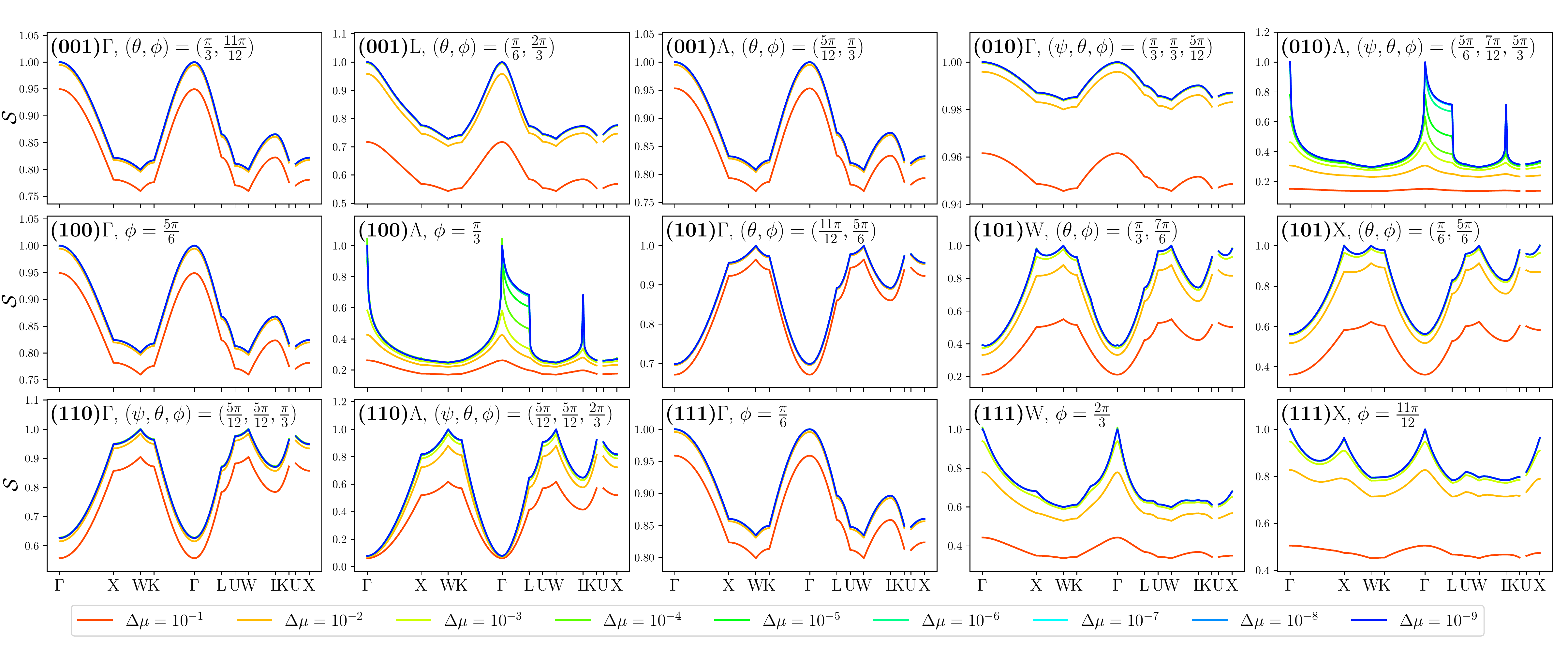}
\caption{Static spin structure factors for representative points in
each of the $15$ paraphases along the high-symmetry path in the
Brillouin zone. The chemical potential $\mu$ is above the critical condensation value by $\Delta \mu = 10^{-1}, 10^{-2},...,10^{-9}$ (in arbitrary units). The vertical axis is the spectral weight $\mathcal{S}$ normalized by the maximum intensity of the $\Delta \mu= 10^{-9}$ line along the path. 
In each paraphase, denoted by its PSG class and condensation
momenta, the representative point is specified by the mean-field
parameters.}\label{SSF_NN_shifted}
\end{figure*}

In this subsection, we present the most direct signatures of our
critical points between magnetic orders and their parent spin
liquids by computing both the static and the dynamic spin structure
factors for the $15$ paraphases. While our calculation is based on
mean-field theory, it still serves as a reference point for
classifying the possible spinon spectra in pyrochlore magnets.

The static structure factor (SSF) is defined as the spatial Fourier
transform of the equal-time spin-spin correlation function,
\begin{equation}
\mathcal{S}
(\vec{q})
=
\frac{1}{N}\sum\limits_{\vec{r}_\mu,\vec{r}'_\nu, \alpha}\left\langle \hat{S}^\alpha_{\vec{r}_\mu} \hat{S}^\alpha_{\vec{r}'_\nu}\right\rangle e^{i\vec{q}\cdot(\vec{r}_\mu-\vec{r}'_\nu)},
\end{equation}
where $\alpha = x,y,z$. We calculate this quantity using the
critical mean-field ans\"atze in Sec.~\ref{secMF}. Writing the $16
\times 16$ matrix $V(\vec{k})$ in Eq.~\eqref{diagonalh} as
\begin{equation}\label{btobeta}
V(\vec{k}) = \left(\begin{array}{cc} V_{11}(\vec{k})& V_{12}(\vec{k})\\
V _{21} (\vec{k}) & V _{22}(\vec{k})\end{array}\right),
\end{equation}
where the $8 \times 8$ blocks generally satisfy
\begin{equation}\label{ph_sym_on_V}
V^*_{11}(\vec{k}) = V_{22}(-\vec{k}),\qquad
V_{12}^*(\vec{k}) = V_{21}(-\vec{k})
\end{equation}
due to charge-conjugation ``symmetry'', the SSF becomes (see
Appendix~\ref{sec:E} for a detailed derivation)
\begin{equation}\label{SSFalpha}
\mathcal{S}(\vec{q})=\frac{1}{2N}
\sum\limits_{\vec{k},\alpha}\mathrm{Tr}\left[U^\alpha(\vec{k},\vec{q})\left(U^{\alpha}(\vec{k},\vec{q})\right)^\dag\right]
\end{equation}
in terms of the auxiliary $8 \times 8$ matrices
\begin{align}\label{SSFalpha1}
U^\alpha(\vec{k},\vec{q}) &= W^\alpha(\vec{k},\vec{q})+\left(W^{\alpha}(-\vec{k}+\vec{q},\vec{q})\right)^T,\notag\\
W^\alpha(\vec{k},\vec{q}) &= V^\dag_{12}(\vec{k})\left(I(\vec{q})\otimes \sigma^\alpha\right)
V_{11}(\vec{k}-\vec{q}),
\end{align}
where $I(\vec{q})$ is the $4 \times 4$ diagonal form-factor matrix
defined in Eq.~\eqref{formfactormatrix}. The resulting SSFs for
representative points in each of the $15$ paraphases are plotted in
Fig.~\ref{SSF_NN_shifted} for chemical potentials $\mu =
\mu_c+10^{-\delta}$, where $\delta = 1,2,\dots,9$, and $\mu_c$ is
the critical value given in Appendix~\ref{sectablemu}. When
numerically computing the SSF, we ensure convergence by taking a
momentum-space grid that does not contain any condensation momenta
$\vec{k}_c$.

For chemical potentials well above the critical value $\mu_c$, the
SSFs of the $15$ paraphases (not shown here) can be partitioned into
two classes, depending on the sum of the $\mathbb{Z}_2$ parameters
$n_{\overline{C}_6S}+n_{ST_1}+n_{\overline{C}_6}$ characterizing the parent spin liquid. Plotted
along the high-symmetry path in the BZ, the SSFs of the
$n_{\overline{C}_6S}+n_{ST_1}+n_{\overline{C}_6} = \text{odd}$ paraphases and those of the
$n_{\overline{C}_6S}+n_{ST_1}+n_{\overline{C}_6}=\text{even}$ paraphases resemble each other after an appropriate reflection in energy. This relation between the two
classes qualitatively survives as the chemical potential approaches
its critical value (see Fig.~\ref{SSF_NN_shifted}). For example,
depending on the the sum $n_{\overline{C}_6S}+n_{ST_1}+n_{\overline{C}_6}$ being even or odd, the SSF has either
a valley or a peak at the $\Gamma$ point. The distinction
between the two behaviors can be traced back to
Eqs.~\eqref{SSFalpha} and \eqref{SSFalpha1}. Since the SSF is the
squared trace norm of the matrix $U^\alpha$, which in turn is the
sum of two matrices $W^\alpha$, there is a cross term from the
product of the two matrices $W^\alpha$, physically corresponding to
the spinon pairing channel $\langle b^\dag_{\vec{k}_1,\mu\sigma_1}
b^\dag_{\vec{k}_3,\nu\sigma_3} \rangle \langle
b_{\vec{k}_1-\vec{q},\mu\sigma_2}b_{\vec{k}_3+\vec{q},\nu\sigma_4}\rangle$
(see Appendix~\ref{sec:E}), and we numerically find this cross term
to be negative for the $n_{\overline{C}_6S}+n_{ST_1}+n_{\overline{C}_6} = \text{even}$ paraphases and positive
for the $n_{\overline{C}_6S}+n_{ST_1}+n_{\overline{C}_6} = \text{odd}$ paraphases. Nevertheless, a deeper understanding of this connection to $n_{\overline{C}_6S}+n_{ST_1}+n_{\overline{C}_6}$ remains to
be found.

Also, there are general differences between the SSFs of the
paraphases governed by $z=1$ and $z=2$ critical theories,
respectively. For most of the $z=1$ paraphases, as the chemical
potential approaches its critical value, the SSF becomes a
non-differentiable function at certain momenta $\vec{q}$. This
feature is clearly observable in Fig. \ref{SSF_NN_shifted} for the
paraphases 
0-(010)$\Lambda$, 0-(100)$\Lambda$, 0-(101)$\mathrm{W}$, 
0-(111)$\mathrm{W}$, 0-(111)$\mathrm{X}$ at the $\Gamma$ point and
for the paraphases 0-(101)$\mathrm{W}$, 0-(101)$\mathrm{X}$,
0-(111)$\mathrm{W}$, 0-(111)$\mathrm{X}$ at the $\mathrm{X}$ point.
However, not all $z=1$ paraphases conform to this rule; the SSFs of
the paraphases 0-(110)$\Gamma$ and 0-(110)$\Lambda$ do not reveal
any singular behavior along the high-symmetry path in the BZ.
Instead, they resemble the SSFs of $z=2$ paraphases, which are
smooth across the entire BZ.

\begin{figure*}[t]
\centering
\includegraphics[width=\textwidth]{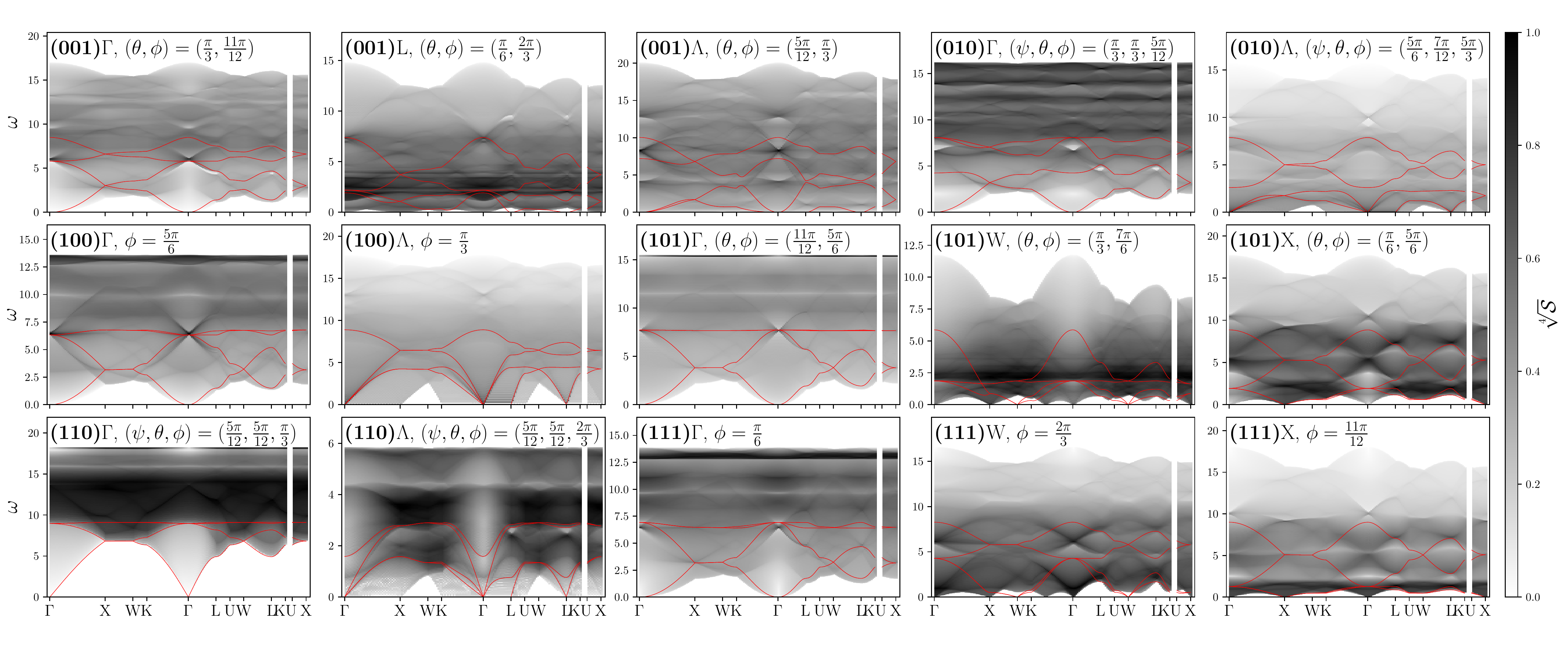}
\caption{Dynamic spin structure factors (gray) and spinon spectra
(red) for representative points in each of the $15$ paraphases along
the high-symmetry path in the Brillouin zone. The vertical axis is
the energy $\omega$ in arbitrary units, while the gray scale is the quartic root of the spectral weight (power is chosen such that maximum resolution is ensured), $\sqrt[4]{\mathcal{S}}$, normalized by its maximum intensity along the path. The chemical potential $\mu$ is $10^{-9}$ above the critical condensation value. In each paraphase, denoted by its PSG class and condensation momenta, the representative point is
specified by the mean-field parameters.}\label{DSF_NN_1En9_shifted}
\end{figure*}

To understand these features, we consider the dynamic structure
factor (DSF), which provides further information on the dynamics of
spinons. This quantity is defined as the spatial and temporal
Fourier transform of the spin-spin correlation function,
\begin{equation}\label{DSF}
\mathcal{S}
(\omega,\vec{q})
=\frac{1}{2\pi N}
\int^\infty_{-\infty} dt \sum\limits_{\vec{r}_\mu,\vec{r}'_\nu,\alpha}\left\langle \hat{S}^\alpha_{\vec{r}_\mu}(t) \hat{S}^\alpha_{\vec{r}'_\nu}\right\rangle e^{i \left(\omega t+\vec{q}\cdot(\vec{r}_\mu-\vec{r}_\nu)\right)},
\end{equation}
and, using the mean-field ans\"atze in Sec.~\ref{secMF}, it takes
the general form (see Appendix~\ref{sec:E} for a detailed
derivation)
\begin{widetext}
\begin{equation}\label{dsf}
\begin{aligned}
\mathcal{S}(\omega,\vec{q})
=&\frac{1}{N}\sum\limits_{\mu,\nu}e^{i\vec{q}\cdot(\hat{\varepsilon}_\mu-\hat{\varepsilon}_\nu)}\sum\limits_{\sigma_1,\sigma_2,\sigma_3,\sigma_4,\alpha}\left(\sigma^\alpha\right)_{\sigma_1,\sigma_2}\left(\sigma^\alpha\right)_{\sigma_3,\sigma_4}
\sum\limits_{\rho_1,\rho_2}\sum\limits_{\tau_1,\tau_2}\sum\limits_{\vec{k}} \delta\left(\omega-\lambda_{-\vec{k},\rho_1\tau_1}-\lambda_{\vec{k}-\vec{q},\rho_2\tau_2}\right)\\
&\quad\cdot \left[\left(V_{12}(\vec{k})\right)^*_{\mu\sigma_1,\rho_1\tau_1}
\left(V_{11}(\vec{k}-\vec{q})\right)_{\mu\sigma_2,\rho_2\tau_2}
\left(V_{11}(-\vec{k})\right)^*_{\nu\sigma_3,\rho_1\tau_1}
\left(V_{12}(-\vec{k}+\vec{q})\right)_{\nu\sigma_4,\rho_2\tau_2}\right.\\
&
\left.\quad\quad+\left(V_{12}(\vec{k})\right)^*_{\mu\sigma_1,\rho_1\tau_1}
\left(V_{11}(\vec{k}-\vec{q})\right)_{\mu\sigma_2,\rho_2\tau_2}
\left(V_{11}(\vec{k}-\vec{q})\right)^*_{\nu\sigma_3,\rho_2\tau_2}
\left(V_{12}(\vec{k})\right)_{\nu\sigma_4,\rho_1\tau_1}\right].
\end{aligned}
\end{equation}
\end{widetext}
The critical ($\mu = \mu_c$) DSFs and the corresponding spinon
spectra are plotted in Fig.~\ref{DSF_NN_1En9_shifted} for
representative points in each of the $15$ paraphases.

Focusing on universal low-energy features, we first observe that
each DSF has characteristic points or regions where it is gapless
(i.e., finite at small $\omega$). Since the DSF describes spin
dynamics, and each spin is decomposed into two spinons, the DSF is
gapless at momenta $\vec{q}$ that are appropriate sums of spinon
condensation momenta $\vec{k}_c$ such that $\vec{q} = \vec{k}_{c,1}
+ \vec{k}_{c,2}$. Consequently, we can establish a one-to-one
correspondence between the potential spinon condensation momenta
described in Table~\ref{table1} and the gapless points or regions of
the DSF plotted in Fig.~\ref{DSF_NN_1En9_shifted}; see Table
\ref{tableSF} for this correspondence.

We also notice that the DSF has different low-energy behavior in the
paraphases governed by $z=1$ and $z=2$ critical theories,
respectively. For the $z=1$ paraphases 0-(010)$\Lambda$,
0-(100)$\Lambda$, 0-(101)$\mathrm{W}$, 0-(111)$\mathrm{W}$, and
0-(111)$\mathrm{X}$, the weight of the low-energy DSF is
concentrated around zero energy, while for the $z=2$ paraphases
0-(001)$\Gamma$, 0-(100)$\Gamma$, 0-(101)$\Gamma$, and 0-(111)$\Gamma$, the
low-energy DSF gradually vanishes as the energy is decreased to
zero.

\begin{table}[t]
\begin{tabular*}{0.4\textwidth}{@{\extracolsep{\fill}} c | c }
\hline \hline
Spinon condensation        & Gapless points or regions in     \\
momenta                    & dynamic structure factor         \\
\hline
$\Gamma$                   & $\Gamma$                         \\
\hline
$\mathrm{L}$               & $\Gamma$, $\mathrm{X}$               \\
\hline
$\mathrm{X}$               & $\Gamma$, $\mathrm{X}$                    \\
\hline
$\mathrm{W}$               & $\Gamma$, $\mathrm{X}$, $\frac{2}{3} \mathrm K$   \\
\hline
$\Lambda$                  & $\Gamma \rightarrow\mathrm  X$, $\mathrm K \rightarrow \Gamma \rightarrow \mathrm L \rightarrow \mathrm U$  \\
\hline \hline
\end{tabular*}
\caption{One-to-one correspondence between the potential set of
momenta at which the spinons condense at the critical point and the
set of momenta at which the corresponding dynamic structure factor
is gapless along the high-symmetry path in
Fig.~\ref{DSF_NN_1En9_shifted}; these set of momenta can be points
$\mathrm A$ or sections $\mathrm A \rightarrow \mathrm B$ between
two points $\mathrm A$ and $\mathrm B$.\label{tableSF}}
\end{table}

These low-energy features in the DSF can be understood from a
scaling analysis of the critical field theories in
Eqs.~\eqref{FTz_2} and \eqref{FTz_1}. The DSF is the expectation
value of a four-point correlation function in the condensation
fields; using Wick's theorem, this expectation value can be written
as the convolution of two Green's functions,
\begin{equation}\label{SSFEFT}
\mathcal{S}_z(\omega,\vec{q}) \sim \int d^3\vec{k}d \omega' \, G_z(\omega', \vec{k})G_z(\omega-\omega',\vec{q}-\vec{k}),
\end{equation}
where $G_z(\omega,\vec{k})$ are labeled by the dynamical critical
exponents $z$ of corresponding field theories:
\begin{eqnarray}
G_1(\omega,\vec{k}) &=& \frac{1}{\omega^2+\mu_{ij}k_ik_j}, \nonumber \\
G_2(\omega,\vec{k}) &=& \frac{1}{\omega+\widetilde \nu_{ij}k_ik_j}.
\end{eqnarray}
Inserting $G_z(\omega,\vec{k})$ into Eq.~\eqref{SSFEFT}, and
evaluating the integrals over $\omega'$ and $\vec{k}$, we obtain the
scaling behaviors
\begin{eqnarray}
\mathcal{S}_{z=1}(\omega,\vec{q}) &\sim& \log (\omega) \,
f_1\left(q/\omega\right), \nonumber \\
\mathcal{S}_{z=2}(\omega,\vec{q}) &\sim& \sqrt{\omega}
\,f_2\left(\sqrt{q}/\omega\right),
\end{eqnarray}
where $f_1$ and $f_2$ are some general functions. At the momentum
with gapless DSF, corresponding to $\vec{q}=0$, the DSF at $\omega
\rightarrow 0$ thus diverges in the $z=1$ case and vanishes in the
$z=2$ case. This result qualitatively explains the low-energy DSF
features described above. Furthermore, it elucidates why the SSFs of
the $z=1$ paraphases have singularities at specific momenta, which
precisely coincide with the gapless momenta of the corresponding
DSFs. We stress again that there are $z=1$ paraphases,
for example, 0-(110)$\Gamma$, which have SSFs and DSFs following
$z=2$ behavior. Such a discrepancy may occur if a coefficient in the
critical theory accidentally vanishes for the NN mean-field ansatz.

\begin{table*}[t]
\begin{tabular*}{1.02\textwidth}{@{\extracolsep{\fill}} c | c | c | c | c | c }
\hline \hline
\multirow{3}{*}{Paraphase}           & \multirow{2}{*}{Condensation}               & Real dimension of             & Decomposition into                                                                                                   & Number of scalars           & General intertwining between                                                                                          \\
                                     & \multirow{2}{*}{fields}                     & order-parameter               & irreducible representations                                                                                          & quadratic in order          & irreducible representations                                                                                           \\
                                     &                                             & space at $\vec{K} = 0$        & (i.e., distinct orders)                                                                                              & parameters                  & $C_1 \oplus \cdots \oplus C_q$                                                                                        \\                                    
\hline
$0$-$(001) \Gamma$                   & $\phi_{1,2}$                                & 10                            & $A_{1 \mathrm{g}} \oplus T_{1 \mathrm{g}} \oplus 2T_{2 \mathrm{u}}$                                                  & $1$                         & $\{ A_{1 \mathrm{g}} \} \oplus \{ T_{1 \mathrm{g}} \} \oplus \{ T_{2 \mathrm{u}} \}$                                  \\
\hline
\multirow{2}{*}{$0$-$(001) L$}       & $\phi_{1-8}$                                & \multirow{2}{*}{40}           & $A_{1 \mathrm{g}} \oplus A_{2 \mathrm{g}} \oplus E_{\mathrm{g}} \oplus 2T_{1 \mathrm{g}} \oplus 2T_{2 \mathrm{g}}$   &                             &                                                                                                                       \\
                                     & ($2$ per $\vec{k}_c$)                       &                               & $\oplus 2A_{1 \mathrm{u}} \oplus 2E_{\mathrm{u}} \oplus 2T_{1 \mathrm{u}} \oplus 4T_{2 \mathrm{u}}$                  &                             &                                                                                                                       \\
\hline
$0$-$(010) \Gamma$                   & $\phi_{1,2}$                                & 10                            & $A_{1 \mathrm{g}} \oplus T_{1 \mathrm{g}} \oplus 2T_{2 \mathrm{g}}$                                                  & $1$                         & $\{ A_{1 \mathrm{g}} \} \oplus \{ T_{1 \mathrm{g}} \} \oplus \{ T_{2 \mathrm{g}} \}$                                  \\
\hline

$0$-$(100) \Gamma$                   & $\phi_{1,2}$                                & 10                            & $A_{1 \mathrm{g}} \oplus 3A_{2 \mathrm{g}} \oplus 3E_{\mathrm{g}}$                                                   & $6$                         & $\{ A_{1 \mathrm{g}} \} \oplus \{ A_{2 \mathrm{g}} \} \oplus \{ E_{\mathrm{g}} \}$                                    \\
\hline

$0$-$(101) \Gamma$                   & $\phi_{1,2}$                                & 10                            & $A_{1 \mathrm{g}} \oplus A_{2 \mathrm{g}} \oplus E_{\mathrm{g}} \oplus 2A_{1 \mathrm{u}} \oplus 2E_{\mathrm{u}}$     & $4$                         & $\{ A_{1 \mathrm{g}} \} \oplus \{ A_{2 \mathrm{g}}, E_{\mathrm{g}} \} \oplus \{ A_{1 \mathrm{u}}, E_{\mathrm{u}} \}$  \\

\hline
\multirow{3}{*}{$0$-$(101) W$}       & \multirow{2}{*}{$\chi_{1-24}$}              & \multirow{3}{*}{48}           & $2A_{1 \mathrm{g}} \oplus A_{2 \mathrm{g}} \oplus 3E_{\mathrm{g}} \oplus T_{1 \mathrm{g}}$                           &                             &                                                                                                                       \\
                                     & \multirow{2}{*}{($8$ per $\pm \vec{k}_c$)}  &                               & $\oplus 2T_{2 \mathrm{g}} \oplus 2A_{1 \mathrm{u}} \oplus A_{2 \mathrm{u}}$                                          &                             &                                                                                                                       \\
                                     &                                             &                               & $\oplus 3E_{\mathrm{u}} \oplus 3T_{1 \mathrm{u}} \oplus 4T_{2 \mathrm{u}}$                                           &                             &                                                                                                                       \\
\hline
\multirow{2}{*}{$0$-$(101) X$}       & $\chi_{1-12}$                               & \multirow{2}{*}{30}           & $A_{1 \mathrm{g}} \oplus 2A_{2 \mathrm{g}} \oplus 3E_{\mathrm{g}} \oplus T_{2 \mathrm{g}} \oplus 2A_{1 \mathrm{u}}$  &                             &
\\
                                     & ($4$ per $\vec{k}_c$)                       &                               & $\oplus A_{2 \mathrm{u}} \oplus 3E_{\mathrm{u}} \oplus T_{1 \mathrm{u}} \oplus 2T_{2 \mathrm{u}}$                    &                             &   \\
\hline
$0$-$(110) \Gamma$                   & $\chi_{1,2}$                                & 3                             & $A_{1 \mathrm{g}} \oplus E_{\mathrm{g}}$                                                                             & $1$                         & $\{ A_{1 \mathrm{g}} \} \oplus \{ E_{\mathrm{g}} \}$                                                                  \\
\hline

$0$-$(111) \Gamma$                   & $\phi_{1,2}$                                & 10                            & $A_{1 \mathrm{g}} \oplus A_{2 \mathrm{g}} \oplus E_{\mathrm{g}} \oplus 2A_{2 \mathrm{u}} \oplus 2E_{\mathrm{u}}$     & $4$                         & $\{ A_{1 \mathrm{g}} \} \oplus \{ A_{2 \mathrm{g}}, E_{\mathrm{g}} \} \oplus \{ A_{2 \mathrm{u}}, E_{\mathrm{u}} \}$  \\
\hline
\multirow{3}{*}{$0$-$(111) W$}       & \multirow{2}{*}{$\chi_{1-24}$}              & \multirow{3}{*}{48}           & $A_{1 \mathrm{g}} \oplus 2A_{2 \mathrm{g}} \oplus 3E_{\mathrm{g}} \oplus 2T_{1 \mathrm{g}}$                          &                             &                                                                                                                       \\
                                     & \multirow{2}{*}{($8$ per $\pm \vec{k}_c$)}  &                               & $\oplus T_{2 \mathrm{g}} \oplus A_{1 \mathrm{u}} \oplus 2A_{2 \mathrm{u}}$                                           &                             &                                                                                                                       \\
                                     &                                             &                               & $\oplus 3E_{\mathrm{u}} \oplus 4T_{1 \mathrm{u}} \oplus 3T_{2 \mathrm{u}}$                                           &                             &                                                                                                                       \\
\hline
\multirow{2}{*}{$0$-$(111) X$}       & $\chi_{1-12}$                               & \multirow{2}{*}{30}           & $2A_{1 \mathrm{g}} \oplus A_{2 \mathrm{g}} \oplus 3E_{\mathrm{g}} \oplus T_{1 \mathrm{g}} \oplus A_{1 \mathrm{u}}$   &                             &                                                                                                                       \\
                                     & ($4$ per $\vec{k}_c$)                       &                               & $\oplus 2A_{2 \mathrm{u}} \oplus 3E_{\mathrm{u}} \oplus 2T_{1 \mathrm{u}} \oplus T_{2 \mathrm{u}}$                   &                             &                                                                                                                       \\
\hline \hline
\end{tabular*}
\caption{Analysis of the zero-momentum ($\vec{K} = 0$) order
parameters for the paraphases characterized by single-point and
multi-point condensation (i.e., excluding line condensation). For
each paraphase, the complex ($\phi$) or real ($\chi$) condensation
fields are specified; the zero-momentum order parameters are then
bilinears of these fields with total momentum $\vec{K} = 0$ and
transform under various irreducible representations of the point
group $O_{\text{h}}$. For some paraphases, distinct order parameters
are intertwined such that they must appear together at condensation;
for each paraphase characterized by single-point condensation, the
order parameters are arranged into classes $C$ (marked by curly
brackets) such that there must be at least one nonzero order
parameter from each class $C$. The number of quadratic scalars in
terms of the order parameters is also specified; if there is only
one such scalar, all order parameters are in different classes and
hence are maximally intertwined. \label{tableGL}}
\end{table*}

Finally, we remark that the DSFs of several paraphases have
high-energy points exhibiting large spectral weights at the $\Gamma$
point. In fact, whenever such points exist, there is a
quasi-mirror-reflection symmetry (in terms of energy) between two
spinon bands, such that the two band energies satisfy
$\lambda_{\vec{k},1} + \lambda_{\vec{k},2} = E$ for all momenta
$\vec{k}$. Due to this ``symmetry'', these two bands can contribute
strongly at the $\Gamma$ point close to energy $E$, resulting in an
increased spectral weight as well as a Dirac-like texture. However,
we emphasize that the high-energy part of the DSF depends on
specific details and is not to be taken too seriously; only the
low-energy part of the DSF captures the universal physics in the
given paraphase.

\subsection{General order parameters: hidden and intertwined orders\label{subsec:hiddenintertwined}}

The na\"ive spin-condensation analysis of magnetic orders in
Sec.~\ref{spincond_1} is far from complete as it presumes that any
zero-momentum order transforms under a representation of
$T_{\text{d}}$ and thus ignores the possibility of hidden orders
transforming under inversion-odd representations of the full
pyrochlore point group $O_{\text{h}} = T_{\text{d}} \times
C_{\text{i}}$, where $C_{\text{i}}$ is the $\mathbb{Z}_2$ group
consisting of inversion and identity. One simple example of such a
hidden order is the alternating expansion and contraction of
tetrahedra realized in the ``breathing'' pyrochlores
\cite{PhysRevLett.110.097203, PhysRevB.90.060414,
PhysRevLett.116.257204}. In this subsection, we analyze
zero-momentum orders more comprehensively by identifying all
possible order parameters in terms of the condensing spinon fields
and constructing the most general Ginzburg-Landau (GL) theory that
is compatible with the point group $O_{\text{h}}$ of the pyrochlore
lattice. Such an analysis has been previously done for several
problems building on the PSG framework \cite{bergman2006ordering,
balents2005putting1}.

When the spinons condense at the critical point, certain bosonic
modes at the condensation momenta $\vec{k}_c$ become macroscopically
occupied, and the expectation values of their bosonic operators thus
become classical condensation fields. For the $z = 2$ critical
points, the condensation fields $\phi_n$ are complex, while for the
$z = 1$ critical points, the condensation fields $\chi_n$ are real.
Importantly, these fields themselves are not valid order parameters
as they carry a $\mathbb{Z}_2$ gauge charge and transform
projectively under the point group. Indeed, the projective
transformation rules of $\phi_n$ and $\chi_n$ under the generators
of the point group can be explicitly obtained from the corresponding
transformation rules of the original bosonic operators
$b_{\vec{k}_c, \mu}$ (see also Appendix \ref{sec:D1}):
\begin{subequations}
\begin{eqnarray}
\label{PSGkI}
I&\colon& b_{\vec{k},\mu} \rightarrow (-1)^{n_{ST_1} \delta_{\mu=0}}
e^{i \vec{k}\cdot \hat{e}_\mu}b_{-\vec{k},\mu},\qquad\\
\label{PSGkC3} C_3&\colon& b_{\vec{k},\mu}\rightarrow
U^\dag_{C_3}
b_{(k_z,k_x,k_y),C_3(\mu)},\qquad\\
S&\colon& b_{\vec{k},\mu}\rightarrow
(-1)^{\delta_{\mu=3}n_{ST_1}+\delta_{\mu=2} n_{\overline{C}_6S}}e^{i \vec{k}\cdot\hat{e}_\mu}\qquad\notag \\
&&\qquad\qquad \qquad\quad\cdot \; U^\dag_S 
b_{(k_y,k_x,-k_z),S(\mu)},\qquad
\end{eqnarray}
\end{subequations}
where $C_3(\mu) = 0,2,3,1$ and $S(\mu) = 3,1,2,0$ for the respective
sublattices $\mu = 0,1,2,3$. The simplest possible order parameters
are then the bilinears of the condensation fields, corresponding to
total momentum $\vec{K} = 0$, which are gauge invariant and
transform as linear, generically reducible, representations of the
point group. For each paraphase, the irrep decomposition of this
reducible representation is given in Table \ref{tableGL}. We now
discuss the physical implications of this decomposition.

The scalar irrep $A_{1 \mathrm{g}}$ corresponds to a quadratic
invariant, i.e., a ``mass'' term in the GL theory, which drives the
phase transition between the spin-liquid phase and the magnetically
ordered phase. For almost all paraphases, it appears only once in
the reducible representation, which indicates that all components of
the condensation occur together by symmetry. The bilinear term
transforming under the scalar irrep is $\sum_n \chi_n^2$, where we
decompose any complex fields into real fields as $\phi_n =
\chi_{2n-1} + i \chi_{2n}$. The effective GL theory governing the
phase transition is then
\begin{equation}
\mathcal{L} = \sum_n (\nabla \chi_n)^2 + r \sum_n \chi_n^2 +
O(\chi^4).
\end{equation}
When $A_{1 \mathrm{g}}$ appears more than once in the reducible
representation [for the paraphases $0$-$(101)\mathrm W$ and
$0$-$(111)\mathrm X$], it signals an accidental degeneracy, which
should be lifted when going beyond the NN level.

The remaining irreps, denoted by standard labels, correspond to
various order parameters that describe distinct scenarios of
symmetry breaking (see Table \ref{tableOP}). Irreps with the
subscript ``$\mathrm{g}$'' are even under inversion and correspond
to the conventional spin orders discussed in
Ref.~\cite{PhysRevB.95.094422}. The single-spin order parameters of
such spin orders are straightforward to detect with neutron
scattering. In contrast, irreps with the subscript ``$\mathrm{u}$''
are odd under inversion and correspond to more unconventional hidden
orders. The order parameters of these inversion-breaking orders
always contain multiple spin operators and are thus harder to
detect~\cite{Harter295}. However, in our case, they are also
accompanied by a spontaneous breaking of inversion symmetry, which
may be observed as a ``breathing'' distortion of the pyrochlore
lattice.

\begin{table}[t]
\begin{tabular*}{0.50\textwidth}{@{\extracolsep{\fill}} c | c | c | c }
\hline \hline
\multirow{2}{*}{Irrep}               & \multirow{2}{*}{Dim.}  & Standard name of                     & Simple example of order                                                                                             \\
                                     &                        & corresponding order                  & parameter in terms of spins                                                                                         \\
\hline
\multirow{2}{*}{$A_{1 \mathrm{g}}$}  & \multirow{2}{*}{$1$}   & \multirow{2}{*}{(N/A)}               & \multirow{2}{*}{$1$}                                                                                                \\
                                     &                        &                                      &                                                                                                                     \\
\hline
\multirow{2}{*}{$A_{2 \mathrm{g}}$}  & \multirow{2}{*}{$1$}   & \multirow{2}{*}{All-in-all-out}      & \multirow{2}{*}{$\sum_i \vec{r}_i \cdot \vec{S}_i$}                                                                 \\
                                     &                        &                                      &                                                                                                                     \\
\hline
\multirow{2}{*}{$E_{\mathrm{g}}$}    & \multirow{2}{*}{$2$}   & \multirow{2}{*}{XY antiferromagnet}  &                                                                                                                     \\
                                     &                        &                                      &                                                                                                                     \\
\hline
\multirow{2}{*}{$T_{1 \mathrm{g}}$}  & \multirow{2}{*}{$3$}   & \multirow{2}{*}{Ferromagnet}         & \multirow{2}{*}{$\sum_i \vec{S}_i$}                                                                                 \\
                                     &                        &                                      &                                                                                                                     \\
\hline
\multirow{2}{*}{$T_{2 \mathrm{g}}$}  & \multirow{2}{*}{$3$}   & \multirow{2}{*}{Palmer-Chalker}      & \multirow{2}{*}{$\sum_i \vec{r}_i \times \vec{S}_i$}                                                                \\
                                     &                        &                                      &                                                                                                                     \\
\hline
\multirow{2}{*}{$A_{1 \mathrm{u}}$}  & \multirow{2}{*}{$1$}   &                                      & \multirow{2}{*}{$\sum_{\langle i,j \rangle} \lambda_{i,j} (\vec{S}_i \cdot \vec{S}_j)$}                             \\
                                     &                        &                                      &                                                                                                                     \\
\hline
\multirow{2}{*}{$A_{2 \mathrm{u}}$}  & \multirow{2}{*}{$1$}   &                                      & \multirow{2}{*}{$\sum_{\langle i,j \rangle} \vec{n}_{i,j} \cdot (\vec{S}_i \times \vec{S}_j)$}                      \\
                                     &                        &                                      &                                                                                                                     \\
\hline
\multirow{2}{*}{$E_{\mathrm{u}}$}    & \multirow{2}{*}{$2$}   &                                      &                                                                                                                     \\
                                     &                        &                                      &                                                                                                                     \\
\hline
\multirow{2}{*}{$T_{1 \mathrm{u}}$}  & \multirow{2}{*}{$3$}   &                                      & \multirow{2}{*}{$\sum_{\langle i,j \rangle} (\vec{r}_i \times \vec{n}_{i,j}) \times (\vec{S}_i \times \vec{S}_j)$}  \\
                                     &                        &                                      &                                                                                                                     \\
\hline
\multirow{2}{*}{$T_{2 \mathrm{u}}$}  & \multirow{2}{*}{$3$}   &                                      & \multirow{2}{*}{$\sum_{\langle i,j \rangle} \vec{n}_{i,j} \times (\vec{S}_i \times \vec{S}_j)$}                     \\
                                     &                        &                                      &                                                                                                                     \\
\hline \hline
\end{tabular*}
\caption{Irreducible representations of the point group
$O_{\mathrm{h}}$ and the corresponding symmetry-breaking orders. For
some representations, simple examples of order parameters are
provided in terms of the spins $\vec{S}_i$ at sites $i$, where
$\vec{r}_i$ is the vector from site $i$ to the center of the nearest
``up'' tetrahedron, $\vec{n}_{i,j}$ is the vector from site $i$ to
site $j$, and $\lambda_{i,j} = \pm 1$ for bonds $\langle i,j
\rangle$ in ``up'' and ``down'' tetrahedra, respectively. Note that
the scalar representation $A_{1 \mathrm{g}}$ does not break any
symmetries and hence does not correspond to any order.
\label{tableOP}}
\end{table}

Table \ref{tableGL} indicates that one paraphase can give rise to
several distinct order parameters. In general, the presence or
absence of a given order parameter is determined by the particular
form of the GL theory governing the phase transition. However, for
some paraphases, we can argue that several distinct orders are
intertwined in the sense that they always accompany one another,
regardless of the GL parameters. This highly nontrivial result
emerges because the magnetically ordered phases are obtained by
condensing fractionalized excitations (spinons) that transform
projectively under symmetries.

To analyze the general intertwining between distinct orders for a
given paraphase, we form an orthogonal basis for the (real) order
parameters $\{ \Psi_{R,1}, \cdots, \Psi_{R, N_R} \}$ that transform
under each distinct irrep $R$. Note that $N_R$ is the product of the
irrep dimension and the multiplicity of the irrep in the reducible
representation. Since each symmetry acts on the vector $(\Psi_{R,1},
\cdots, \Psi_{R, N_R})$ by an orthogonal matrix, the quadratic term
$W_R = \sum_{j=1}^{N_R} \Psi_{R,j}^2$ must be a scalar transforming
under $A_{1 \mathrm{g}}$. This scalar can be interpreted as the
``weight'' of the given irrep; since it is a function of the
condensation fields $\chi_n$, it may vanish for some special
configurations of these fields, indicating the absence of the
corresponding order. In contrast, the total weight of all irreps,
\begin{equation}
W_0 = \sum_R W_R = \sum_R \sum_{j=1}^{N_R} \Psi_{R,j}^2 \propto
(\sum_n \chi_n^2)^2,
\end{equation}
is nonzero for all field configurations, indicating that at least
one order must always be present.

For each paraphase, however, the irreps $R$ may be partitioned into
classes $C_1 \oplus \cdots \oplus C_q$ (see Table \ref{tableGL})
such that the total weight of each class $C = \{ R_1, \cdots,
R_{N_C} \}$, containing some nontrivial subset of all irreps, is
proportional to the total weight of all irreps,
\begin{equation}
W_C = \sum_{R \in C} W_{R} = \sum_{R \in C} \sum_{j=1}^{N_R}
\Psi_{R,j}^2 \propto W_0,
\end{equation}
and is thus nonzero for all configurations of the condensation
fields. Consequently, at least one order from each class $C$ must
always be present, regardless of the GL parameters. In the most
extreme scenario, when each irrep forms its own class, such that
$W_R \propto W_0$ for all irreps $R$, the orders are maximally
intertwined, i.e., all of them must appear together. For certain
paraphases, one can argue for this scenario by counting all possible
quadratic scalars that can be formed from the order parameters or,
equivalently, all possible fourth-order scalars that can be formed
from the condensation fields. There is always at least one such
scalar, $(\sum_n \chi_n^2)^2$; however, if there is only one such
scalar, it is clear that the weight $W_R$ of each irrep $R$ must be
proportional to this scalar, and all orders must therefore be
simultaneously present.

While we do not analyze the general intertwining between distinct
orders in all paraphases, we observe from the particular examples
studied (see Table \ref{tableGL}) that the presence of intertwined
orders is a common feature of magnetically ordered phases obtained
by spinon condensation on the pyrochlore lattice. In particular, for
parent spin liquids with $n_{\overline{C}_6} = 1$, where inversion
symmetry acts projectively on the spinons, we generically anticipate
the (already intertwined) spin orders to be also accompanied by
inversion-breaking hidden orders.

\section{Discussion}
\label{sec:discussion}

\subsection{Summary}
\label{sec:summary}

In this paper, we gave a complete classification of
spin-orbit-coupled $\mathbb{Z}_2$ spin liquids on the pyrochlore
lattice by using the PSG method for Schwinger bosons. We studied the
mean-field Hamiltonians of the six $0$-flux spin liquids at the NN
level and examined the critical field theories that describe phase
transitions to ordered phases via spinon condensation. We found two
crucially different classes of critical field theories,
characterized by dynamical exponents $z=1$ and $z=2$, respectively,
which have distinct properties ranging from Hamiltonian
diagonalizability to experimental observables. Moreover, we
investigated the zero-momentum orders obtained from spinon
condensation, both by a na\"ive spin-condensation analysis and by
the representation theory of the full pyrochlore point group
$O_{\text{h}}$. We found that seemingly unrelated orders are
generically intertwined with each other and that conventional spin
orders are often accompanied by more exotic inversion-breaking
``hidden'' orders. Finally, we calculated several physical
observables for our critical theories, including the heat capacity,
as well as the static and dynamic spin structure factors, which may
be compared with experimental data.

\subsection{Possible implications}
\label{sec:poss-impl}

Many pyrochlore materials have been experimentally confirmed to
possess one of the spin orders discussed in this paper. For example,
Yb${}_2$Pt${}_2$O${}_7$ has ferromagnetic order
\cite{PhysRevB.93.014443}, while Nd${}_2$Zr${}_2$O${}_7$ possesses
all-in-all-out order \cite{PhysRevLett.115.197202}. Since all of
these spin orders can appear as a result of spinon condensation from
one of our $\mathbb{Z}_2$ spin liquids, one can contemplate the
possibility that some of these materials are proximate to such a
spin liquid.

As a particular example, one may consider Er${}_2$Ti${}_2$O${}_7$,
which is confirmed to have a $\Psi_2$ antiferromagnetic ground
state. The $\Psi_2$ ground state is selected from the $\Gamma_5$
irrep, containing both $\Psi_2$ and $\Psi_3$ states, as a result of
order-by-disorder mechanism, possibly aided by virtual crystal-field
effects \cite{PhysRevLett.62.2056, PhysRevLett.109.167201,
PhysRevB.88.220404, PhysRevB.91.174424, PhysRevB.93.184408}. This
$\Psi_2$ ground state is quite stable, which suggests that, if it is
obtained from an instability of a spin liquid, such an instability
should uniquely prefer $E_{\rm g}$ order. Consulting Table
\ref{tableGL}, we see that the paraphase $\Gamma$ of the PSG class
0-(110) has a single nontrivial irrep $E_{\text{g}}$, which is not
intertwined with any other orders. Hence, if Er$_2$Ti$_2$O$_7$ is
proximate to a spin liquid, a natural candidate for its parent spin
liquid is the one corresponding to the PSG class 0-(110).

One motivation of this paper was to understand the puzzling
experiments on Yb$_2$B$_2$O$_7$, where B = Ge, Ti, Sn.  These three
compounds have distinct ground states: the Ge compound is
antiferromagnetic \cite{PhysRevB.93.104405}, while the Ti
\cite{PhysRevB.93.064406} and Sn \cite{PhysRevLett.110.127207}
compounds are ferromagnetic, at least when any order can be clearly
identified.  The Ti compound is also sensitive to disorder.  Despite
the disparate ground states, inelastic neutron scattering gives very
similar spectra for all three materials \cite{PhysRevB.93.100403},
consisting of continuum weight over the entire Brillouin zone down
to the lowest energies resolvable in the measurements.  This
observation suggests that the relevant excitations are
characteristic of some common underlying structure, which is
distinct from the usual spin waves tied to the individual ordered
states. The approach in this paper gives one possible explanation:
the excitations may be the spinons of a parent spin-liquid state.

To identify a potential parent spin liquid, we seek a PSG class from
which both antiferromagnetic and ferromagnetic orders can be
obtained through the same condensation paraphase. It is clear from
Table \ref{six0fluxclasses} that such classes exist; the classes
0-(001), 0-(101) and 0-(111) all satisfy this criterion. Therefore,
the proximity to a spin liquid corresponding to either of these
classes can potentially explain the observed excitation spectra.
Looking at the dynamic spin structure factors in
Fig.~\ref{DSF_NN_1En9_shifted}, we indeed see that many of the
critical structure factors in these classes [e.g., 0-(001)L,
0-(101)W, and 0-(111)X] have a large scattering continuum over the
entire Brillouin zone down to a very small fraction of the
spin-excitation bandwidth. It would be interesting to attempt a more
quantitative comparison with the experimental data, which would
require, at the very least, a careful consideration of effects
beyond mean-field theory.

If the scattering continua in the Yb pyrochlores are reflections of
a parent spin liquid, it also suggests that hidden order may be
present in these materials \cite{PhysRevB.93.100403}. Indeed, from
the last column of Table \ref{six0fluxclasses}, we see that the
paraphases 0-(001), 0-(101) and 0-(111) all include hidden orders
breaking inversion symmetry. Searching for such inversion-breaking
orders may be an incisive test of the physical picture presented in
this paper; if such an order is identified, a full characterization
may be assisted by the associated order parameters in Table
\ref{tableOP}. We note that hidden order may also participate in the
specific-heat anomalies of the Yb pyrochlores
\cite{PhysRevB.93.100403}.

\subsection{Future directions}
\label{sec:future-directions}

The present paper explored the physics of proximity to a broad class
of quantum spin liquids on the pyrochlore lattice. Nevertheless,
several assumptions in the analysis could be modified or relaxed in
future work.  We focused on $\mathbb{Z}_2$ spin liquids and used the
framework of bosonic spinons; it would be interesting to consider
$U(1)$ spin liquids and explore fermionic spinons as well.  The
fermionic approach does not, however, provide a simple mean-field
way to study magnetic instabilities, which is straightforward with
bosonic spinons by condensing them.

In addition, the PSG results may be further exploited even within
the framework of bosonic spinons. We concentrated on the $0$-flux NN
mean-field Hamiltonians for simplicity, assuming that NNN terms do
not qualitatively change our results. This assumption, however, is
not necessarily true; in certain cases, a NNN term one-tenth as
strong as a NN term can already change the condensation momenta.
Moreover, the $\pi$-flux PSG classes may exhibit interesting physics
of their own. These PSG classes have a fourfold enlarged unit cell
due to nontrivial translational PSG along the $\hat{e}_2$ and
$\hat{e}_3$ directions, which leads to a $64\times 64$ mean-field
Hamiltonian in terms of the parameters in Table \ref{MFTparameters}.
Multi-spinon condensation may further enlarge the magnetic unit
cell. In turn, this enlargement results in a complex spinon spectrum
that probably requires a more computational approach.

We also presumed that the full symmetry group of the pyrochlore
lattice is preserved at the level of the spin Hamiltonian.  However,
there is a large family of ``breathing'' pyrochlore materials
\cite{PhysRevLett.110.097203, PhysRevB.90.060414,
PhysRevLett.116.257204} that explicitly break inversion symmetry
($Fd\overline{3}m\rightarrow F\overline{4}3m$) by expansion and
contraction of alternating tetrahedra.  One material from this
family, Ba${}_3$Yb${}_2$Zn${}_5$O${}_{11}$, was reported to remain
disordered down to 0.38 K \cite{PhysRevB.90.060414}, and a gauge
mean-field theory, distinct from the spinon approach in this paper,
predicts that this material may experience a non-symmetry-breaking
transition between a paramagnet and a quantum spin ice
\cite{PhysRevB.94.075146}. It would be interesting to see how this
material (and the phase transition predicted for it) fits into a
pyrochlore PSG classification.

Finally, the PSG method can be connected to the energetics of
realistic spin Hamiltonians.  Indeed, our mean-field spinon states
can in principle be used as variational wave functions, as can their
so far unexplored fermionic counterparts.  Calculating variational
energies for these wave functions would require a major effort in
variational Monte Carlo in three dimensions; it is well beyond the
present work but is quite worthwhile to explore.

\begin{acknowledgments}

We acknowledge Yi-Zhuang You, Yuan-Ming Lu, and Bill Jacob for useful discussions. The work of G.B.H. was
supported at ORNL by Laboratory Director's Research and Development
funds and at the KITP by the Gordon and Betty Moore Foundation's
EPiQS Initiative through Grant No.~GBMF4304.  C.L. and L.B. were
supported by the DOE Office of Science’s Basic Energy Sciences program under Award No. DE-FG02-08ER46524.
\end{acknowledgments}

\appendix

\section{\label{app:A}Point-group structure}

The space group $Fd\overline{3}m$ belongs to the cubic crystal system with point group $O_{\text{h}}$. The point group $O_{\text{h}}$ has order 48 and is the symmetry group of a pyrochlore primitive cell -- a pair of corner sharing tetrahedra. It has a direct product structure $O_{\text{h}}\cong S_4\times \mathbb{Z}_2$, which can be understood as following.

We label the seven vertices by $\mu^\pm$, where $\mu=0,1,2,3$ is the sublattice index and ``$+$'' (``$-$'') denotes the upper (lower) tetrahedon (where $0^+=0^-$ is the shared corner), then the symmetry operations in $O_{\text{h}}$ are permutations over two sets $\{0,1,2,3\}$ and $\{+,-\}$. The generators include a 3-fold rotation $C_3 = (123)$, a screw operation (modding out translations) $S = (03)(+-)$ and an inversion $I = (+-)$, written in terms of the cycle notation for permutations. We also define the operation $\Sigma = S\circ I = (03)$ for future convenience. (We can also define $\overline{C}_6 = C_3\circ I$ to reduce the number of generators, since equivalently $C_3 = \overline{C}_6^4$ and $I = \overline{C}^3_6$.) The inversion $I$ is the generator of the $\mathbb{Z}_2$ group, therefore we can write $O_{\text{h}} \cong S_4 \cup (I\circ S_4)$, where $I\circ S_4$ is the coset of $S_4$ left-composed by $I$. The subgroup $S_4$ corresponds exactly to the tetrahedron group, $T_{\text{d}}$. The 24 elements of the group $S_4\simeq T_{\text{d}}$ are generated by $\Sigma$ and $C_3$ as (where it is understood $4\equiv 0$)

\begin{subequations}\label{S4group}
\begin{eqnarray}
(1)&=&C_3\circ C_3\circ C_3, \\
(12)&=&\Sigma\circ C_3\circ \Sigma\circ C^{-1}_3\circ \Sigma\circ C_3,\qquad\\
(13)&=&\Sigma\circ C_3\circ \Sigma\circ C^{-1}_3\circ \Sigma, \qquad\\
(14)&=&\Sigma\circ C_3\circ \Sigma\circ C^{-1}_3\qquad\notag\\
&&\qquad \circ \Sigma\circ C^{-1}_3 \circ \Sigma\circ C_3\circ \Sigma\circ C_3, \qquad\\
(23)&=&\Sigma\circ C_3\circ \Sigma\circ C^{-1}_3\circ \Sigma \circ C^{-1}_3, \qquad\\
(24)&=&C_3\circ \Sigma\circ C^{-1}_3\circ \Sigma\circ C^{-1}_3\circ \Sigma, \qquad\\
(34)&=&\Sigma, \qquad\\
(123)&=&C_3, \qquad\\
(132)&=&C^{-1}_3, \qquad\\
(124)&=&\Sigma\circ C_3\circ \Sigma, \qquad\\
(142)&=&\Sigma\circ C^{-1}_3\circ \Sigma, \qquad\\
(134)&=&\Sigma\circ C_3\circ \Sigma\circ C^{-1}_3, \qquad\\
(143)&=&\Sigma\circ C_3\circ \Sigma\circ C^{-1}_3\qquad\notag\\
&&\qquad \circ \Sigma\circ C_3\circ \Sigma\circ C^{-1}_3,\qquad \\
(234)&=&C^{-1}_3\circ \Sigma\circ C_3\circ \Sigma\qquad\notag\\
&&\qquad\circ C^{-1}_3\circ \Sigma\circ C_3\circ \Sigma, \qquad\\
(243)&
=&C^{-1}_3\circ \Sigma\circ C_3\circ \Sigma, \qquad\\
(1243)&=&\Sigma\circ C_3,\qquad\\
(14)(23)&=&\Sigma\circ C_3\circ \Sigma\circ C_3, \qquad\\
(1342)&=&\Sigma\circ C_3\circ \Sigma\circ C_3\circ \Sigma\circ C_3, \qquad\\
(1234)&=&C_3\circ \Sigma, \qquad\\
(13)(24)&=&C_3\circ \Sigma\circ C_3\circ \Sigma, \qquad\\
(1432)&=&C_3\circ \Sigma\circ C_3\circ \Sigma\circ C_3\circ \Sigma, \qquad\\
(1324)&=&C_3\circ \Sigma\circ C_3, \qquad\\
(12)(34)&=&C_3\circ \Sigma\circ C^{-1}_3\circ \Sigma\circ C_3, \qquad\\
(1423)&=&C_3\circ \Sigma\circ C^{-1}_3\circ \Sigma\circ C^{-1}_3\circ \Sigma\circ C_3.\qquad
\end{eqnarray}
\end{subequations}

Eqs.~\eqref{S4group} will be useful in determining the mean field ans\"atze parameter constraints for the PSG classes.

\section{\label{app:B}Solving PSG equations}

The space group part of the PSG equations are
\begin{subequations}\label{simpfiedgrprelations}
\begin{eqnarray}
(G_{T_i}T_i)(G_{T_{i+1}}T_{i+1})(G_{T_i}T_i)^{-1} (G_{T_{i+1}}T_{i+1})^{-1} \qquad\notag\\\in \mathbb{Z}_2,\qquad\label{PSG123}\\
(G_{\overline{C}_6} \overline{C}_6)^6 \in \mathbb{Z}_2,\qquad\label{PSGc6}\\
(G_SS)^2 (G_{T_3} T_3)^{-1} \in \mathbb{Z}_2,\qquad\label{PSGs}\\
(G_{\overline{C}_6}\overline{C}_6)(G_{T_i}T_i)(G_{\overline{C}_6}\overline{C}_6)^{-1} (G_{T_{i+1}}T_{i+1})\qquad\notag\\
 \in \mathbb{Z}_2,\qquad\label{PSG789}\\
(G_SS)(G_{T_i}T_i)(G_SS)^{-1}(G_{T_3}T_3)^{-1} (G_{T_i}T_i)\qquad\notag\\
 \in \mathbb{Z}_2,\qquad\label{PSG1314}\\
(G_SS)(G_{T_3}T_3)(G_S S)^{-1}(G_{T_3}T_3)^{-1} \in\mathbb{Z}_2,\qquad\label{PSG15}\\
{}[(G_{\overline{C}_6}\overline{C}_6)(G_SS)]^4 \in \mathbb{Z}_2,\qquad\label{PSG18}\\
{}[(G_{\overline{C}_6}\overline{C}_6)^3(G_SS)]^2 \in \mathbb{Z}_2.\qquad\label{PSGhashtag}
\end{eqnarray}
\end{subequations}

The corresponding phase equations are
\begin{widetext}
\begin{subequations}\label{PSGphieqs}
\begin{eqnarray}
\phi_{T_i}(\vec{r}_\mu)+ \phi_{T_{i+1}}[T^{-1}_i(\vec{r}_\mu)]
-\phi_{T_i}[T^{-1}_{i+1}(\vec{r}_\mu)]
-\phi_{T_{i+1}}(\vec{r}_\mu) &=& n_i \pi,\label{PSGn123}\\
\phi_{\overline{C}_6}(\vec{r}_\mu)
+\phi_{\overline{C}_6}[\overline{C}_6^{-1}(\vec{r}_\mu)]
+\phi_{\overline{C}_6}[\overline{C}_6^{-2}(\vec{r}_\mu)]
+\phi_{\overline{C}_6}[\overline{C}_6^{-3}(\vec{r}_\mu)]
+\phi_{\overline{C}_6}[\overline{C}_6^{-4}(\vec{r}_\mu)]
+\phi_{\overline{C}_6}[\overline{C}_6^{-5}(\vec{r}_\mu)]
 &=& n_{\overline{C}_6} \pi,\label{PSGnc6}\\
 \phi_S(\vec{r}_\mu)+ \phi_S[S^{-1}(\vec{r}_\mu)]-\phi_{T_3}(\vec{r}_\mu) &=& n_{S}\pi,\label{PSGns}\\
\phi_{\overline{C}_6}(\vec{r}_\mu)+\phi_{T_i}[\overline{C}^{-1}_6(\vec{r}_\mu)]
-\phi_{\overline{C}_6}[T_{i+1}(\vec{r}_\mu)] + \phi_{T_{i+1}}[T_{i+1}(\vec{r}_\mu)] &=& n_{\overline{C}_6T_i} \pi,\label{PSGn789}\\
\phi_S(\vec{r}_\mu)+ \phi_{T_i}[S^{-1}(\vec{r}_\mu)]-\phi_S[T^{-1}_3 T_i(\vec{r}_\mu)]
-\phi_{T_3}[T_i(\vec{r}_\mu)]+\phi_{T_i}[T_i(\vec{r}_\mu)] &=& n_{ST_i} \pi,\label{PSGn1314}\\
\phi_S(\vec{r}_\mu)+\phi_{T_3}[S^{-1}(\vec{r}_\mu)]
-\phi_S[T^{-1}_3(\vec{r}_\mu)]-\phi_{T_3}(\vec{r}_\mu) &=& n_{ST_3} \pi,\label{PSGn15}\\
\phi_{\overline{C}_6}(\vec{r}_\mu)+\phi_S[\overline{C}_6^{-1}(\vec{r}_\mu)]
+\phi_{\overline{C}_6}[(\overline{C}_6 S)^{-1}(\vec{r}_\mu)]
+\phi_S[(\overline{C}_6S \overline{C}_6)^{-1}(\vec{r}_\mu)]
+\phi_{\overline{C}_6}[(\overline{C}_6S \overline{C}_6S)^{-1}(\vec{r}_\mu)]&&\notag\\
+\phi_S[(\overline{C}_6S \overline{C}_6S\overline{C}_6)^{-1}(\vec{r}_\mu)]
+\phi_{\overline{C}_6}[(\overline{C}_6S \overline{C}_6S\overline{C}_6S)^{-1}(\vec{r}_\mu)]
+
\phi_S[(\overline{C}_6S \overline{C}_6S\overline{C}_6S\overline{C}_6)^{-1}(\vec{r}_\mu)] &=& n_{\overline{C}_6S} \pi,\label{PSGn18}\\
\phi_{\overline{C}_6}(\vec{r}_\mu)+\phi_{\overline{C}_6}[\overline{C}_6^{-1}(\vec{r}_\mu)]
+\phi_{\overline{C}_6}[\overline{C}_6^{-2}(\vec{r}_\mu)]+\phi_S[\overline{C}_6^{-3}(\vec{r}_\mu)]&&\notag\\
+\phi_{\overline{C}_6}[(\overline{C}_6^3S)^{-1}(\vec{r}_\mu)]
+\phi_{\overline{C}_6}[(\overline{C}_6^3S\overline{C}_6)^{-1}(\vec{r}_\mu)]
+\phi_{\overline{C}_6}[(\overline{C}_6^3S\overline{C}^2_6)^{-1}(\vec{r}_\mu)]+\phi_S[S(\vec{r}_\mu)] &=& n_{S\overline{C}_6}\pi. \label{PSGnhashtag}
 \end{eqnarray}
 \end{subequations}
\end{widetext}
where both Eq.~\eqref{PSGn123} and Eq.~\eqref {PSGn789} stand for three equations $i=1,2,3$, and Eq.~\eqref{PSGn1314} stands for two equations $i=1,2$.

First we solve Eq.~\eqref{PSGn123}. Due to gauge freedom of second type, we can use a gauge transformation to achieve $\phi_{T_1}(r_1,r_2,r_3)_\mu=\phi_{T_2}(0,r_2,r_3)_\mu=\phi_{T_3}(0,0,r_3)=0$. Then Eq.~\eqref{PSGn123} gives
\begin{equation}\label{Transsol}
\begin{aligned}
\phi_{T_1}(\vec{r}_\mu) =0,\quad \phi_{T_2}(\vec{r}_\mu)=n_1 \pi r_1,\\
\phi_{T_3}(\vec{r}_\mu) = n_3 \pi r_1+n_2 \pi r_2\quad (\text{mod } 2\pi).
\end{aligned}
\end{equation}
Using Eq.~\eqref{Transsol} to solve Eq.~\eqref{PSGn789}
\begin{subequations}
\begin{eqnarray}
\phi_{\overline{C}_6}(r_1,r_2,r_3)_\mu-\phi_{\overline{C}_6}(r_1,r_2+1,r_3)_\mu&&\qquad\notag\\
+n_1 \pi r_1&=& n_{\overline{C}_6T_1} \pi,\qquad\quad\\
\phi_{\overline{C}_6}(r_1,r_2,r_3)_\mu - n_1 \pi (r_2+\delta_{\mu=2})&&\qquad\notag\\ - \phi_{\overline{C}_6}(r_1,r_2,r_3+1)_\mu
+n_3 \pi r_1+ n_2 \pi r_2 &=&n _{\overline{C}_6T_2} \pi,\qquad\quad \\
\phi_{\overline{C}_6}(r_1,r_2,r_3)_\mu - n_3 \pi (r_2+\delta_{\mu=2})&&\qquad\notag\\ - n_2 \pi( r_3+\delta_{\mu=3})- \phi_{\overline{C}_6}(r_1+1,r_2,r_3)_\mu &=& n_{\overline{C}_6T_3} \pi \qquad\quad
\end{eqnarray}
\end{subequations}
we get $n_1 =n_2 =n_3$, and
\begin{equation}\label{C6sol}
\begin{aligned}
\phi_{\overline{C}_6}(\vec{r}_\mu) &= \phi_{\overline{C}_6} (\vec{0}_\mu) + (n_{\overline{C}_6T_3}+n_1\delta_{\mu=2,3})  \pi r_1\\
 &\quad+n_{\overline{C}_6T_1} \pi r_2 + (n_{\overline{C}_6T_2}+n_1\delta_{\mu=2}) \pi r_3\\
 &\quad+n_1\pi (r_1r_2+r_1r_3).
\end{aligned}
\end{equation}
Then using Eq.~\eqref{Transsol} to solve Eq.~\eqref{PSGn1314} and \eqref{PSGn15}, we get
\begin{equation}\label{Ssol}
\begin{aligned}
\phi_S(\vec{r}_\mu)
&=\phi_S(\vec{0}_\mu)+(n_{ST_3}+n_1 \delta_{\mu=1,2}-n_{ST_1}) \pi r_1\\
&\quad +(n_{ST_3}+n_1 \delta_{\mu=2}-n_{ST_2}) \pi r_2\\
&\quad +(n_{ST_3}+n_1 \delta_{\mu=1,2}) \pi r_3\\
&\quad-\frac{1}{2}n_1 \pi (r_1+r_2)(r_1+r_2+1).
\end{aligned}
\end{equation}

Using Eqs.~\eqref{Transsol},~\eqref{C6sol} and~\eqref{Ssol} to solve Eq.~\eqref{PSGn18} and Eq.~\eqref{PSGnhashtag} we get
\begin{equation}
\sum\limits_{\mu=0}^3 \phi_{\overline{C}_6}(\vec{0}_\mu)+\phi_S(\vec{0}_\mu)=\left(n_{\overline{C}_6S}+\sum\limits_{i=1}^3 n_{\overline{C}_6 T_i}\right)\pi,
\end{equation}
and
\begin{subequations}
\begin{eqnarray}
n_{ST_3}+\sum_{i=1}^3 n_{\overline{C}_6 T_i}& =& 0,\label{ST3C6}\\
3\phi_{\overline{C}_6}(\vec{0}_0)+\sum\limits_{j=1}^3\phi_{\overline{C}_6}(\vec{0}_j)&&\quad\notag\\
+\phi_S(\vec{0}_0)+\phi_S(\vec{0}_3)&=&\left(n_{S\overline{C}_6}+\sum\limits_{j=2}^3 n_{\overline{C}_6 T_j}\right)\pi,\qquad\\
2\phi_S(\vec{0}_i)+2\sum\limits_{j=1}^3\phi_{\overline{C}_6}(\vec{0}_j) &=& n_{S\overline{C}_6} \pi,\quad i=1,2.\qquad
\end{eqnarray}
\end{subequations}
Then, from Eq.~\eqref{PSGnc6} we get
\begin{subequations}
\begin{eqnarray}
6\phi_{\overline{C}_6}(\vec{0}_0) = n_{\overline{C}_6}\pi,\\
2\sum\limits_{j=1}^3\phi_{\overline{C}_6}(\vec{0}_j)
+\sum\limits_{i=1}^3 n_{\overline{C}_6 T_i}\pi =  n_{\overline{C}_6}\pi,
\end{eqnarray}
\end{subequations}
and Eq.~\eqref{PSGns} gives
\begin{equation}\label{nST3}
n_{ST_3}=0
\end{equation}
and
\begin{subequations}\label{S0312}
\begin{eqnarray}
\phi_S(\vec{0}_0)+\phi_S(\vec{0}_3)&=& n_S\pi,\\
2\phi_S(\vec{0}_1)+(n_1+n_{ST_1})\pi &=& n_S\pi,\\
2\phi_S(\vec{0}_2)+(n_1+n_{ST_2})\pi &=& n_S\pi.
\end{eqnarray}
\end{subequations}
Eqs.~\eqref{ST3C6} and~\eqref{nST3} further imply that
\begin{equation}\label{C6T123}
n_{\overline{C}_6T_1}+n_{\overline{C}_6T_2}+n_{\overline{C}_6T_3}=0.
\end{equation}
This completes solving the inter-unit cell part of the space group PSG equations. We can use some of the remaining gauge freedom to simplify results. In order to use the IGG freedom we notice that Eq.~\eqref{PSGs}, \eqref{PSG789} and \eqref{PSG1314} have operators that appear odd number of times. According to our analysis in the main text, we can set $n_S=0$, and two out of the three parameters $n_{\overline{C}T_i}$ to be zero, which together with Eq.~\eqref{C6T123} means that $n_{\overline{C}_6T_1}=n_{\overline{C}_6T_2} = n_{\overline{C}_6T_3} = 0$. The independent $\mathbb{Z}_2$ parameters at this point are
\begin{equation}\label{6nparamsafterinterunitcell}
n_1,\quad n_{\overline{C}_6},\quad n_{ST_1},\quad n_{ST_2},\quad n_{\overline{C}_6S},\quad n_{S\overline{C}_6}.
\end{equation}

Then we add time reversal operation. From
\begin{equation}
(G_{\mathcal{T}}\mathcal{T})(\mathcal{G}_{\mathcal{O}}\mathcal{O})
(G_{\mathcal{T}}\mathcal{T})^{-1}(\mathcal{G}_{\mathcal{O}}\mathcal{O})^{-1} \in \mathbb{Z}_2
\end{equation}
where $\mathcal{O}\in \{ T_1,T_2,T_3,\overline{C}_6,S\}$, we get
\begin{subequations}
\begin{eqnarray}
\phi_{\mathcal{T}}(\vec{r}_\mu)-\phi_{\mathcal{T}}[T_i^{-1}(\vec{r}_\mu)]
-2\phi_{T_i}(\vec{r}_\mu) &=& n_{\mathcal{T}T_i}\pi,\label{PSGn19}\qquad\\
\phi_{\mathcal{T}}(\vec{r}_\mu)-\phi_{\mathcal{T}}[\overline{C}_6^{-1}(\vec{r}_\mu)]
-2\phi_{\overline{C}_6}(\vec{r}_\mu) &=& n_{\mathcal{T}\overline{C}_6}\pi,\label{PSGn22}\qquad\\
\phi_{\mathcal{T}}(\vec{r}_\mu)-\phi_{\mathcal{T}}[S^{-1}(\vec{r}_\mu)]
-2\phi_S(\vec{r}_\mu) &=& n_{\mathcal{T}S}\pi.\qquad\label{PSGn23}
\end{eqnarray}
\end{subequations}
where Eq.~\eqref{PSGn19} stands for three equations $i=1,2,3$. From Eq.~\eqref{PSGn19} we get
\begin{equation}
\phi_{\mathcal{T}}(\vec{r}_\mu) = \phi_{\mathcal{T}}(\vec{0}_\mu)+\pi \sum\limits_{i=1}^3 n_{\mathcal{T} T_i} r_i.
\end{equation}
From Eq.~\eqref{PSGn22} we get $n_{\mathcal{T}T_1}=n_{\mathcal{T}T_2}=n_{\mathcal{T}T_3}\equiv n_{\mathcal{T}T}$, and
\begin{subequations}
\begin{eqnarray}
2\phi_{\overline{C}_6}(\vec{0}_0) &=& n_{\mathcal{T}\overline{C}_6}\pi,\qquad \\
\phi_{\mathcal{T}}(\vec{0}_i)-\phi_{\mathcal{T}}(\vec{0}_{i-1})\quad\quad&& \notag\\
+ n_{\mathcal{T}T}\pi-2\phi_{\overline{C}_6}(\vec{0}_i) &=& n_{\mathcal{T}\overline{C}_6}\pi,\quad i=1,2,3,\qquad 
\end{eqnarray}
\end{subequations}
therefore $n_{\mathcal{T}\overline{C}_6} + n_{\mathcal{T}T} = n_{\overline{C}_6}$. Finally Eq.~\eqref{PSGn23}  gives $n_{\mathcal{T}T} = 0$, $n_{ST_1}=n_{ST_2} = n_{\mathcal{T}S}-n_1$, and
\begin{equation}
\phi_{\mathcal{T}}(\vec{0}_0)-\phi_{\mathcal{T}}(\vec{0}_3)-2\phi_S(\vec{0}_0)=(n_1+n_{ST_1})\pi.
\end{equation}

Lastly the equation $\mathcal{T}^2 = -1$ gives no constraint. 

Now we have solved all the inter-unit cell part of the PSG equations. The intra-unit cell part gives
\begin{subequations}\label{intraunitcell}
\begin{eqnarray}
 2\phi_{\overline{C}_6}(\vec{0}_0) &=& n_{\overline{C}_6}\pi,\qquad\\
 2\sum\limits_{j=1}^3\phi_{\overline{C}_6}(\vec{0}_j)&=& n_{\overline{C}_6}\pi,\qquad\\
\phi_S(\vec{0}_0)+\phi_S(\vec{0}_3)&=&0,\qquad\\
2\phi_S(\vec{0}_i)+(n_1 + n_{ST_1})\pi &=&0,\quad i=1,2,\qquad\\
\sum\limits_{\mu=0}^3 \phi_{\overline{C}_6}(\vec{0}_\mu)+\phi_S(\vec{0}_\mu)&=&n_{\overline{C}_6S}\pi,\qquad\\
3\phi_{\overline{C}_6}(\vec{0}_0)+\sum\limits_{j=1}^3\phi_{\overline{C}_6}(\vec{0}_j)&=& n_{S\overline{C}_6} \pi,\qquad\\
2\sum\limits_{j=1}^3\phi_{\overline{C}_6}(\vec{0}_j)+2\phi_S(\vec{0}_i) &=& n_{S\overline{C}_6} \pi,\quad i=1,2,\qquad\\
\phi_{\mathcal{T}}(\vec{0}_i)-\phi_{\mathcal{T}}(\vec{0}_{i-1})\quad\quad &&\qquad\notag \\
-2\phi_{\overline{C}_6}(\vec{0}_i)&=&n_{\overline{C}_6}\pi,\quad i=1,2,3,\qquad\\
 \phi_{\mathcal{T}}(\vec{0}_0)-\phi_{\mathcal{T}}(\vec{0}_3) -2\phi_S(\vec{0}_0)&=&(n_1+n_{ST_1})\pi.\qquad
\end{eqnarray}
\end{subequations}

Then we use the gauge freedom of second type. Note under gauge transformation
\begin{equation}
\phi(\vec{r}_\mu) = \phi_\mu,\quad \mu = 0,1,2,3,
\end{equation}
we have $\phi_{\mathcal{O}}(\vec{r}_\mu) \rightarrow \phi_{\mathcal{O}}(\vec{r}_\mu)+\phi(\vec{r}_\mu)-\phi[\mathcal{O}^{-1}(\vec{r})_\mu]$, where $\phi$ is an artitrary $U(1)$ phase, we have $\phi_{\overline{C}_6}(\vec{0})_0\rightarrow \phi_{\overline{C}_6}(\vec{0})_0$, $\phi_{\overline{C}_6}(\vec{0})_i\rightarrow \phi_{\overline{C}_6}(\vec{0})_i+\phi_i-\phi_{i+1}$, $\phi_S(\vec{0})_0\rightarrow \phi_S(\vec{0})_0+\phi_0-\phi_3$, $\phi_S(\vec{0})_{1,2}\rightarrow \phi_S(\vec{0})_{1,2}$, $\phi_S(\vec{0})_3\rightarrow \phi_S(\vec{0})_3+\phi_3-\phi_0$, and $\phi_{\mathcal{T}}(\vec{0})_\mu\rightarrow \phi_{\mathcal{T}}(\vec{0})_\mu+2\phi_\mu$.
Then, we can choose the value of $\phi_\mu$ to fix
\begin{equation}\label{phitime0}
\phi_{\mathcal{T}}(\vec{0}_\mu) = 0,
\end{equation}
and
\begin{subequations}
\begin{eqnarray}
\phi_{\overline{C}_6}(\vec{0}_\mu)&=&(\frac{n_{\overline{C}_6}}{2}+p_\mu)\pi,\qquad\\
\phi_S(\vec{0}_0)=-\phi_S(\vec{0}_3)&=&(\frac{n_1+n_{ST_1}}{2}+m_0)\pi,\qquad\\
\phi_S(\vec{0}_{1,2})&=&(-\frac{n_1+n_{ST_1}}{2}+m_{1,2})\pi,\qquad
\end{eqnarray}
\end{subequations}
where $p_\mu, m_0$ and $m_{1,2}$ are all $\mathbb{Z}_2$ parameters.

Note we still have a discrete gauge freedom: we can choose a particular sublattice $\nu$ and define gauge transformation
\begin{equation}\label{gaugenu}
\phi(\vec{r}_\mu) = \pi \delta_{\mu,\nu},
\end{equation} then Eq.~\eqref{phitime0} is preserved but the relative phase of $\phi_{\overline{C}_6}$ can be changed. By choosing $\nu=1,2,3$ we can use gauge \eqref{gaugenu} to fix $p_1=p_2=p_3\equiv p$. Furthermore, we can use the global $\mathbb{Z}_2$ freedom for $\phi_{\overline{C}_6}(\vec{r}_\mu)$ and $\phi_S(\vec{r}_\mu)$ to fix $p_0=0$ and $m_1=0$. Then, let $\nu=0,3$, we can use gauge \eqref{gaugenu} to fix $m_0=0$. By checking Eqs.~\eqref{intraunitcell}, we have $n_{S\overline{C}_6}= n_{\overline{C}_6}+n_1+n_{ST_1}$, $p = n_1+n_{ST_1}$ and $m_2 = n_{\overline{C}_6S}$. The final solution is presented in Eq.~\eqref{PSGsolution}.

\section{\label{app:Yan}Basis for irreps of $T_d$}

This appendix gives the representation analysis result for the spins
$\mathbf{S}$ on a single tetrahedron, which can be equally applied
to pyrochlore lattices with a $\Gamma$ point order. The
twelve-component spin $\mathbf{S}$ form a 12-dimensional
representation of the tetrahedon group $T_{\text{d}}$. The group
$T_{\text{d}}$ has irreducible representation
(irrep)$A_1,A_2,E,T_1,T_2$. $\mathbf{S}$ can be decomposed into
irreps $A_2,E,T_{1,A},T_{1,B}$ and $T_2$. The corresponding basis
and orders are listed in Table.~\ref{tableyan}. This is simply a
reproduction of TABLE III in Ref.~\cite{PhysRevB.95.094422}.

The basis are
\begin{equation}
\begin{aligned}
\mathbf{S}_1 &= \frac{1}{2 \sqrt{3}}(1,1,1,1,-1,-1,-1,1,-1,-1,-1,1),\\
\mathbf{S}_2 &= \frac{1}{2 \sqrt{6}}(-2,1,1,-2,-1,-1,2,1,-1,2,-1,1),\\
\mathbf{S}_3 &= \frac{1}{2 \sqrt{2}}(0,-1,1,0,1,-1,0,-1,-1,0,1,1),\\
\mathbf{S}_4 &= \frac{1}{2} (1,0,0,1,0,0,1,0,0,1,0,0),\\
\mathbf{S}_5 &= \frac{1}{2} (0,1,0,0,1,0,0,1,0,0,1,0),\\
\mathbf{S}_6 &= \frac{1}{2} (0,0,1,0,0,1,0,0,1,0,0,1),\\
\mathbf{S}_7 &= -\frac{1}{2 \sqrt{2}}(0,1,1,0,-1,-1,0,-1,1,0,1,-1),\\
\mathbf{S}_8 &= -\frac{1}{2 \sqrt{2}}(1,0,1,-1,0,1,-1,0,-1,1,0,-1),\\
\mathbf{S}_9 &= -\frac{1}{2 \sqrt{2}}(1,1,0,-1,1,0,1,-1,0,-1,-1,0),\\
\mathbf{S}_{10} &= \frac{1}{2 \sqrt{2}}(0,-1,1,0,1,-1,0,1,1,0,-1,-1),\\
\mathbf{S}_{11} &= \frac{1}{2 \sqrt{2}}(1,0,-1,-1,0,-1,-1,0,1,1,0,1),\\
\mathbf{S}_{12} &= \frac{1}{2 \sqrt{2}}(-1,1,0,1,1,0,-1,-1,0,1,-1,0).
\end{aligned}
\end{equation}
\begin{table}[!thb]
\begin{tabular}{c|c|c}
\hhline{===}Irrep&Basis&Orders\\
\hline
$A_2$ & $\mathbf{S}_1$ &  all in-all out\\
$E$ & $\mathbf{S}_2,\mathbf{S}_3$ & $\Psi_2$ and $\Psi_3$ \\
$T_{1,A}$ & $\mathbf{S}_4, \mathbf{S}_5,\mathbf{S}_6$ & collinear FM\\
$T_{1,B}$ & $\mathbf{S}_7,\mathbf{S}_8,\mathbf{S}_9$ & non-collinear FM\\
$T_2$ & $\mathbf{S}_{10},\mathbf{S}_{11},\mathbf{S}_{12}$ & Palmer-Chalker\\
\hhline{===}
\end{tabular}
\caption{Correspondence between orders, irreps and basis of
irreps}\label{tableyan}
\end{table}

\section{\label{app:D}Derivation of the mean-field Hamiltonians}

In this section we present the solution of mean field constraints from PSG classes, up to NNN level.

For the on-site bond $\vec{0}_0\rightarrow \vec{0}_0$, the 12 group elements that map the bond back are
\begin{equation}\notag
\begin{aligned}
&(1),~(12),~(13),~(23),~(123),~(13),~(+-),~(12)(+-),\\
&(13)(+-),~(23)(+-),~(123)(+-),~(13)(+-),
\end{aligned}
\end{equation}
which give constraints
\begin{equation}
\begin{aligned}
(\alpha,\beta,\gamma,\delta)&=(\alpha,-\gamma,-\beta,-\delta)\\
&=(\alpha,-\delta,-\gamma,-\beta)\\
&=(\alpha,-\beta,-\delta,-\gamma)\\
&=(\alpha,\delta,\beta,\gamma)\\
&=(\alpha,\gamma,\delta,\beta),\\
(0,\beta',\gamma',\delta')
&=(-1)^{n_1+n_{ST_1}}( 0,-\gamma',-\beta',-\delta')\\
&=(-1)^{n_1+n_{ST_1}}( 0,-\delta',-\gamma',-\beta')\\
&=(-1)^{n_1+n_{ST_1}}(0,-\beta',-\delta',-\gamma')\\
&=(0,\delta',\beta',\gamma')\\
&=(0,\gamma',\delta',\beta')\\
&=(-1)^{n_{\overline{C}_6}}(0,\beta',\gamma',\delta')\\
&=(-1)^{n_1+n_{ST_1}+n_{\overline{C}_6}}( 0,-\gamma',-\beta',-\delta')\\
&=(-1)^{n_1+n_{ST_1}+n_{\overline{C}_6}}(0,-\delta',-\gamma',-\beta')\\
&=(-1)^{n_1+n_{ST_1}+n_{\overline{C}_6}}(0,-\beta',-\delta',-\gamma')\\
&=(-1)^{n_{\overline{C}_6}}(0,\delta',\beta',\gamma')\\
&=(-1)^{n_{\overline{C}_6}}(0,\gamma',\delta',\beta').
\end{aligned}
\end{equation}
The only allowed onsite hopping term is $\alpha$ which is simply the chemical potential $\mu$. The allowed pairing terms are $\beta'=\gamma'=\delta'\equiv \nu$ when and only when $(n_1+n_{ST_1},n_{\overline{C}_6})=(1,0)$. Note the singlet pairing term is not allowed, $\alpha'=0$, due to hermiticity.

Consider the case for the NN bond $\vec{0}_0\rightarrow \vec{0}_1$. The four group elements that map the bond back are [see Appendix \ref{app:A} for details]
$$(1), ~(14), ~(23), ~(14)(23),$$
using Eqs.\eqref{S4group}, \eqref{bonds_under_o}, \eqref{su2_under_o} and \eqref{uhupNN}, we get constraints

\begin{subequations}
\begin{eqnarray}
(a,b,c,d)&=&(-1)^{n_{\overline{C}_6S}}(a,-b,c,d)\notag\\
         &=&(a,b,-d,-c)\notag\\
         &=& (-1)^{n_{\overline{C}_6S}}(a,-b,-d,-c),\qquad \\
(a',b',c',d')&=&(-1)^{n_{\overline{C}_6S}}(-a',b',-c',-d')\notag\\
             &=&(-1)^{n_1+n_{ST_1}+n_{\overline{C}_6}}(-a',-b',d',c')\notag\\
             &=&(-1)^{n_1+n_{ST_1}+n_{\overline{C}_6}+n_{\overline{C}_6S}}(a',-b',-d',-c'),\notag\\
\end{eqnarray}
\end{subequations}
we get
\begin{itemize}
\item $n_{\overline{C}_6S}=0$: $b=0$, $c = -d$, $a$, $c$ free independent; $a'=c'=d'=0$, we have
\begin{itemize}
\item $n_1+n_{ST_1}=n_{\overline{C}_6}$: no NN pairing term allowed;
\item $n_1+n_{ST_1}=n_{\overline{C}_6}+1$: $b'$ free.
\end{itemize}
\item $n_{\overline{C}_6S}= 1$: $a=c=d=0$, $b$ free; $b'=0$, and
\begin{itemize}
\item $n_1+n_{ST_1} = n_{\overline{C}_6}$: $a'=0$, $c'=d'$, $c'$ free;
\item $n_1+n_{ST_1} = n_{\overline{C}_6}+1$: $c' = -d'$, $a'$, $c'$ free independent.
\end{itemize}
\end{itemize}

For the NNN bond $\vec{0}_1\rightarrow \vec{0}_2-\hat{e}_2$, it can be checked that only the identity $(1)$ and the element $(12)(+-) = S\circ C_3\circ S\circ C^{-1}_3\circ S\circ C_3$ map the bond back, which gives
\begin{subequations}
\begin{eqnarray}
(A,B,C,D)&=&(-1)^{n_1}(A,C,B,D),\\
(A',B',C',D')&=&(-1)^{1+n_{ST_1}}(A',C',B',D').\qquad
\end{eqnarray}
\end{subequations}
Therefore for hopping
\begin{itemize}
\item $n_1=0$: $B=C$, $A,B,D$ free independent;
\item $n_1=1$: $A=D=0$, $B = -C$, $B$ free,
\end{itemize}
and for pairing
\begin{itemize}
\item $n_{ST_1}=0$: $A'=D'=0$, $B'=-C'$, $B'$ free;
\item $n_{ST_1} = 1$: $B'=C'$, $A',B',D'$ free independent.
\end{itemize}
These results are listed in Table~\ref{MFTparameters}.

\section{\label{sectablemu}Critical chemical potential $\mu$}

The critical chemical potential $\mu_c$ for the 15 paraphases is
listed in Table \ref{tablemu}.
\begin{table}[!thb]
\begin{ruledtabular}
\begin{tabular}{c|c}
Paraphase & Critical $\mu_{\mathrm{c}}$ \\
\hline
0-(001)$\Gamma$& $\mu_{\mathrm{c}}=\max\{-6a,2a-8c\}$\\
\hline
\multirow{2}{*}{0-(001)$\mathrm{L}$}&
Largest root of $\mu^3+2(a+2c)\mu^2$\\
&$-4(2a^2-4ac+2c^2+3b'^2)\mu-24b'^2(a+2c)=0$\\
\hline
0-(001)$\Lambda$& $\mu_c = 2a+4c$ \\
\hline 0-(010)$\Gamma$& $\mu_{\mathrm{c}}=2 \left(a+2 c+ \sqrt{4
(a-c)^2+3 (\nu -b')^2}\right)$
\\
\hline
0-(010)$\Lambda$& $\mu_{\mathrm{c}}=2\left(-a-2c+\sqrt{3}|\nu+b'|\right)$\\
\hline

0-(100)$\Gamma$& $\mu_{\mathrm{c}}=-6b$\\
\hline 0-(100)$\Lambda$&
$\mu_{\mathrm{c}}=2b+4\sqrt{2}|c'|$\\
\hline
0-(101)$\Gamma$& $\mu_{\mathrm{c}}=-6b$\\
\hline
\multirow{3}{*}{0-(101)$\mathrm{W}$}
& Largest root of
$\mu^4-8(b^2+2a'^2+4c'^2)\mu^2$\\
&$+64(2a'+c')bc'\mu-64a'c'(b^2+3c'^2)$\\
&$+16(b^2-3c'^2)^2-32b^2a'^2+48a'^4=0$\\\hline
0-(101)$\mathrm{X}$& $\mu_{\mathrm{c}}=2b+2\sqrt{2}|a'-c'|$\\
\hline 0-(110)$\Gamma$&
$\mu_{\mathrm{c}}=6b+2\sqrt{3}|\nu+a'+2c'|$ \\
\hline 0-(110)$\Lambda$&
$\mu_{\mathrm{c}}=-2b+2\sqrt{(\nu+a'-2c')^2+2(\nu - a')^2}$\\
\hline

 0-(111)$\Gamma$&
$\mu_{\mathrm{c}}=6b$ \\
\hline 0-(111)$\mathrm{W}$&$\mu_{\mathrm{c}}=\max \left\{\pm\sqrt{2} w+
\sqrt{2} \sqrt{2 b^2\mp 4
\sqrt{2} b w+7 w^2} \right\}$\\
\hline 0-(111)$\mathrm{X}$& $\mu_{\mathrm{c}}=-2b+2\sqrt{6}|c'|$
\end{tabular}
\end{ruledtabular}
\caption{\label{tablemu}%
Critical chemical potential $\mu$ for the 15 paraphases.}
\end{table}

\onecolumngrid

\section{\label{sec:D1}Condensation results}

The three vectors $\mathbf{S}^r$, $\mathbf{S}^c$, and $\mathbf{S}^s$ for the paraphase 0-$(100)\Gamma$, mentioned in Eq.~\eqref{Srcs}, are
\begin{subequations}
\begin{eqnarray}
\mathbf{S}^r &=& (0,0,-1,0,0,1,0,0,1,0,0,-1),\\
\mathbf{S}^c &=& (-1,0,0,-1,0,0,1,0,0,1,0,0),\\
\mathbf{S}^s &=& (0,-1,0,0,1,0,0,-1,0,0,1,0).
\end{eqnarray}
\end{subequations}

The three vectors $\mathbf{S}^r$, $\mathbf{S}^c$, and $\mathbf{S}^s$ for the paraphase 0-$(001)\Gamma$, mentioned in Sec.~\ref{subsubsec:ferri}, are
\begin{subequations}
\begin{eqnarray}
\mathbf{S}^r &=& (4,4,7,-8,-4,-1,-4,-8,-1,0,0,9),\\
\mathbf{S}^c &=& (1,-8,4,1,-4,8,-7,4,-4,9,0,0),\\
\mathbf{S}^s &=& (-8,1,4,4,-7,-4,-4,1,8,0,9,0).
\end{eqnarray}
\end{subequations}

In writing down the Ginzburg-Landau theory for the paraphase 0-(010)$\Gamma$,
the transformation rules of $\phi_1,\phi_2,\overline{\phi}_1,\overline{\phi}_2$ are
\begin{equation}\label{010L1C6}
\overline{C}_6\colon
\left(\begin{array}{c}\phi_1\\\phi_2\\\overline{\phi}_1\\\overline{\phi}_2\end{array}\right)\rightarrow
\left(
\begin{array}{cccc}
 \frac{\left(\frac{1}{6}-\frac{i}{6}\right) ((1+2 i) \delta +(1-i) \Delta )}{\delta } & \frac{\left(\frac{1}{6}+\frac{i}{6}\right) (\delta +\Delta )}{\delta } & -\frac{\left(\frac{1}{2}-\frac{i}{2}\right) \zeta }{\delta } & 0 \\
 -\frac{\left(\frac{1}{6}-\frac{i}{6}\right) (\delta +\Delta )}{\delta } & \frac{\left(\frac{1}{6}+\frac{i}{6}\right) ((1-2 i) \delta +(1+i) \Delta )}{\delta } & 0 & -\frac{\left(\frac{1}{2}+\frac{i}{2}\right) \zeta }{\delta } \\
 -\frac{\left(\frac{1}{2}+\frac{i}{2}\right) \zeta }{\delta } & 0 & \frac{\left(\frac{1}{6}+\frac{i}{6}\right) ((1-2 i) \delta +(1+i) \Delta )}{\delta } & \frac{\left(\frac{1}{6}-\frac{i}{6}\right) (\delta +\Delta )}{\delta } \\
 0 & -\frac{\left(\frac{1}{2}-\frac{i}{2}\right) \zeta }{\delta } & -\frac{\left(\frac{1}{6}+\frac{i}{6}\right) (\delta +\Delta )}{\delta } & \frac{\left(\frac{1}{6}-\frac{i}{6}\right) ((1+2 i) \delta +(1-i) \Delta )}{\delta } \\
\end{array}
\right)
\left(\begin{array}{c}\phi_1\\\phi_2\\\overline{\phi}_1\\\overline{\phi}_2\end{array}\right),
\end{equation}
\begin{equation}\label{010L1S}
S\colon
\left(\begin{array}{c}\phi_1\\\phi_2\\\overline{\phi}_1\\\overline{\phi}_2\end{array}\right)\rightarrow
\left(
\begin{array}{cccc}
 \frac{1}{\sqrt{2}} & -\frac{1}{\sqrt{2}} & 0 & 0 \\
 -\frac{1}{\sqrt{2}} & -\frac{1}{\sqrt{2}} & 0 & 0 \\
 0 & 0 & \frac{1}{\sqrt{2}} & -\frac{1}{\sqrt{2}} \\
 0 & 0 & -\frac{1}{\sqrt{2}} & -\frac{1}{\sqrt{2}} \\
\end{array}
\right)
\left(\begin{array}{c}\phi_1\\\phi_2\\\overline{\phi}_1\\\overline{\phi}_2\end{array}\right),
\end{equation}
where the definition of $\delta, \Delta$ and $\zeta$ has been given in Section~\ref{spincond_1}. We see that in this case the fields transform to their complex conjugates under $O_{\text{h}}$.

The only quartic term invariant under $O_{\text{h}}$ is
\begin{equation}
\Phi = \left[4(|\phi_1|^2+|\phi_2|^2)+\left((-1+i) \phi^2_1-(1+i)\phi^2_2-i \phi_1\phi_2+c.c.\right)\right]^2.
\end{equation}

In writing down the Ginzburg-Landau theory for the paraphase 0-(100)$\Gamma$, the transformation rules of $\phi_{1,2}$ under $\overline{C}_6$ and $S$ are recorded by the following matrices:
\begin{equation}
U^{(100)\Gamma}_{\overline{C}_6} = \frac{1}{2}\left(\begin{array}{cc} 1-i & 1-i\\ -1-i & 1+i\end{array}\right),\quad
U^{(100)\Gamma}_S = \frac{1}{\sqrt{2}}\left(\begin{array}{cc} 0 & -1-i\\1-i & 0 \end{array}\right).
\end{equation}
There are six quartic terms that are invariant under $O_{\text{h}}$. Three of them can be written as $\Phi_i^2$, where
\begin{equation}
\begin{aligned}
\Phi_1 &= |\phi_1|^2+|\phi_2|^2,\\
\Phi_2 &= \frac{1}{2}\left(|\phi_1|^2-|\phi_2|^2\right)+\left(\frac{1+3i}{4}\phi_1^2+\frac{1-2i}{2}\phi_1\phi_2+\frac{1-i}{2}\phi_1\phi_2^*+\frac{3-i}{4}\phi_2^2+c.c.\right),\\
\Phi_3 &= |\phi_1|^2-|\phi_2|^2+ \left(\frac{\phi_1^2}{2}-\frac{1+i}{2}\phi_1\phi_2+(1-i)\phi_1\phi_2^*-\frac{i}{2}\phi_2^2 + c.c\right).
\end{aligned}
\end{equation}

In writing down the Ginzburg-Landau theory for the paraphase 0-(101)$\Gamma$, the transformation rules of $\phi_{1,2}$ under $\overline{C}_6$ and $S$ are recorded by the following matrices:
\begin{equation}
\widetilde{U}^{(101)\Gamma}_{\overline{C}_6} = iU^{(100)\Gamma}_{\overline{C}_6},\quad
\widetilde{U}^{(101)\Gamma}_S =  U^{(100)\Gamma}_S.
\end{equation}
the extra factor of $i$ for $\overline{C}_6$ is due to $n_{\overline{C}_6}=1$.

In writing down the Ginzburg-Landau theory for the paraphase 0-(110)$\Gamma$, the transformation rules of $\chi_{1,2}$ under $\overline{C}_6$ and $S$ are recorded by the following matrices:
\begin{equation}
U^{(110)\Gamma}_{\overline{C}_6}
=\left(
\begin{array}{cc}
 \frac{1}{2} & -\frac{\sqrt{3}}{2} \\
 \frac{\sqrt{3}}{2} & \frac{1}{2} \\
\end{array}
\right),\quad
U^{(110)\Gamma}_S = \left(
\begin{array}{cc}
 \frac{\sqrt{\frac{3}{2}}}{\sqrt{3}+3} & -\frac{\sqrt{6}+\frac{3}{\sqrt{2}}}{\sqrt{3}+3} \\
 -\frac{\sqrt{6}+\frac{3}{\sqrt{2}}}{\sqrt{3}+3} & -\frac{\sqrt{\frac{3}{2}}}{\sqrt{3}+3} \\
\end{array}
\right).
\end{equation}
the only quadratic and quartic order parameter are the trivial one: $(\chi_1^2+\chi_2^2)^i$, $i=1,2$. At sextic order, there are two terms allowed:
\begin{equation}
(\chi_1^2+\chi_2^2)^3,\quad \frac{1}{3}(\chi_1-\chi_2)\chi_2(3\chi_1^2-\chi_2^2)(\chi_1^2+4\chi_1\chi_2+\chi_2^2).
\end{equation}

In writing down the Ginzburg-Landau theory for the paraphase 0-(111)$\Gamma$, the transformation rules of $\phi_{1,2}$ under $\overline{C}_6$ and $S$ are recorded by the following matrices:
\begin{equation}
\overline{U}^{(111)\Gamma}_{\overline{C}_6} = \frac{1}{2}\left(\begin{array}{cc} 1-i & -1+i\\ -1-i & -1-i\end{array}\right),\quad
\overline{U}^{(111)\Gamma}_S = \frac{1}{\sqrt{2}}\left(\begin{array}{cc} 0 & -1-i\\-1+i & 0 \end{array}\right).
\end{equation}

\section{\label{sec:E}Structure factor}

We present the calculation of the static and dynamic structure factors. Importantly, we must distinguish the phase factors in the above two expressions: the $e^{i\vec{q}\cdot (\vec{r}_\mu-\vec{r}'_\nu)}$ in structure factor $\mathcal{S}^\alpha$ is \emph{global}, namely $\vec{r}_\mu = \vec{r}+\hat{\varepsilon}_\mu$, which keeps track the relative displacement between sublattices, and this comes from the definition of structor factor. The phase factor in Fourier transformation, on the other hand, must agree with the convention we choose in Fourier transforming the Hamiltonian into $\vec{k}$ space: remember that Block Hamiltonian sets the displacement between sublattices to \emph{zero}, therefore we must also set the displacement between sublattices to zero in the Fourier transforms, i.e. $\vec{r}_\mu = \vec{r}$.
\begin{equation}
\begin{aligned}
\mathcal{S}^{\alpha}(\vec{q})
&=
\frac{1}{N^3} \sum\limits_{\vec{r}_\mu,\vec{r}'_\nu} \sum\limits_{\vec{k}_1,\vec{k}_2,\vec{k}_3,\vec{k}_4}
e^{i\vec{q}\cdot (\vec{r}_\mu-\vec{r}'_\nu)} e^{i(\vec{k}_2-\vec{k}_1)\cdot\vec{r}_\mu}e^{i(\vec{k}_4-\vec{k}_3)\cdot \vec{r}'_\nu} \sum\limits_{\sigma_1,\sigma_2,\sigma_3,\sigma_4}\left(\sigma^\alpha\right)_{\sigma_1,\sigma_2}\left(\sigma^\alpha\right)_{\sigma_3,\sigma_4}\left\langle b^\dag_{\vec{k}_1,\mu\sigma_1} b_{\vec{k}_2,\mu\sigma_2}b^\dag_{\vec{k}_3,\nu\sigma_3}b_{\vec{k}_4,\nu\sigma_4}\right\rangle\\
&=\frac{1}{N}\sum\limits_{\vec{k}_1,\vec{k}_3}\sum\limits_{\mu,\nu}e^{i\vec{q}\cdot(\hat{\varepsilon}_\mu-\hat{\varepsilon}_\nu)}\sum\limits_{\sigma_1,\sigma_2,\sigma_3,\sigma_4}
\left(\sigma^\alpha\right)_{\sigma_1,\sigma_2}\left(\sigma^\alpha\right)_{\sigma_3,\sigma_4}\left\langle b^\dag_{\vec{k}_1,\mu\sigma_1} b_{\vec{k}_1-\vec{q},\mu\sigma_2}b^\dag_{\vec{k}_3,\nu\sigma_3}b_{\vec{k}_3+\vec{q},\nu\sigma_4}\right\rangle.
\end{aligned}
\end{equation}
From Eq.~\eqref{btobeta} we write
\begin{equation}
b_{\vec{k},\mu\sigma} = \sum\limits_{\rho=0}^3 \sum\limits_{\tau=\uparrow,\downarrow}
\left( V_{11}(\vec{k})\right)_{\mu\sigma,\rho\tau}\widetilde{b}_{\vec{k},\rho\tau}+
\left( V_{12}(\vec{k})\right)_{\mu\sigma,\rho\tau}\widetilde{b}^\dag_{-\vec{k},\rho\tau},\label{component1101}
\end{equation}

one can show using Wick's theorem that
\begin{equation}
\begin{aligned}
&\left\langle b^\dag_{\vec{k}_1,\mu\sigma_1} b_{\vec{k}_1-\vec{q},\mu\sigma_2}b^\dag_{\vec{k}_3,\nu\sigma_3}b_{\vec{k}_3+\vec{q},\nu\sigma_4}\right\rangle\\
&=
\sum\limits_{\rho_1,\rho_2}\sum\limits_{\tau_1,\tau_2}
\delta_{\vec{k}_1,-\vec{k}_3}
\left(V_{12}(\vec{k}_1)\right)^*_{\mu\sigma_1,\rho_1\tau_1}
\left(V_{11}(\vec{k}_1-\vec{q})\right)_{\mu\sigma_2,\rho_2\tau_2}
\left(V_{11}(-\vec{k}_1)\right)^*_{\nu\sigma_3,\rho_1\tau_1}
\left(V_{12}(-\vec{k}_1+\vec{q})\right)_{\nu\sigma_4,\rho_2\tau_2}\\
&\qquad\qquad+\delta_{\vec{k}_1,\vec{k}_3+\vec{q}}\left(V_{12}(\vec{k}_1)\right)^*_{\mu\sigma_1,\rho_1\tau_1}
\left(V_{11}(\vec{k}_1-\vec{q})\right)_{\mu\sigma_2,\rho_2\tau_2}
\left(V_{11}(\vec{k}_1-\vec{q})\right)^*_{\nu\sigma_3,\rho_2\tau_2}
\left(V_{12}(\vec{k}_1)\right)_{\nu\sigma_4,\rho_1\tau_1}\\
&\qquad\qquad+
\delta_{\vec{q},\vec{0}}
\left(V_{12}(\vec{k}_1)\right)^*_{\mu\sigma_1,\rho_1\tau_1}
\left(V_{12}(\vec{k}_1)\right)_{\mu\sigma_2,\rho_1\tau_1}\left(V_{12}(\vec{k}_3)\right)^*_{\nu\sigma_3,\rho_2\tau_2}\left(V_{12}(\vec{k}_3)\right)_{\nu\sigma_4,\rho_2\tau_2},
\end{aligned}
\end{equation}
where the first, second, and third terms come from the channels $\left\langle b^\dag_{\vec{k}_1,\mu\sigma_1} b^\dag_{\vec{k}_3,\nu\sigma_3} \right\rangle \left\langle b_{\vec{k}_1-\vec{q},\mu\sigma_2}b_{\vec{k}_3+\vec{q},\nu\sigma_4}\right\rangle$, $\left\langle b^\dag_{\vec{k}_1,\mu\sigma_1} b_{\vec{k}_3+\vec{q},\nu\sigma_4}\right\rangle \left\langle b^\dag_{\vec{k}_3,\nu\sigma_3}b_{\vec{k}_1-\vec{q},\mu\sigma_2}\right\rangle$, and $\left\langle b^\dag_{\vec{k}_1,\mu\sigma_1} b_{\vec{k}_1-\vec{q},\mu\sigma_2}\right\rangle \left\langle b^\dag_{\vec{k}_3,\nu\sigma_3}b_{\vec{k}_3+\vec{q},\nu\sigma_4}\right\rangle$, respectively. In the expression for structure factors the third term becomes $\left\langle \hat{S}^\alpha_{\vec{r}_\mu}\right\rangle$ and vanishes due to time reversal symmetry. Therefore
\begin{equation}
\begin{aligned}
&\mathcal{S}^{\alpha}(\vec{q})\\
&=\frac{1}{N}\sum\limits_{\mu,\nu}e^{i\vec{q}\cdot(\hat{\varepsilon}_\mu-\hat{\varepsilon}_\nu)}\sum\limits_{\sigma_1,\sigma_2,\sigma_3,\sigma_4}\left(\sigma^\alpha\right)_{\sigma_1,\sigma_2}\left(\sigma^\alpha\right)_{\sigma_3,\sigma_4}\cdot\\
\quad&\qquad\qquad\qquad\qquad
\sum\limits_{\rho_1,\rho_2}\sum\limits_{\tau_1,\tau_2}\sum\limits_{\vec{k}_1}\left[\left(V_{12}(\vec{k}_1)\right)^*_{\mu\sigma_1,\rho_1\tau_1}
\left(V_{11}(\vec{k}_1-\vec{q})\right)_{\mu\sigma_2,\rho_2\tau_2}
\left(V_{11}(-\vec{k}_1)\right)^*_{\nu\sigma_3,\rho_1\tau_1}
\left(V_{12}(-\vec{k}_1+\vec{q})\right)_{\nu\sigma_4,\rho_2\tau_2}\right.\\
\quad&\qquad\qquad\qquad\qquad\qquad\qquad
\left.\quad\quad+\left(V_{12}(\vec{k}_1)\right)^*_{\mu\sigma_1,\rho_1\tau_1}
\left(V_{11}(\vec{k}_1-\vec{q})\right)_{\mu\sigma_2,\rho_2\tau_2}
\left(V_{11}(\vec{k}_1-\vec{q})\right)^*_{\nu\sigma_3,\rho_2\tau_2}
\left(V_{12}(\vec{k}_1)\right)_{\nu\sigma_4,\rho_1\tau_1}\right]\\
&=\frac{1}{N}\sum\limits_{\mu,\nu}e^{i\vec{q}\cdot(\hat{\varepsilon}_\mu-\hat{\varepsilon}_\nu)}
\sum\limits_{\rho_1,\rho_2}\sum\limits_{\tau_1,\tau_2}
\sum\limits_{\vec{k}_1}\left[\left(V^\dag_{12}(\vec{k}_1)\right)_{\rho_1\tau_1,\mu}\sigma^\alpha
\left(V_{11}(\vec{k}_1-\vec{q})\right)_{\mu,\rho_2\tau_2}
\left(V^\dag_{11}(-\vec{k}_1)\right)_{\rho_1\tau_1,\nu}\sigma^\alpha
\left(V_{12}(-\vec{k}_1+\vec{q})\right)_{\nu,\rho_2\tau_2}\right.\\
&
\left.\quad\qquad\qquad\quad\qquad\qquad\qquad\qquad+\left(V^\dag_{12}(\vec{k}_1)\right)_{\rho_1\tau_1,\mu}\sigma^\alpha
\left(V_{11}(\vec{k}_1-\vec{q})\right)_{\mu,\rho_2\tau_2}
\left(V^\dag_{11}(\vec{k}_1-\vec{q})\right)_{\rho_2\tau_2,\nu}
\sigma^\alpha
\left(V_{12}(\vec{k}_1)\right)_{\nu,\rho_1\tau_1}\right]\\
&=\frac{1}{N}
\sum\limits_{\rho_1,\rho_2}\sum\limits_{\tau_1,\tau_2}
\sum\limits_{\vec{k}_1}\left[\left(V^\dag_{12}(\vec{k}_1)\right)_{\rho_1\tau_1,\colon}\left(I(\vec{q})\otimes \sigma^\alpha\right)
\left(V_{11}(\vec{k}_1-\vec{q})\right)_{\colon,\rho_2\tau_2}
\left(V^\dag_{11}(-\vec{k}_1)\right)_{\rho_1\tau_1,\colon}
\left(I^*(\vec{q})\otimes \sigma^\alpha\right)
\left(V_{12}(-\vec{k}_1+\vec{q})\right)_{\colon,\rho_2\tau_2}\right.\\
&\left.\qquad\quad\qquad\qquad\quad+\left(V^\dag_{12}(\vec{k}_1)\right)_{\rho_1\tau_1,\colon}
\left(I(\vec{q})\otimes \sigma^\alpha\right)
\left(V_{11}(\vec{k}_1-\vec{q})\right)_{\colon,\rho_2\tau_2}
\left(V^\dag_{11}(\vec{k}_1-\vec{q})\right)_{\rho_2\tau_2,\colon}
\left(I^*(\vec{q})\otimes \sigma^\beta\right)
\left(V_{12}(\vec{k}_1)\right)_{\colon,\rho_1\tau_1}\right]\\
&=\frac{1}{N}
\sum\limits_{\vec{k}_1}\mathrm{Tr}\left[V^\dag_{12}(\vec{k}_1)\left(I(\vec{q})\otimes \sigma^\alpha\right)
V_{11}(\vec{k}_1-\vec{q})
\left(V^\dag_{21}(\vec{k}_1-\vec{q})
\left(I^*(\vec{q})\otimes \sigma^\alpha\right)^T
V_{22}(\vec{k}_1)
+
V^\dag_{11}(\vec{k}_1-\vec{q})
\left(I_{4\times4}\otimes \sigma^\alpha\right)
V_{12}(\vec{k}_1)\right)\right],
\end{aligned}
\end{equation}
where we have used Eq.~\eqref{ph_sym_on_V} and defined $I(\vec{q})$ as in Eq.~\eqref{formfactormatrix}. We used the notation ``$\colon$'' to denote that the corresponding rows (columns) are retained in the matrix: for example, the notaton $\left(V^\dag_{12}(\vec{k}_1)\right)_{\rho_1\tau_1,\colon}$ denotes the $\rho_1\tau_1$-row of the matrix $V^\dag_{12}(\vec{k}_1)$ (where all columns in this row are retained), and the notation $\left(V_{11}(\vec{k}_1 - \vec{q})\right)_{\colon,\rho_2\tau_2}$ denotes the $\rho_2\tau_2$-column of the matrix $V_{11}(\vec{k}_1 - \vec{q})$ (where all rows in this column are retained). 

Furthermore,
Define $W^\alpha(\vec{k},\vec{q}) = V^\dag_{12}(\vec{k}_1)\left(I(\vec{q})\otimes \sigma^\alpha\right)
V_{11}(\vec{k}_1-\vec{q})$, we have
\begin{equation}\label{h55}
\mathcal{S}^{\alpha}(\vec{q})
=\frac{1}{N} \sum_{\vec{k}_1}\mathrm{Tr}\left[ W^\alpha(\vec{k}_1,\vec{q})\left(W^\alpha(-\vec{k}_1+\vec{q},\vec{q})\right)^*+ W^\alpha(\vec{k}_1,\vec{q}) \left(W^\alpha(\vec{k}_1,\vec{q})\right)^\dag\right].
\end{equation}
Notice that if we pick the four terms relevant for a given $\vec{k}_1$ (we omit the $\alpha$ index for simplicity):
\begin{equation}
\begin{aligned}
&\mathrm{Tr}\left[ W(\vec{k}_1,\vec{q})W^*(-\vec{k}_1+\vec{q},\vec{q})+ W(\vec{k}_1,\vec{q}) W^\dag(\vec{k}_1,\vec{q})\right]
+
\mathrm{Tr}\left[ W(-\vec{k}_1+\vec{q},\vec{q})W^*(\vec{k}_1,\vec{q})+ W(-\vec{k}_1+\vec{q},\vec{q}) W^\dag(-\vec{k}_1+\vec{q},\vec{q})\right]\\
&=\mathrm{Tr}\left[ W(\vec{k}_1,\vec{q})W^*(-\vec{k}_1+\vec{q},\vec{q})+ W(\vec{k}_1,\vec{q}) W^\dag(\vec{k}_1,\vec{q})\right]
+
\mathrm{Tr}\left[ \left(W(-\vec{k}_1+\vec{q},\vec{q})W^*(\vec{k}_1,\vec{q})+ W(-\vec{k}_1+\vec{q},\vec{q}) W^\dag(-\vec{k}_1+\vec{q},\vec{q})\right)^T\right]\\
&=\mathrm{Tr}\left[ W(\vec{k}_1,\vec{q})W^*(-\vec{k}_1+\vec{q},\vec{q})+ W(\vec{k}_1,\vec{q}) W^\dag(\vec{k}_1,\vec{q})\right]
+
\mathrm{Tr}\left[ W^\dag(\vec{k}_1,\vec{q})W^T(-\vec{k}_1+\vec{q},\vec{q})+ W^*(-\vec{k}_1+\vec{q},\vec{q}) W^T(-\vec{k}_1+\vec{q},\vec{q})\right]\\
&=\mathrm{Tr}\left[ W(\vec{k}_1,\vec{q})W^*(-\vec{k}_1+\vec{q},\vec{q})+ W(\vec{k}_1,\vec{q}) W^\dag(\vec{k}_1,\vec{q})\right]
+
\mathrm{Tr}\left[ W^T(-\vec{k}_1+\vec{q},\vec{q})W^\dag(\vec{k}_1,\vec{q})+W^T(-\vec{k}_1+\vec{q},\vec{q}) W^*(-\vec{k}_1+\vec{q},\vec{q}) \right]\\
&=\mathrm{Tr}\left[\left( W(\vec{k}_1,\vec{q}) + W^T(-\vec{k}_1+\vec{q},\vec{q})\right)\left( W(\vec{k}_1,\vec{q}) + W^T(-\vec{k}_1+\vec{q},\vec{q})\right)^\dag\right],
\end{aligned}
\end{equation}
therefore Eq.~\eqref{h55} can be written as Eqs.~\eqref{SSFalpha}--\eqref{SSFalpha1} in the main text.

The dynamic structure factor is defined in Eq.~\eqref{DSF}. We have
\begin{equation}\label{dsf1}
\sum\limits_{\vec{r}_\mu,\vec{r}'_\nu}\left\langle \hat{S}^\alpha_{\vec{r}_\mu}(t) \hat{S}^\alpha_{\vec{r}'_\nu}\right\rangle e^{i\vec{q}\cdot(\vec{r}_\mu-\vec{r}'_\nu)}
=
\sum\limits_{\vec{k}_1,\vec{k}_3}\sum\limits_{\mu,\nu}e^{i\vec{q}\cdot(\hat{\varepsilon}_\mu-\hat{\varepsilon}_\nu)}\sum\limits_{\sigma_1,\sigma_2,\sigma_3,\sigma_4}
\left(\sigma^\alpha\right)_{\sigma_1,\sigma_2}\left(\sigma^\alpha\right)_{\sigma_3,\sigma_4}\left\langle b^\dag_{\vec{k}_1,\mu\sigma_1}(t) b_{\vec{k}_1-\vec{q},\mu\sigma_2}(t)b^\dag_{\vec{k}_3,\nu\sigma_3}b_{\vec{k}_3+\vec{q},\nu\sigma_4}\right\rangle,
\end{equation}
in Heisenberg representation, $\widetilde{b}(t) = e^{iHt}\widetilde{b} e^{-iHt}
=e^{i\lambda \widetilde{b}^\dag \widetilde{b} t} \widetilde{b} e^{-i\lambda \widetilde{b}^\dag \widetilde{b} t}
= \widetilde{b} e^{-i \lambda t}$,
we have
\begin{equation}
\left\langle
(\widetilde{b}^\dag_{i_1}(t)+\widetilde{b}_{i_1}(t))(\widetilde{b}^\dag_{i_2}(t)+\widetilde{b}_{i_2}(t))
(\widetilde{b}^\dag_{i_3}+\widetilde{b}_{i_3})(\widetilde{b}^\dag_{i_4}+\widetilde{b}_{i_4})\right\rangle
=\delta_{i_1,i_2}\delta_{i_3,i_4}+\delta_{i_1,i_3}e^{-i(\lambda_{i_1}+\lambda_{i_2})t}(\delta_{i_2,i_4}+\delta_{i_1,i_4}\delta_{i_2,i_3}),
\end{equation}
again we neglect the first term which vanishes in dynamical structure factor due to time reversal symmetry. So that
\begin{equation}
\begin{aligned}
&\left\langle b^\dag_{\vec{k}_1,\mu\sigma_1}(t) b_{\vec{k}_1-\vec{q},\mu\sigma_2}(t)b^\dag_{\vec{k}_3,\nu\sigma_3}b_{\vec{k}_3+\vec{q},\nu\sigma_4}\right\rangle\\
&=
\sum\limits_{\rho_1,\rho_2,\rho_3,\rho_4}\sum\limits_{\tau_1,\tau_2,\tau_3,\tau_4}\left(V_{12}(\vec{k}_1)\right)^*_{\mu\sigma_1,\rho_1\tau_1}
\left(V_{11}(\vec{k}_1-\vec{q})\right)_{\mu\sigma_2,\rho_2\tau_2}
\left(V_{11}(\vec{k}_3)\right)^*_{\nu\sigma_3,\rho_3\tau_3}
\left(V_{12}(\vec{k}_3+\vec{q})\right)_{\nu\sigma_4,\rho_4\tau_4}\cdot\\
&\qquad\qquad\qquad\qquad\quad\qquad\left\langle \widetilde{b}_{-\vec{k}_1,\rho_1\tau_1}(t)\widetilde{b}_{\vec{k}_1-\vec{q},\rho_2\tau_2}(t) \widetilde{b}_{\vec{k}_3,\rho_3\tau_3}^\dag \widetilde{b}^\dag_{-(\vec{k}_3+\vec{q}),\rho_4\tau_4}\right\rangle
\\
&=
\sum\limits_{\rho_1,\rho_2,\rho_3,\rho_4}\sum\limits_{\tau_1,\tau_2,\tau_3,\tau_4}\left(V_{12}(\vec{k}_1)\right)^*_{\mu\sigma_1,\rho_1\tau_1}
\left(V_{11}(\vec{k}_1-\vec{q})\right)_{\mu\sigma_2,\rho_2\tau_2}
\left(V_{11}(\vec{k}_3)\right)^*_{\nu\sigma_3,\rho_3\tau_3}
\left(V_{12}(\vec{k}_3+\vec{q})\right)_{\nu\sigma_4,\rho_4\tau_4}\cdot\\
&\quad e^{-i (\lambda_{-\vec{k}_1,\rho_1\tau_1}+\lambda_{\vec{k}_1-\vec{q},\rho_2\tau_2}) t}
\left(\delta_{(-\vec{k}_1,\rho_1\tau_1),(\vec{k}_3,\rho_3\tau_3)}
\delta_{(\vec{k}_1-\vec{q},\rho_2\tau_2),(-(\vec{k}_3+\vec{q}),\rho_4\tau_4)}+
\delta_{(-\vec{k}_1,\rho_1\tau_1),(-(\vec{k}_3+\vec{q}),\rho_4\tau_4)}\delta_{(\vec{k}_1-\vec{q},\rho_2\tau_2),(\vec{k}_3,\rho_3\tau_3)}\right),
\end{aligned}
\end{equation}

plug this equation into Eq.~\eqref{dsf1} we obatin Eq.~\eqref{dsf}.

We see that the dynamic structure factor simply ``disperses'' the static structure factor according to the energy level of each matrix element of $V^\dag_{12}(\vec{k}_1)\left(I(\vec{q})\otimes \sigma^\alpha\right) V_{11}(\vec{k}_1-\vec{q})$.


\twocolumngrid

\bibliography{pyrochlore_PSG} 

\end{document}